\documentclass[12pt]{iopart}

\usepackage{iopams}

\usepackage{graphicx}
\usepackage{color}
\usepackage{amsfonts}
\usepackage{amssymb}
\usepackage{amsthm}
\usepackage[T1]{fontenc}
\usepackage[utf8]{inputenc}
\usepackage{lmodern}
\usepackage{hyperref}
\usepackage[sort&compress,numbers]{natbib}
\def\newblock{\hskip .11em plus .33em minus .07em}

\ifx\bibfont\relax
\newcommand{\bibfont}{\footnotesize}
\else
\renewcommand{\bibfont}{\footnotesize}
\fi

\newcommand{\codename}[1]{\texttt{#1}}
\newcommand{\beq}[1]{\begin{equation} #1 \end{equation}}
\newcommand{\dF}{{^{^*}\!\!F}}
\newcommand{\Bcon}{{\mathcal B}}
\newcommand{\Econ}{{\mathcal E}}
\newcommand{\GRHydro}{\codename{GRHydro}~}

\newcommand{\eqref}[1]{\eref{#1}}
\newcommand{\lVert}{\left|}
\newcommand{\rVert}{\right|}
\newcommand{\text}[1]{{\rm #1}}
\newcommand{\operatorname}[1]{\mathop{\rm#1}}
\def\aap{Astron. Astrophys. }
\def\prl{Phys. Rev. Lett.}
\def\prd{Phys. Rev. D} 
\def\apj{Astrophys. J.}

\begin{document}

\title[GRHydro: A new open source GRMHD code for the Einstein Toolkit]{GRHydro:\\ A new open source general-relativistic magnetohydrodynamics code for the Einstein \makebox[0pt][l]{Toolkit}}

\date{\today}

\author{Philipp M\"osta$^1$,
        Bruno C. Mundim$^{2,3}$,
        Joshua A. Faber$^3$,
        Roland Haas$^{1,4}$,
        Scott C. Noble$^3$,
        Tanja Bode$^{8,4}$,
        Frank Löffler$^5$,
        Christian D. Ott$^{1,5}$,
        Christian Reisswig$^1$,
        Erik Schnetter$^{6,7,5}$}
\address{$^1$ TAPIR, California Institute of Technology, Pasadena, CA 91125, USA}
\address{$^2$ Max-Planck-Institut f\"{u}r Gravitationsphysik, Albert-Einstein-Institut, D-14476 Golm, Germany}
\address{$^3$ Center for Computational Relativity and Gravitation and  School of Mathematical Sciences, Rochester Institute of Technology, Rochester, NY 14623, USA}
\address{$^4$ Center for Relativistic Astrophysics, School of Physics, Georgia Institute of Technology, Atlanta, GA 30332, USA}
\address{$^5$ Center for Computation \& Technology, Louisiana State University, Baton Rouge, LA 70803, USA}
\address{$^6$ Perimeter Institute for Theoretical Physics, Waterloo, ON N2L 2Y5, Canada}
\address{$^7$ Department of Physics, University of Guelph, Guelph, ON N1G 2W1, Canada}
\address{$^8$ Theoretische Astrophysik, Institut f\"ur Astronomie und Astrophysik, Universit\"at  T\"ubingen, 72076 T\"ubingen, Germany}
\ead{pmoesta@tapir.caltech.edu}

\begin{abstract}
We present the new general-relativistic magnetohydrodynamics (GRMHD)
capabilities of the Einstein Toolkit, an open-source community-driven
numerical relativity and computational relativistic astrophysics
code. The GRMHD extension of the Toolkit builds upon previous releases
and implements the evolution of relativistic magnetised fluids in the
ideal MHD limit in fully dynamical spacetimes using the same
shock-capturing techniques previously applied to hydrodynamical
evolution.  In order to maintain the divergence-free character of the
magnetic field, the code implements both hyperbolic divergence
cleaning and constrained transport schemes.  We present test results
for a number of MHD tests in Minkowski and curved spacetimes.
Minkowski tests include aligned and oblique planar shocks, cylindrical
explosions, magnetic rotors, Alfv\'en waves and advected loops, as
well as a set of tests designed to study the response of the
divergence cleaning scheme to numerically generated monopoles.  We
study the code's performance in curved spacetimes with spherical
accretion onto a black hole on a fixed background spacetime and in
fully dynamical spacetimes by evolutions of a magnetised polytropic
neutron star and of the collapse of a magnetised stellar core.  Our
results agree well with exact solutions where these are
available and we demonstrate convergence.  All code and input files
used to generate the results are available on
\url{http://einsteintoolkit.org}. This makes our work fully
reproducible and provides new users with an introduction to
applications of the code.
\end{abstract}


\maketitle

\section{Introduction}

The recent years have seen rapid developments in the field of
numerical relativity.  Beginning with the first fully
general-relativistic (GR) simulations of binary neutron star (NS)
mergers by Shibata and Uryu in 1999~\cite{Shibata:1999wm}, fully
dynamical general-relativistic hydrodynamics (GRHD) has been explored
by a growing number of research groups.  A major step forward occurred
2005, when three independent groups developed two different
techniques, the generalised harmonic gauge
formalism~\cite{Pretorius:2005gq} and the so-called ``moving
puncture'' method~\cite{Baker:2005vv,Campanelli:2005dd}, to evolve
vacuum spacetimes containing black holes (BH) without encountering
numerical instabilities that mark the end of a numerical simulation.
Since then, these evolution schemes have been incorporated in
hydrodynamic simulations and now one or the other is used in
essentially every full GR calculation of merging binaries, as well as
other astrophysical phenomena involving dynamical spacetimes, such
as stellar collapse and BH formation (e., \cite{ott:11}).

A more recent advance concerns the incorporation of magnetic fields,
particularly in the ideal magnetohydrodynamics (MHD) limit, into
dynamical simulations.  Building upon work that originally focused on
astrophysical systems in either Newtonian gravity or fixed GR
backgrounds, as is appropriate for studies of accretion disks
(see~\cite{Font:2007zz,Balbus:1998zz,Hawley2009apss} for reviews of
the topic), MHD evolution techniques have been incorporated into
dynamical GR codes and used to study the evolution of NS-NS
(e.g., \cite{Anderson:2008zp,Giacomazzo:2009mp,Giacomazzo:2010bx,Liu:2008xy})
and BH-NS mergers~(e.g., \cite{Chawla:2010sw,Etienne:2011ea}),
self-gravitating tori around black
holes~\cite{Shibata:2007gp,Shibata:2012ty}, neutron star collapse
(e.g.,~\cite{Duez:2005cj,Lehner:2011aa}), stellar core collapse
\cite{Shibata:2006hr}, and the evolution of magnetised plasma around
merging binary BHs~(e.g.,~\cite{Giacomazzo:2012iv,Farris:2012ux}).
Given their rapidly maturing capabilities and widespread use, many
numerical relativity codes have been made public, either in total or
in part.  The largest community-based effort to do so is the Einstein
Toolkit consortium~\cite{Loffler:2011ay,EinsteinToolkit:web}, which
has as one of its goals to provide a free, publicly available,
community-driven general-relativistic MHD (GRMHD) code capable of
performing simulations that include realistic treatments of
matter, electromagnetic fields, and gravity.  The code is built upon
several open-source components that are widely used throughout the
numerical relativity community, including the {\tt Cactus}
computational infrastructure
\cite{Cactuscode:web,Goodale:2002a,CactusUsersGuide:web}, the {\tt
  Carpet} adaptive mesh refinement (AMR) code~\cite{Schnetter:2003rb,
  Schnetter:2006pg,CarpetCode:web}, the {\tt McLachlan} GR evolution
code~\cite{Brown:2008sb,Reisswig:2010cd}, 
and the GRHD code {\tt GRHydro}~\cite{Loffler:2011ay}, which has been
developed starting from a public variant of the {\tt Whisky}
code~\cite{Baiotti:2004wn,Hawke:2005zw,Baiotti:2010zf,Whisky:web}. Many
of the details of the Einstein Toolkit may be found
in~\cite{Loffler:2011ay}, which describes the routines used to provide
the supporting computational infrastructure for grid setup and
parallelization, constructing initial data, evolving dynamical GRHD
configurations, and analysing simulation outputs. An extension of the
Einstein Toolkit to general multi-patch grids with Cartesian and
curvilinear geometry is discussed in~\cite{reisswig:13a}.

In this paper, we describe the newest component of the Einstein
Toolkit: a MHD evolution scheme incorporated into the {\tt GRHydro}
module.  Using what is known as the \emph{Valencia} formulation
\cite{Marti:1991wi,Banyuls:1997zz,Ibanez:2001:godunov} of the GRMHD
equations, the new code evolves magnetic fields in fully dynamical GR
spacetimes under the assumptions of ideal MHD, i.e., the resistivity
is taken to be zero and electric fields vanish in the comoving frame
of the fluid.  While the evolution equations themselves are easily
cast into the same flux-conservative form as the other hydrodynamics
equations, the primary challenge for evolving magnetic fields is
numerically maintaining the divergence-free constraint of the magnetic
field.
For this, we have implemented a variant of the constrained transport
approach~\cite{Toth:00} and a hyperbolic ``divergence cleaning''
technique, similar to that discussed in
\cite{Liebling:2010bn,Penner:2010px}.

The {\tt GRHydro} module of the Einstein Toolkit is a new,
independently written code, and---in particular---the development is
completely independent of the {\tt WhiskyMHD}
code~\cite{Giacomazzo:2007ti}, since the {\tt GRHydro} development
started from the GRHD (non-MHD) version of the {\tt Whisky} code.
Given its close interaction with the hydrodynamic evolution, the MHD
routines are contained within the {\tt GRHydro} package (or ``thorn'',
in the language of {\tt Cactus}-based routines). It is developed in a
public repository, along with other components of the Einstein
Toolkit, and it is publicly available under the same open-source
licensing terms as other components in the Toolkit.  Given the public nature
of the project, the code release includes the subroutines themselves,
complete documentation for their use, and parameter files needed to
reproduce the tests shown here. Code and parameter files are available
on the Einstein Toolkit web page, \url{http://einsteintoolkit.org}.

In order to validate our GRMHD implementation, we include a number of
flat-space tests whose solutions are either known exactly or
approximately, and which have frequently been used as a testbed for
other GRMHD codes.  These include planar and cylindrical shocktubes, a
rotating bar threaded by a magnetic field (the ``magnetic rotor''),
propagating Alfv\'en waves, and the advection of a flux loop.  We test
the code's ability to handle static curved spacetime by studying Bondi
accretion onto a Schwarzschild BH and perform GRMHD simulations with
dynamical spacetime evolution for a magnetised TOV star and for the
collapse of a rotating magnetised stellar core to a rotating neutron
star.

This paper is structured as follows: in section~\ref{sec:codes}, we
describe the Einstein Toolkit and briefly discuss other relativistic
MHD codes.  In section~\ref{sec:valencia}, we describe the Valencia
formulation of the GRMHD equations, and in section~\ref{sec:numerical}
the numerical techniques used in \codename{GRHydro}.  In
section~\ref{sec:tests}, we describe the tests we have carried out
with the new code.  Finally, in section~\ref{sec:conclusions}, we
summarise and discuss future directions of the Einstein Toolkit.

\section{Numerical Relativistic Hydrodynamics and GRMHD
codes}\label{sec:codes}

The {\tt GRHydro} code is, to the best of our knowledge, the first
publicly released 3D GRMHD code capable of evolving
configurations in fully dynamical spacetimes.  Still, its development
makes use of code and techniques from a number of other public codes,
which we summarise briefly here.  The most obvious of these is the
Einstein Toolkit, within which its development has taken place.  As we
discuss below, the Toolkit itself is composed of tens of different
modules developed by a diverse set of authors whose number is
approaching 100.
The MHD techniques, while independently implemented in nearly all
cases, rely heavily on both the numerical techniques found within the
{\tt GRHydro} package~\cite{Loffler:2011ay}, and thus the original
{\tt Whisky} code~\cite{Baiotti:2004wn,Hawke:2005zw,Baiotti:2010zf,Whisky:web},
as well as the publicly available {\tt HARM} code~\cite{Gammie:2003rj}.

\subsection{The Einstein Toolkit}

The Einstein Toolkit includes components to aid in performing
relativistic astrophysical simulations that range from physics
modelling (initial conditions, evolution, analysis) to infrastructure
modules (grid setup, parallelization, I/O) and include related tools
(workflow management, file converters, etc.). The overall goal is to
provide a set of well-documented, well-tested, state-of-the-art
components for these tasks, while allowing users to replace these
components with their own and/or add additional ones.

Many components of the Einstein Toolkit use the \texttt{Cactus}
Computational
Toolkit~\cite{Cactuscode:web,Goodale:2002a,CactusUsersGuide:web}, a
software framework for high-performance computing. \codename{Cactus}
simplifies designing codes in a modular (``component-based'') manner,
and many existing \codename{Cactus} modules provide infrastructure
facilities or basic numerical algorithms such as coordinates, boundary
conditions, interpolators, reduction operators, or efficient I/O in
different data formats.

The utilities contained in Einstein Toolkit, e.g., help manage
components~\cite{Seidel:2010aa,Seidel:2010bb}, build code and submit
simulations on supercomputers~\cite{Thomas:2010aa,SimFactory:web}, or
provide remote debuggers~\cite{Korobkin:2011tg} and post-processing
and visualisation interfaces for \codename{VisIt}~\cite{Childs:2005ACS}.

Adaptive mesh refinement (AMR) and multi-block methods are implemented
by \codename{Carpet}~\cite{Schnetter:2003rb, Schnetter:2006pg,
  CarpetCode:web} which also provides MPI
parallelization. \codename{Carpet} supports Berger-Oliger style
(``block-structured'') adaptive mesh refinement~\cite{Berger:1984zza}
with sub-cycling in time as well as certain additional features
commonly used in numerical relativity (see \cite{Schnetter:2003rb} for
details). \codename{Carpet} provides also the respective prolongation
(interpolation) and restriction (projection) operators to move data
between coarse and fine grids.  \codename{Carpet} has been
demonstrated to scale efficiently up to several thousand
cores~\cite{Loffler:2011ay}.

\codename{Carpet} supports both vertex-centred and cell-centred
AMR\@. In vertex-centred AMR, each coarse grid point (cell Center)
coincided with a fine grid point (cell Center), which simplifies
certain operations (e.g.\ restriction, and also visualisation). In
this paper, we present results obtained with vertex-centred AMR only.
In cell-centred AMR, coarse grid \emph{cell faces} are aligned with
fine grid faces, allowing in particular exact conservation across AMR
interfaces. For a more detailed discussion of cell-centred AMR in
hydrodynamics simulations with the Einstein Toolkit we refer
the reader to~\cite{reisswig:13a}.
  
The evolution of the spacetime metric in the Einstein Toolkit is
handled by the \codename{McLachlan}
package~\cite{Brown:2008sb,Reisswig:2010cd}. This code is
autogenerated by Mathematica using the \codename{Kranc}
package~\cite{Husa:2004ip,Lechner:2004cs,Kranc:web}, implementing the
Einstein equations via a $3+1$-dimensional split using the BSSN
formalism~\cite{Nakamura:1987zz,Shibata:1995we,Baumgarte:1998te,Alcubierre:2000xu,Alcubierre:2002kk}.
The BSSN equations are finite-differenced at a user-specified order of
accuracy, and coupling to hydrodynamic variables is included via the
stress-energy tensor.  The time integration and coupling with
curvature are carried out with the Method of Lines (MoL)
\cite{Hyman:1976cm}, implemented in the \codename{MoL} package.
  
Hydrodynamic evolution techniques are provided in the Einstein Toolkit
by the \GRHydro package, a code derived from the public
\codename{Whisky} GRHD
code~\cite{Baiotti:2004wn,Hawke:2005zw,Baiotti:2010zf,Whisky:web}. The
code is designed to be modular, interacting with the vacuum metric
evolution only by contributions to the stress-energy tensor and by the
local values of the metric components and extrinsic curvature. It uses
a high-resolution shock capturing finite-volume scheme to evolve
hydrodynamic quantities, with several different methods for
reconstruction of states on cell interfaces and Riemann solvers. It
also assumes an atmosphere to handle low-density and vacuum
regions. In particular, a density floor prevents numerical errors from
developing near the edge of matter configurations.  Boundaries and
symmetries are handled by registering hydrodynamic variables with the
appropriate \codename{Cactus} routines.  Passive ``tracers'' or local
scalars that are advected with the fluid flow may be used, though
their behaviour is generally unaffected by the presence or absence of
magnetic fields.
  
In adding MHD to the pre-existing hydrodynamics code, our design
philosophy has been to add the new functionality in such a way that
the original code would run without alteration should MHD not be
required.  For nearly all of the more complicated routines in the
code, this involved creating parallel MHD and non-MHD routines, while
for some of the more basic routines we simply branch between different
code sections depending on whether MHD is required or not.  While this
requires care when the code is updated, since some changes may need to
be implemented twice, it does help to insure that users performing
non-MHD simulations will be protected against possible errors
introduced in the more frequently changing MHD routines.
 
\subsection{Other relativistic MHD codes}
 
In extending the Einstein Toolkit to include MHD functionality, we
incorporated techniques that have previously appeared in the literature.
Of particular importance is the {\tt HARM} code
\cite{Harmcode:web,Gammie:2003rj,Noble:2008tm}, a free, publicly
available GRMHD code that can operate on fixed background metrics,
particularly those describing spherically symmetric or rotating black
holes.  Our routines for converting conservative variables back into
primitive ones are adapted directly from {\tt HARM} (\cite{Noble:2005gf};
see section~\ref{sec:numerical_c2p} below), as is the approximate technique
used to calculate wave speeds for the Riemann solver
(\cite{Gammie:2003rj}; see section~\ref{sec:numerical_riemann} below).
 
 Since many groups have introduced relativistic MHD codes, there are a
 number of standard tests that may be used to gauge the performance of
 a particular code.  Many of these are compiled in the descriptions of
 the {\tt HARM} code mentioned above, as well as papers describing the
 {\tt Athena} MHD code~\cite{Stone:2008mh,Athenacode:web}, the {\tt Echo}
 GRMHD code~\cite{DelZanna:2007pk}, the Tokyo/Kyoto group's GRMHD
 code~\cite{Shibata:2005gp}, the UIUC GRMHD
 code~\cite{Duez:2005sf,Etienne:2010ui}, the {\tt WhiskyMHD}
 code~\cite{Giacomazzo:2007ti}, and the LSU GRMHD
 code~\cite{Anderson:2006ay}.  We have implemented several of these
 tests, covering all aspects of our code, as we describe in
 section~\ref{sec:tests} below.
 
\section{The Valencia formulation of ideal MHD} \label{sec:valencia}

While \codename{GRHydro} does not technically assume a particular
evolution scheme to evolve the field equations for the GR metric, the
most widely used choice for the ET code is the BSSN formalism
\cite{Shibata:1995we,Baumgarte:1998te}, the particular implementation of
which is provided by the {\tt McLachlan} code
\cite{Brown:2008sb,Reisswig:2010cd}.  We do assume that the metric is
known in the ADM (Arnowitt-Deser-Misner) form~\cite{Arnowitt:1962hi}, in
which we have
\begin{equation}
\rmd s^2=g_{\mu\nu}\rmd x^\mu \rmd x^\nu\equiv (-\alpha^2+\beta_i\beta^i)\rmd t^2+2\beta_i \rmd t~\rmd x^i+\gamma_{ij} \rmd x^i\rmd x^j\,\,,\label{eq:adm}
\end{equation}
with $g_{\mu\nu}$, $\alpha$, $\beta^i$, and $\gamma_{ij}$ being the
spacetime 4-metric, lapse function, shift vector, and spatial
3-metric, respectively. Note that we are assuming spacelike signature,
so the Minkowski metric in flat space reads $\eta_{\mu\nu} =
\mathrm{diag}(-1,1,1,1)$. Roman indices are used for 3-quantities and
4-quantities are indexed with Greek characters. We work in units of
$c = G = M_\odot = 1$ unless explicitly stated otherwise.

\codename{GRHydro} employs the {\em ideal MHD} approximation --
fluids have infinite conductivity and there is no charge separation.
Thus, electric fields $E_\nu = u_{\mu} F^{mu\nu}$ in the rest frame 
of the fluid vanishes and
ideal MHD corresponds to imposing the following conditions:
\begin{equation}
u_\mu F^{\mu\nu}=0\,\,.
\end{equation}
Note that throughout this paper we rescale the magnitude of the relativistic 
Faraday tensor $F^{\mu\nu}$ and its dual $\dF^{\mu\nu}\equiv
\frac{1}{2}\epsilon^{\mu\nu\kappa\lambda}F_{\kappa\lambda}$, as well as the
magnetic and electric fields, by a factor of ${1}/{\sqrt{4\pi}}$ to
eliminate the need to include the permittivity and permeability of free space
in cgs-Gaussian units. 

Magnetic fields enter into the equations of hydrodynamics by their
contributions to the stress-energy tensor and through the solutions of
Maxwell's equations.  The hydrodynamic and electromagnetic contributions
to the stress-energy tensor are given, respectively, by 
\beq{
T^{\mu \nu}_{\rm H} = \rho h u^\mu u^\nu + P g^{\mu \nu}  = \left( \rho + \rho \epsilon + P \right) u^\mu u^\nu + P g^{\mu \nu}  
\quad , \label{hydro-stress-tensor}
}
and 
\beq{
T^{\mu \nu}_{\rm EM} = F^{\mu \lambda} {F^{\nu}}_\lambda  - \frac{1}{4} g^{\mu \nu} F^{\lambda \kappa} F_{\lambda \kappa} 
= b^2 u^\mu u^\nu - b^\mu b^\nu  + \frac{b^2}{2} g^{\mu \nu} \quad , \label{em-stress-tensor} 
}
where $\rho$, $\epsilon$, $P$, $u^\mu$, and $h\equiv
1+\epsilon+P/\rho$ are the fluid rest mass density, specific
internal energy, gas pressure, 4-velocity, and specific enthalpy, respectively, and
$b^\mu$ is the magnetic $4$-vector (the projected component of the
Maxwell tensor parallel  to the 4-velocity of the fluid):
\beq{
b^\mu  = u_\nu \dF^{\mu \nu} \quad . \label{magnetic-4-vector}
}
Note that  $b^2 = b^\mu b_\mu = 2 P_m$,   where $P_m$ is the magnetic pressure. 
Combined, the stress-energy tensor takes the form:
\begin{eqnarray}
T^{\mu \nu} 
  &=&  \left( \rho +  \rho \epsilon + P + b^2 \right) u^\mu u^\nu + 
       \left(P + \frac{b^2}{2} \right) g^{\mu \nu} - b^\mu b^\nu  
  \\\nonumber
  &\equiv& \rho h^*u^\mu u^\nu + P^* g^{\mu \nu} - 
           b^\mu b^\nu,  \label{mhd-stress-energy-tensor}
\end{eqnarray}
where we define the magnetically modified pressure and enthalpy, 
$P^*=P+P_m=P+b^2/{2}$ and $h^*\equiv 1+\epsilon+\left(P + b^2\right)/\rho$,
respectively.

The spatial magnetic field (living on the spacelike 3-hypersurfaces),
is defined as the Eulerian component of the Maxwell tensor \beq{ B^i =
  n_\mu \dF^{i \mu} = - \alpha \dF^{i 0} \,\,
  ,\label{spatial-magnetic-field-vector} } where
$n_\mu=[-\alpha,0,0,0]$ is the normal vector to the hypersurface.

The equations of ideal GRMHD evolved by \codename{GRHydro} are
derived from the local GR conservation laws of mass and
energy-momentum,
\begin{equation}
  \nabla_{\!\mu} J^\mu = 0\,\,, \qquad \nabla_{\!\mu} T^{\mu \nu} = 0\,\,,
  \label{eq:equations_of_motion_gr}
\end{equation}
where $ \nabla_{\!\mu} $ denotes the covariant derivative with respect
to the 4-metric and $J^{\,\mu} = \rho u^{\,\mu} $ is the mass current, and from Maxwell's equations,
\begin{eqnarray}
\nabla_\nu \dF^{\mu\nu} = 0\,\,. \label{eq:maxwell_equations}
\end{eqnarray}

The \codename{GRHydro} scheme is written in a first-order
hyperbolic flux-conservative evolution system for the conserved
variables $D$, $S^i$, $\tau$, and $\Bcon^i$, defined in terms of the
primitive variables $\rho, \epsilon, v^i$, and $B^i$ such that
\begin{eqnarray}
  D &=& \sqrt{\gamma} \rho W\,\,,\label{eq:p2c1}\\
  S_j &=& \sqrt{\gamma} \left(\rho h^* W^{\,2} v_j-\alpha b^0b_j\right)\,\,,\label{eq:p2c2}\\
  \tau &=& \sqrt{\gamma} \left(\rho h^* W^2 - P^*-(\alpha b^0)^2\right) - D\label{eq:p2c3}\,\,, \\
\Bcon^k&=&\sqrt{\gamma}B^k\,\,,\label{eq:p2c4}
\end{eqnarray}
where $ \gamma $ is the determinant of $\gamma_{ij} $. We choose a
definition of the 3-velocity $v^i$ that corresponds to the velocity seen
by an Eulerian observer at rest in the current spatial
3-hypersurface~\cite{York:1983aa},
\begin{equation}
v^i = \frac{u^i}{W} + \frac{\beta^i}{\alpha}\,\,,
\label{eq:vel}
\end{equation}
and $W \equiv (1-v^i v_i)^{-1/2}$ is the Lorentz factor.  
Note that
$v^i,~B^i,~S^i$, and $\beta^i$ are 3-vectors,
and their indices are raised and lowered with the 3-metric, e.g.,
$v_i\equiv \gamma_{ij}v^j$.

The evolution scheme used in \codename{GRHydro} is often referred
to as the Valencia formulation
\cite{Marti:1991wi,Banyuls:1997zz,Ibanez:2001:godunov,Font:2007zz}.
Our notation here most closely follows that found
in~\cite{Anton:2005gi}.  The evolution system for the conserved
variables, representing~\eref{eq:equations_of_motion_gr} and the
spatial components of~\eref{eq:maxwell_equations}, is
\begin{equation}
  \frac{\partial \mathbf{U}}{\partial t} +
  \frac{\partial \mathbf{F}^{\,i}}{\partial x^{\,i}} =
  \mathbf{S}\,\,,
  \label{eq:conservation_equations_gr}
\end{equation}
with
\begin{eqnarray}
  \mathbf{U}  ~=~ & &[D, S_j, \tau,\Bcon^k]\,\,, \nonumber\\
  \mathbf{F}^{\,i} ~ = ~&& \alpha\times
  \left[\begin{array}{c} 
          D \tilde{v}^{\,i}\\ 
          S_j \tilde{v}^{\,i} + \sqrt{\gamma} P^* \delta^{\,i}_j - b_j\Bcon^i/W\\
          \tau \tilde{v}^{\,i} + \sqrt{\gamma} P^* v^{i}-\alpha b^0 \Bcon^i/W\\
          \Bcon^k\tilde{v}^i-\Bcon^i\tilde{v}^k 
        \end{array} \right]\,\,,   \label{eq:definition-of-flux} \\
  \mathbf{S} ~ =~ && \alpha \sqrt{\gamma} \times\left[\begin{array}{c}
         0, \\ 
         T^{\mu \nu} \left( \frac{\partial g_{\nu j}}{\partial x^{\,\mu}} - 
                      \Gamma^{\,\lambda}_{\mu \nu} g_{\lambda j} \right) \\
         \qquad\alpha \left( T^{\mu 0}
         \frac{\partial \ln \alpha}{\partial x^{\,\mu}} -
         T^{\mu \nu} \Gamma^{\,0}_{\mu \nu} \right)\\
         \vec{0} \end{array}\right]\,\,.
  \label{eq:definition-of-source}
\end{eqnarray}%
Here, $ \tilde{v}^{\,i} = v^{\,i} - \beta^i / \alpha $ and $
\Gamma^{\,\lambda}_{\mu \nu} $ are the 4-Christoffel symbols.  

The time component of~\eref{eq:maxwell_equations} yields the 
condition that the magnetic field is divergence-free, the
``no-monopoles'' constraint:
\beq{\nabla \cdot B  \equiv \frac{1}{\sqrt{\gamma}} \, \partial_i \left( \sqrt{\gamma} B^i \right) =  0\,\,,\label{eq:nomonopole}}
which also implies
\beq{\partial_i  \Bcon^i = 0\,\,.\label{eqn:divB_constraint}}
In practice, we implement two different methods to actively enforce this
constraint.
In the ``divergence cleaning'' technique we
include the ability to modify
the magnetic field evolution by introducing a new field variable that
dissipates away numerical divergences,
 which we discuss in detail in
section~\ref{sec:numerical_divclean} below.
An alternative method, commonly called ``constrained transport'', instead
carefully constructs a numerical method so that the
constraint~\eqref{eqn:divB_constraint} is satisfied to round-off error at the
discrete level. We discuss this method in
section~\ref{sec:constrained_transport}.

\section{Numerical Methods}\label{sec:numerical}

{\tt GRHydro}'s GRMHD code  uses the same infrastructure and backend routines as
its
pure general relativistic hydrodynamics variant~\cite{Loffler:2011ay}.  In the
following, we focus on the discussion of the numerical methods used in
the extension to GRMHD.

{\tt GRHydro}'s GRMHD code implements reconstruction of fluid and magnetic
field variables to cell
interfaces for all of the methods present in the
original {\tt GRHydro} code: TVD (total variation diminishing) (e.g.,~\cite{toro:99}),
PPM (piecewise parabolic method)~\cite{Colella:1982ee}, and ENO
(essentially non-oscillatory)~\cite{Harten:1987un,Shu:1999ho}. In
addition, we added enhanced PPM (ePPM, as described
in~\cite{reisswig:13a,mccorquodale:11}), WENO5 (5\textsuperscript{th} order
weighted-ENO)~\cite{shu:98}, and MP5 (5\textsuperscript{th} order monotonicity
preserving)~\cite{suresh:97} both in GRMHD and pure GR hydrodynamics
simulations. We discuss the different reconstruction methods in
section~\ref{sec:numerical_reconstruction}.  The code computes the solution of
the local
Riemann problems at cell interfaces using the HLLE
(Harten-Lax-van Leer-Einfeldt)~\cite{Harten:1983on,Einfeldt:1988og}
approximate Riemann solver discussed in
section~\ref{sec:numerical_riemann}. We implement the conversion of conserved
to
primitive variables for arbitrary equations of state
(EOS), including polytropic, $\Gamma$-law, hybrid
polytropic/$\Gamma$-law, and microphysical finite-temperature
EOS. We summarise the new methods for GRMHD in
section~\ref{sec:numerical_c2p}.  An important aspect of any (GR)MHD
scheme is the numerical method used to preserve the divergence free
constraint. {\tt GRHydro} implements both the hyperbolic divergence
cleaning method (e.g.,~\cite{Liebling:2010bn,Penner:2010px}) and a
variant of the constrained transport method (e.g.,
\cite{Toth:00,Giacomazzo:2007ti,Giacomazzo:2007fe}), which are both
discussed in detail in section~\ref{sec:divtreatment}.

\subsection{Evaluation of magnetic field expressions}

The MHD code stores the values of both the primitive B-field vector,
$B^i$ ({\tt Bvec} in the code), and the evolved conservative field
$\Bcon^i$ ({\tt Bcons} in the code) in the frame of the Eulerian
observers. For analysis purposes we include options to compute the
magnetic field in fluid's rest frame 
\begin{eqnarray}
b^0&=&\frac{WB^kv_k}{\alpha}\,\,,\\
b^i&=&\frac{B^i}{W}+W(B^kv_k)\left(v^i-\frac{\beta^i}{\alpha}\right)\,\,,\\
b^2&=&\frac{B^iB_i}{W^2}+(B^iv_i)^2\,\,.\\
\end{eqnarray}

\subsection{Reconstruction}
\label{sec:numerical_reconstruction}

In a finite-volume scheme, one evaluates fluxes at cell faces by
solving Riemann problems involving potentially discontinuous
hydrodynamic states on either side of the interface.
To construct these Riemann
problems, one must first obtain the fluid state on the left and right
sides of the interface.  The fluid state is known within cells in the
form of cell-averages of the hydrodynamical variables.
The {\em reconstruction} step interpolates the fluid state
from cell averaged values to values at cell interfaces without introducing
oscillations at shocks and other discontinuities.  It is possible to
reconstruct either primitive or conserved fluid variables, however
using the former makes it much easier to guarantee physically valid
results for which the pressure is positive and the fluid velocities
sub-luminal.

By assuming an orthogonal set of coordinates,
it is possible to reconstruct each coordinate direction
independently.  Thus, the fluid reconstruction reduces to a
one-dimensional problem.

We define $U^L_{i+1/2}$ to be the value of an element of our conservative variable state vector $\mathbf{U}$
 on the left side of the face between $U_i\equiv U(x_i,y,z)$
and $U_{i+1}\equiv U(x_{i+1},y,z)$, where $x_i$ is the $i$\textsuperscript{th} point in
the $x$-direction, and $U^R_{i+1/2}$  the value on the right side of the same face (coincident in space, but with a potentially different value).  
These quantities are computed directly from the primitive values on the face.

\subsubsection{TVD reconstruction.}
For total variation
diminishing (TVD) methods, we let 
\begin{equation}
U^L_{i+1/2} = U_i+\frac{F(U_i)\Delta x}{2};
\qquad
U^R_{i+1/2} = U_{i+1}-\frac{F(U_{i+1})\Delta x}{2}
\end{equation}
where $F(U_i)$ is a slope-limited gradient function, typically
determined by the values of $U_{i+1}-U_i$ and $U_i-U_{i-1}$, with a
variety of different forms of the slope limiter available.  In practice,
all try to accomplish the same task of preserving monotonicity and
removing the possibility of spuriously creating local extrema.
\codename{GRHydro} includes minmod, superbee~\cite{Roe:1986cb}, and
monotonized central~\cite{vanLeer:1977aa} limiters. 
TVD methods are second-order accurate in regions of smooth, monotonic flows.
At extrema and shocks, they reduce to first order.

\subsubsection{PPM reconstruction.}
The original piecewise parabolic method (oPPM,~\cite{Colella:1982ee})
uses quadratic functions to represent cell-averages from which new
states at cell interfaces are constructed.  The fluid state is
interpolated using a fourth-order polynomial.  A number of subsequent
limiter steps constrain parabolic profiles and preserve monotonicity
so that no new extrema can form.  The version implemented in \GRHydro
includes the steepening and flattening routines described in the
original PPM papers, with a simplified flattening procedure that
allows for fewer ghost points~\cite{Baiotti:2004wn,reisswig:13a}.  
The original PPM
always reduces to first order near local extrema and shocks.  The
enhanced PPM (ePPM) maintains high-order at local extrema that are
smooth~\cite{mccorquodale:11,reisswig:13a}.

\subsubsection{ENO reconstruction.}
Essentially non-oscillatory (ENO) methods use a divided differences
approach to achieve high-order accuracy via polynomial
interpolation~\cite{Harten:1987un,Shu:1999ho} without reducing the
order near extrema or shocks.  In the third-order case, two
interpolation polynomials with different stencil points are used.
Based on the smoothness of the field, one of the interpolation
polynomials is selected.

\subsubsection{WENO reconstruction}
Weighted essentially non-oscillatory (WENO) reconstruction
\cite{shu:98} is an improved algorithm based on the ENO approach.  The
drawback of ENO methods is that the order of accuracy is not maximal
for the available number of stencil points.  Furthermore, since a
single stencil is selected, ENO reconstruction is not continuous.
In addition, the
large number of required \emph{if statements} make ENO methods
unnecessary slow.  WENO reconstruction, on the other hand, attempts to
overcome all these drawbacks.  The essential idea is to combine all
possible ENO reconstruction stencils by assigning a weight to each
stencil.  The weight is determined by the local smoothness of the
flow.  When all weights are non-zero, the full set of stencil points
is used, and the maximum allowable order of accuracy for a given
number of stencil points is achieved. When the flow becomes less
smooth, some weights are suppressed, and a particular lower order
interpolation stencil dominates the reconstruction.  Using the same
number of stencil points as third-order ENO, WENO is fifth-order
accurate when the flow is smooth, and reduces to third order near
shocks and discontinuities.

We implement fifth-order WENO reconstruction based on an improved version presented in~\cite{Tchekhovskoy:2007zn}.
We introduce three interpolation polynomials that approximate $U^L_{i+1}$ from cell-averages of a given quantity $U_i$:
\begin{eqnarray} \label{eq:weno_stencils}
U^{L,1}_{i+1/2} &=& \frac{3}{8} U_{i-2} - \frac{5}{4}U_{i-1} + \frac{15}{8}U_i\,\,, \\
U^{L,2}_{i+1/2} &=-& \frac{1}{8}U_{i-1} + \frac{3}{4}U_{i} - \frac{3}{8}U_{i+1}\,\,, \\
U^{L,3}_{i+1/2} &=& \frac{3}{8}U_{i} + \frac{3}{4}U_{i+1} - \frac{1}{8}U_{i+2}\,\,.
\end{eqnarray}
Each of the three polynomials yields a third-order accurate approximation of $\mathbf{U}$ at the cell-interface.
By introducing a convex linear combination of the three interpolation polynomials,
\begin{equation}
U^L_{i+1/2} = w^{1}U^{L,1}_{i+1/2} + w^2U^{L,2}_{i+1/2} + w^{3}U^{L,3}_{i+1/2}\,\,,
\end{equation}
where the weights $w^i$ satisfy $\sum_i w^i=1$,
it is possible to obtain a fifth-order interpolation polynomial that spans all five stencil points $\left\{U_{i-2},\ldots,U_{i+2}\right\}$.
The weights $w^i$ are computed using so-called smoothness indicators $\beta^i$.
In the original WENO algorithm~\cite{shu:98}, they are given by
\begin{eqnarray}
\beta^1 &=& \frac{1}{3}(4u_{i-2}^2 - 19u_{i-2}u_{i-1} + 25u^2_{i-1} + 11u_{i-2}u_i - 31u_{i-1}u_i + 10u_i^2 )\,\,, \label{eq:weno-smoothness1} \\
\beta^2 &=& \frac{1}{3}(4u_{i-1}^2 - 13u_{i-1}u_{i} + 13u^2_{i} + 5u_{i-1}u_{i+1} - 13u_{i}u_{i+1} + 4u_{i+1}^2 )\,\,, \label{eq:weno-smoothness2} \\
\beta^3 &=& \frac{1}{3}(10u_{i}^2 - 31u_{i}u_{i+1} + 25u^2_{i+1} + 11u_{i}u_{i+2} - 19u_{i+1}u_{i+2} + 4u_{i+2}^2 )\,\,. \label{eq:weno-smoothness3}
\end{eqnarray}
Note that, in this section only, $\beta^i$ refers to the smoothness
indicators~\eref{eq:weno-smoothness1} and not to the shift
vector~\eref{eq:adm}.
The weights are then obtained via
\begin{equation}
w^i = \frac{\bar{w}^i}{\bar{w}^1+\bar{w}^2+\bar{w}^3}\,\,,\qquad \mathrm{with}\qquad \bar{w}^i=\frac{\gamma^i}{(\epsilon+\beta^i)^2}\,\,,
\end{equation}
where $\gamma^i=\left\{1/16,5/8,5/16\right\}$, and $\epsilon$ is a
small constant to avoid division by zero.  Unfortunately, the choice
of $\epsilon$ is scale dependent, and choosing a fixed number is
inappropriate for cases with large variation in scales.  The
improvement made by~\cite{Tchekhovskoy:2007zn} overcomes this problem
by modifying the smoothness indicators
\eqref{eq:weno-smoothness1}-\eqref{eq:weno-smoothness3}.
Following~\cite{Tchekhovskoy:2007zn}, we compute \textit{modified}
smoothness indicators via
\begin{equation}
\bar{\beta}^i=\beta^i + \epsilon \lVert U^2 \rVert + \delta\,\,,
\end{equation}
where $\lVert U^2 \rVert$ is the sum of the $U_j^2$ in the $i$-th
stencil, and $\delta$ is the smallest number that the chosen floating
point variable type can hold.  Now, the smoothness indicators depend
on the scale of the reconstructed field.  In our case, we set
\begin{equation}
\bar{\beta}^i=\beta^i + \epsilon (\lVert U^2 \rVert + 1)\,\,,
\end{equation}
with $\epsilon=10^{-26}$ and use these values instead of $\beta^i$ in
\eref{eq:weno-smoothness1}-\eref{eq:weno-smoothness3}.

\subsubsection{MP5 reconstruction}

Monotonicity-preserving fifth-order (MP5) reconstruction is based on a
geometric approach to maintain high-order at maxima that are
smooth. In contrast to WENO or ENO, MP5 makes use of limiters to a
high-order reconstruction polynomial similar to PPM to avoid
oscillations at shocks and discontinuities. The key advantages of MP5
reconstruction are that it preserves monotonicity and accuracy, and is
fast. In practice, it compares favourable to enhanced PPM and WENO in
terms of accuracy (see \ref{sec:recon-comp}).  MP5 reconstruction is carried out in two steps. In
a first step, a fifth-order polynomial is used to interpolate
cell-averages $U_i$ on cell interfaces $U^L_{i+1/2}$:
\begin{equation}
U^L_{i+1/2}=(2U_{i-2} - 13U_{i-1} + 47U_i + 27U_{i+1} - 3U_{i+2})/60\,\,.
\end{equation}
In a second step, the interpolated value $U^L_{i+1/2}$ is limited. 
Whether a limiter is applied is determined via the following condition.
First we compute 
\begin{equation}
U^{MP} = U_i + \operatorname{minmod}(U_{i+1}-U_{i}, \tilde{\alpha}\,(U_i-U_{i-1}))\,\,,
\end{equation}
where $\tilde{\alpha}$ is a constant, which we set to $4.0$, and where
\begin{equation}
\operatorname{minmod}(x,y) = \frac{1}{2} \left(\operatorname{sign}(x)+\operatorname{sign}(y)\right)\min(|x|,|y|)\,\,.
\end{equation} 
A limiter is \emph{not} applied when
\begin{equation} \label{eq:mp5-cond}
(U^L_{i+1/2}-U_i)(U^L_{i+1/2}-U^{MP}) \leq \epsilon \lVert U \rVert\,,
\end{equation}
where $\epsilon$ is a small constant, and $\lVert U \rVert$ is the
$L_2$ norm of $U_i$ over the stencil points $\left\{U_{i-2},\ldots,
U_{i+2}\right\}$.  Note that the norm factor is not
present in the original algorithm. We have added this to take into
account the scale of the reconstructed field.  Following~\cite{suresh:97}, we set
$\epsilon=10^{-10}$, though in some cases, it may be necessary to set
$\epsilon=0$ to avoid oscillations at strong shocks or contact discontinuities such as the
surface of a neutron star.

In case \eqref{eq:mp5-cond} does not hold, the following limiter
algorithm is applied.  First, we compute the second derivatives
\begin{eqnarray}
D_i^{-} &=& U_{i-2} - 2U_{i-1} + U_i\,\,, \\
D_i^{0} &=& U_{i-1} - 2U_{i} + U_{i+1}\,\,, \\
D_i^{+} &=& U_{i} - 2U_{i+1} + U_{i+2}\,\,.
\end{eqnarray}
Next, we compute
\begin{eqnarray}
D^{M4}_{i+1/2} &=& \operatorname{minmod}(4D_i^0-D_i^+, 4D_i^+-D_i^0, D_i^0, D_i^+)\,\,, \\
D^{M4}_{i-1/2} &=& \operatorname{minmod}(4D_i^0-D_i^-, 4D_i^--D_i^0, D_i^0, D_i^-)\,\,,
\end{eqnarray}
where the four-argument $\operatorname{minmod}$ function is given by
\begin{eqnarray}
\operatorname{minmod}(w,x,y,z) &=& \frac{1}{8}(\operatorname{sign}(w)+\operatorname{sign}(x)) \times
\\\nonumber 
&&|(\operatorname{sign}(w)+\operatorname{sign}(y))(\operatorname{sign(w)}+\operatorname{sign}(z))| \times
\\\nonumber
&&\min(|w|,|x|,|y|,|z|))\,\,.
\end{eqnarray}
We then compute
\begin{eqnarray}
U^{UL} &=& U_i + \alpha(U_i-U_{i+1})\,\,, \\
U^{AV} &=& \frac{1}{2}(U_i+U_{i+1})\,\,, \\
U^{MD} &=& U^{AV} - \frac{1}{2} D^{M4}_{i+1/2}\,\,,\\
U^{LC} &=& U_i + \frac{1}{2}(U_i-U_{i-1}) + \frac{4}{3}D^{M4}_{i-1/2}\,\,.
\end{eqnarray}
Using these expressions, we compute
\begin{eqnarray}
U_{\min} = \max(\min(U_i, U_{i+1}, U^{MD}), \min(U_i, U^{UL}, U^{LC}))\,\,,\\
U_{\max} = \min(\max(U_i, U_{i+1}, U^{MD}), \max(U_i, U^{UL}, U^{LC}))\,\,.\\
\end{eqnarray}
Finally, a new limited value for the cell interface $U^L_{i+1/2}$ is obtained via
\begin{equation}
U_{i+1/2}^{L,\rm{limited}} = U_{i+1/2}^L + \operatorname{minmod}(U_{\min}-U_{i+1/2}^L, U_{\max}-U_{i+1/2}^L)\,\,.
\end{equation}

To obtain the reconstructed value at the right interface
$U_{i-1/2}^R$, the values at $U_{i-2},\ldots,U_{i+2}$
are replaced by the values $U_{i+2},\ldots,U_{i-2}$.

For a detailed description and derivation of the MP5 reconstruction
algorithm, we refer the reader to the original paper~\cite{suresh:97}.

\subsection{Riemann Solver}\label{sec:numerical_riemann}

We implement the Harten-Lax-van Leer-Einfeldt (HLLE) approximate
solver~\cite{Harten:1983hr,Einfeldt:1988og}.  The more accurate Roe
and Marquina solvers require determining the eigenvalues
characterising the linearizing hydrodynamic evolution scheme, which is
extremely resource intensive while not providing a decisive advantage in
accuracy.  In contrast, HLLE uses a two-wave approximation to
compute the update terms across the discontinuity at the cell interface.  
With $\xi_-$ and
$\xi_+$ the most negative and most positive wave speed eigenvalues
present on either side of the interface (including magnetic field modes but not those associated with divergence cleaning subsystem, 
as discussed in section~\ref{sec:numerical_divclean}), the solution state vector $\mathbf{U}$ is assumed to take the
form
\begin{equation}
  \label{hlle1}
  \mathbf{U} = \left\{ \begin{array}[c]{r c l} \mathbf{U}^L & {\rm if} & 0
        < \xi_-\,\,, \\  \mathbf{U}_* & {\rm if} & \xi_- < 0 < \xi_+\,\,, \\
        \mathbf{U}^R & {\rm if} & 0   > \xi_+\,\,, \end{array}\right. 
\end{equation}
\noindent with the intermediate state $\mathbf{U}_*$ given by
\begin{equation}
  \label{hlle2}
  \mathbf{U}_* = \frac{\xi_+ \mathbf{U}^R - \xi_- \mathbf{U}^L - \mathbf{F}(
  \mathbf{U}^R) + \mathbf{F}(\mathbf{U}^L)}{\xi_+ - \xi_-}\,\,.
\end{equation}
\noindent The numerical flux along the interface
takes the form
\begin{equation}
  \label{eq:hlleflux}
  \mathbf{F}(\mathbf{U}) = \frac{\widehat{\xi}_+\mathbf{F}(\mathbf{U}^L) -
  \widehat{\xi}_-\mathbf{F}(\mathbf{U}^R) + \widehat{\xi}_+ \widehat{\xi}_-
  (\mathbf{U}^R - \mathbf{U}^L)}{\widehat{\xi}_+ - \widehat{\xi}_-}\,\,,
\end{equation}
\noindent where
\begin{equation}
  \label{hlle3}
  \widehat{\xi}_- = {\rm min}(0, \xi_-), \quad \widehat{\xi}_+ =
  {\rm max}(0, \xi_+)\,\,. 
\end{equation}
We use the flux terms~\eref{eq:hlleflux} to evolve the hydrodynamic
quantities within our Method of Lines scheme.

Our calculations of the wave speeds is approximate, following the
methods outlined in~\cite{Gammie:2003rj} to increase the speed of
the calculation at the cost of increased diffusivity.  The method
overstates the true wavespeeds by no more than a factor of $\sqrt{2}$,
and then only for certain magnetic field and fluid velocity
configurations.  When computing the wavespeeds, we replace the full MHD
dispersion relation
by the approximate quadratic form (see 27 and 28
of~\cite{Gammie:2003rj}),
\begin{equation}
\omega_d^2=k_d^2\left[v_A^2+c_s^2\left(1-\frac{v_A^2}{c^2}\right)\right]\,\,,\label{eq:dispersion}
\end{equation}
where we define the wave vector $k_\mu\equiv(-\omega, k_i)$, the fluid
sound speed $c_s$, the Alfv\'en velocity, 
\beq{v_A\equiv
\sqrt{\frac{b^2}{\rho h+b^2}}=\sqrt{\frac{b^2}{\rho h^*}}\,\,,}
the projected wave vector
\beq{K_\mu\equiv (\delta_\mu^\nu+u_\mu u^\nu)k_\nu\,\,,}
and the dispersion relation between frequency and (squared) wave number as
\begin{eqnarray}
\omega_d &=& k_\mu u^\mu = -\omega u^0 + ku^i\,\,,\\
k_d^2 &=& K_\mu K^\mu = \omega_d^2+g^{\nu\sigma}k_\nu k_\sigma\,\,.
\end{eqnarray}
The resulting quadratic may be written 

\begin{eqnarray}
\xi^2\left[W^2(V^2-1)-V^2\right]-2\xi \left[ \alpha W^2 \tilde{v}^i (V^2-1)+V^2\beta^i\right] + &&
\\\nonumber
\left[(\alpha W \tilde{v}^i)^2(V^2-1)+V^2\left(\alpha^2 \gamma^{ii} - \beta^i \beta^i\right)\right]&=&0\,\,,
\end{eqnarray}
where $V^2\equiv v_A^2+c_s^2(1-v_A^2)$ and $\xi$ is the resulting
wavespeed.  Note that the indices are not to be summed over. Instead, we
find different wavespeeds in different directions.  

When divergence cleaning is used to dissipate spurious numerical
constraint violations that appear in the magnetic field, the characteristic
wavespeeds include two additional (\emph{luminal}) modes of the divergence cleaning subsystem.
Since the divergence cleaning subsystem
decouples from he remainder of the evolution system, its wavespeeds must be handled slightly
differently~\cite{Mignone:2010br,Penner:2011zz}, which we discuss
below in section~\ref{sec:numerical_divclean}.

\subsection{Conservative to Primitive variable transformations}
\label{sec:numerical_c2p}

In GRHD, converting the conservative variables back to the primitives
is a relatively straightforward task, which can be accomplished by
inverting~\eref{eq:p2c1} -- \eref{eq:p2c3} with all of the magnetic
field quantities set to zero.  \GRHydro accomplishes the task through
a 1D Newton-Raphson scheme (summarised in section~5.5.4
of~\cite{Loffler:2011ay} and described in detail in the code
documentation~\cite{GRHydro_Manual:web}). The scheme iterates by estimating
the
fluid pressure, determining the density, internal energy and the conservative
quantities given this estimate, and using those in turn to
calculate a new value for the pressure and its residual.  The method
works for any EOS, so long as one can calculate the thermodynamic
derivatives ${\rmd P}/{\rmd\rho}|_\epsilon$ and ${\rmd P}/{\rmd\epsilon|_\rho}$.

MHD adds several complications to the inversion, although the
additional equation, \eref{eq:p2c4}, is immediately invertible,
yielding $B^i=\Bcon^i/\sqrt{\gamma}$.  Instead, the primary difficulty
is that $b^\mu$ cannot be immediately determined, and there is no
simple analogue to the GRHD expressions that allow us to calculate the
density easily once the pressure is specified.  As a reminder, if we
consider the (known) values of the undensitised conservative
variables,
\begin{eqnarray}
\hat{D} &\equiv& \frac{D}{\sqrt{\gamma}}=\rho W\,\,,
\qquad
\hat{S}^i \equiv \frac{S^i}{\sqrt{\gamma}}=\rho h^*W^2v^i-\alpha b^0b^i\,\,,
\\\nonumber
\hat{\tau}&\equiv&\frac{\tau}{\sqrt{\gamma}} = \rho h^* W^2-P^*-(\alpha b^0)^2-\hat{D}\,\,,
\end{eqnarray}
the GRHD system (in which $P^*=P$, $h^*=h$, and $b^\mu=0$) allows us to define
\begin{equation}
Q\equiv \hat{\tau}+\hat{D}+P = \rho h W^2\,\,,\label{eq:Q_grhd}
\end{equation} 
and then determine the density as a function of pressure through the relation
\begin{equation}
\rho = \frac{\hat{D}\sqrt{Q^2-\gamma_{ij}\hat{S}^i\hat{S}^j}}{Q}\,\,,
\end{equation}
and thus the Lorentz factor and internal energy as well (see, e.g.,
\cite{GRHydro_Manual:web}).  In GRMHD, the most efficient approach for
inverting the conservative variable set is often to use a
multi-dimensional Newton-Raphson solver, with simplifications possible
for barotropic EOS, for which the internal energy is assumed to be a
function of the density only, eliminating the need to evolve the
energy equation.  Our methods follow very closely those
of~\cite{Noble:2005gf}, particularly the $2D$ and $1D_W$ solvers they
discuss.

We define a few auxiliary quantities for use in our numerical
calculations. Unfortunately, it is impossible to construct a
consistent notation that agrees with both the Einstein Toolkit release
paper~\cite{Loffler:2011ay} and the paper that describes the
conservative to primitive variable inversion
scheme~\cite{Noble:2005gf}, so we choose to be consistent with the
former here, noting the key differences in \ref{app:notation}.

From the values of the conservative variables 
and the metric, we construct the momentum density 
\begin{equation}
\mathcal{S}_\mu \equiv-n_\nu {T^\nu}_\mu= \alpha T^0_\mu\,\,,
\end{equation}
whose spatial components are given by the relation
$\mathcal{S}_i\equiv \hat{S}_i$ and its normal projection
$\tilde{\mathcal{S}}$ given by
\begin{equation}
\tilde{\mathcal{S}}_\mu = (\delta_\mu^\nu+n_\mu n^\nu)\mathcal{S}_\nu\,\,.
\end{equation}
Inside the Newton-Raphson scheme we make use of an auxiliary variable
$Q$ defined analogously to~\eref{eq:Q_grhd} by the expression
\begin{equation}
Q\equiv \rho h W^2\,\,.
\end{equation}
Since the inversion methods for general EOS and barotropic ones differ in
some of the details, we discuss each in turn.

\subsubsection{General EOS}

The 2D Newton-Raphson approach implemented solves the
following two equations for the unknown quantities $Q$ and $v^2$, with
all other terms known from the given conserved set: 
\begin{eqnarray}
\tilde{\mathcal{S}}^2 =\hat{S}_i \hat{S}^i&=& v^2(B^2+Q)^2-(\mathcal{S}\cdot \mathbf{B})^2\frac{B^2+2Q}{Q^2}\,\,,\label{eq:nreq1}\\
\mathcal{S}\cdot n =-(\hat{\tau}+\hat{D})&=&-\frac{B^2}{2}(1+v^2)+\frac{(\mathcal{S}\cdot \mathbf{B})^2}{2}Q^{-2}-Q+P\,\,,\label{eq:nreq2}
\end{eqnarray}
where all dot products are understood as four-dimensional when
involving four-vectors and three-dimensional when not: e.g., $\mathcal{S}\cdot
\mathbf{B}=\mathcal{S}_\mu B^\mu$, which is equivalent to $\mathcal{S}_i B^i$
since $B^\mu$ is a three-vector and thus $B^0=0$.

For a polytropic/Gamma-law EOS where $P=(\Gamma-1)u$ and 
$u\equiv \rho\epsilon$ is the internal energy density, we may calculate $P$
from the conserved variables and the current guess for $Q$ and $v^2$
by noting that
\begin{equation}
P=\frac{\Gamma - 1}{\Gamma}\left[(1-v^2)Q-\hat{D}\sqrt{1-v^2}\right]\,\,.\label{eq:gammalaw_p}
\end{equation}
The iteration updates $Q$ and $v^2$ subject to the consistency
conditions that $0\le v^2\le 1$ and $Q>0$, and a post-iteration check is performed to ensure that $\epsilon>0$.

For a more general EOS where~\eref{eq:gammalaw_p} does not apply,
and given the structure of the EOS interface in the Einstein Toolkit,
it is easier to solve for the internal energy density $u$, and 
use it as a variable in a three-dimensional Newton scheme along with $Q$
and $v^2$.  First, noting that
\begin{equation}
P = (1-v^2)Q-\hat{D}\sqrt{1-v^2}-u\,\,,
\end{equation}
we may rewrite~\eref{eq:nreq2} as
\begin{equation}
\mathcal{S}\cdot \mathbf{n} =-\frac{B^2}{2}(1+v^2)+\frac{(\mathcal{S}\cdot B)^2}{2}Q^{-2}-v^2Q-\hat{D}\sqrt{1-v^2}-u\,\,.\label{eq:nreq2a}
\end{equation}
In the Newton-Raphson steps, we first set
$\rho=\hat{D}\sqrt{1-v^2}$ and solve
\begin{equation}
u+P(\rho,u) = Q(1-v^2)-\hat{D}\sqrt{1-v^2}\,\,,
\end{equation}
where the left-hand side will, in general,
be a monotonic function of $\rho$ and $u$, along with~\eref{eq:nreq1}
and~\eref{eq:nreq2a}.

Noting that the density depends only on the constant $\hat{D}$ and the value of $v^2$ within the Newton-Raphson scheme, the partial derivatives of the pressure with respect to the Newton-Raphson variables are given by
\begin{eqnarray}
\frac{\partial P}{\partial (v^2)} &=& \left(\frac{\partial P}{\partial \rho}\right)_u \frac{\rmd \rho}{\rmd (v^2)} =  \left[\left(\frac{\partial P}{\partial \rho}\right)_\epsilon-\frac{\epsilon}{\rho}\left(\frac{\partial P}{\partial \epsilon}\right)_{\rho}\right] \frac{\rmd \rho}{\rmd (v^2)}\,\,,\\
\frac{\partial P}{\partial u}&=&\frac{1}{\rho}\left(\frac{\partial P}{\partial \epsilon}\right)_{\rho}\,\,.
\end{eqnarray}
We write these terms in this way, since the Einstein Toolkit EOS
interface uses the pressure as a function of density and the {\em
  specific} internal energy $\epsilon$, $P=P(\rho,\epsilon)$, and
calculates partial derivatives against those variables, rather than
$\rho$ and $u$.

\subsubsection{Barotropic EOS} 

For cases where the pressure and internal energy are functions of the
rest mass density only, the EOS is barotropic (a polytropic $P= K
\rho^\Gamma$ EOS is a special case of a barotropic EOS), we need a
different inversion technique, since the quantity $\mathcal{S}\cdot \mathbf{n}$
in (\ref{eq:nreq2}) requires knowledge of $\hat{\tau}$, which is
not evolved in these cases.  Instead, the inversion uses
(\ref{eq:nreq1}) only, using it to eliminate the variable $v^2$
from the Newton-Raphson scheme by solving for $v^2(Q)$:
\begin{equation}
v^2(Q) = \frac{Q^2\tilde{\mathcal{S}}^2+(\mathcal{S}\cdot B)^2(B^2+2Q)}{Q^2(B^2+Q)^2}\label{eq:v2q}\,\,.
\end{equation}

In the special case of a polytropic EOS, we proceed by first solving
for $\rho(Q)$ through an independent Newton-Raphson loop over the
equation\begin{equation} \rho Q =\hat{D}^2\left(1+\frac{\Gamma
  K\rho^{\Gamma-1}}{\Gamma-1}\right)\,\,.\label{eq:wrho}
\end{equation}

Next, we use~\eref{eq:v2q} and the fact that
\begin{equation}
v^2=\frac{\rho^2}{\hat{D}^2}-1\,\,,
\end{equation}
to replace (\ref{eq:nreq1}) by an expression given only in terms
of $Q$ and $\rho(Q)$:
\begin{eqnarray}
0&=&Q^2(B^2+Q)^2v^2-Q^2(B^2+Q)^2v^2\,\,,\nonumber\\
&=&Q^2\tilde{\mathcal{S}}^2+(\mathcal{S}\cdot B)^2(B^2+2Q) - \left(\frac{\rho^2}{\hat{D}^2}-1\right)Q^2(B^2+Q)^2\,\,.
\label{eq:nreq-barotropic}
\end{eqnarray}
When performing the Newton-Raphson step, all quantities in~\eref{eq:nreq-barotropic}
are known a priori except the iteration variable
$Q$ and $\rho(Q)$, which depends upon it.  The loop over $Q$ makes use
of the derivative of~\eref{eq:wrho}, given by
\begin{equation}
\frac{d\rho}{dQ}=\frac{\rho}{\hat{D}^2\gamma K \rho^{\Gamma-2}-Q}\,\,.
\end{equation}

\subsection{Divergence-free Constraint Treatment}\label{sec:divtreatment}

One of the main difficulties in numerical MHD simulations is the
appearance of non-zero divergence of the magnetic field due to numerical
errors,
violations of~\eref{eq:nomonopole} that would be interpreted as
magnetic monopoles.  A number of techniques have been designed to
combat these, including ``divergence
cleaning''~\cite{Dedner:2002aa,Liebling:2010bn,Penner:2010px}, a
method which introduces a new field that both damps and advects
divergences off the grid, ``constrained transport''
algorithms~\cite{Toth:00} that balance out fluxes exactly to
maintain divergence-free magnetic fields to round-off accuracy, and
magnetic vector potential methods that seek to achieve the same
result~\cite{Etienne:2010ui,Etienne:2011re}. Constrained transport
methods are often difficult to implement in simulations that employ
mesh refinement, since maintaining balance at refinement boundaries is
algorithmically complex. In the initial \GRHydro MHD release we
implement both divergence cleaning and constrained transport, both of
which are detailed below.

\subsubsection{Divergence cleaning}\label{sec:numerical_divclean}

Divergence cleaning works by introducing a new field variable that
both damps divergences and advects them off the grid through a hyperbolic 
equation modelled after the telegraph equation, driving numerical solutions 
towards zero divergence. Our implementation follows closely that of 
\cite{Liebling:2010bn,Penner:2010px}, with some minor differences.
The new field variable $\psi$ satisfies the evolution equation 
\beq{
\nabla_\mu \left( \dF^{\mu\nu} + g^{\mu \nu} \psi \right) =  \kappa n^\nu \psi\,\,,
\label{general-div-cleaning-eq}
} a modification of Maxwell's equations
\eref{eq:maxwell_equations} that reduces to the familiar form as
$\psi\rightarrow 0$. Our parameter $\kappa$, as we note below,
determines the damping rate of the divergence cleaning field $\psi$,
and incorporates the ratio of the parabolic
to the hyperbolic damping speed dependence for the scheme, often
denoted respectively by $c_p$ and $c_h$ in other works.

The evolution equation for $\psi$ is given by the time component of
\eref{general-div-cleaning-eq}.  Noting that~\cite{Anton:2005gi}
\beq{\dF^{\mu\nu} = \frac{1}{W} \left( u^\mu B^\nu - u^\nu B^\mu \right)\,\,,
}
the first term on the left-hand-side of~\eqref{general-div-cleaning-eq}
yields, after some algebra and use of the fact that 
$B^0=-\alpha F^{00}=0$,
\beq{\nabla_\mu  \dF^{\mu 0}   =  - \frac{1}{\alpha \sqrt{\gamma}} \partial_i \sqrt{\gamma} B^i\,\,,
}
and the second term
\beq{\nabla_\mu g^{\mu 0} \psi = g^{\mu 0} \partial_\mu \psi = \frac{1}{\alpha^2} \left[ -\partial_t \psi + \beta^i \partial_i \psi \right]\,\,.}
Combining the two, we find
\beq{\partial_t \psi + \partial_i \left( \alpha B^i - \psi \beta^i \right)  
=  \psi \left( -\kappa \alpha - \partial_i \beta^i  \right)
+ \sqrt{\gamma} B^i \partial_i \left( \frac{\alpha}{\sqrt{\gamma}} \right)  \label{dphi-dt-eq}\,\,,}
where we have grouped the derivative terms to respect the
flux-conservative form.  

The spatial part of~\eref{general-div-cleaning-eq} is the
modified evolution equation for the magnetic field.  The two terms
on the left hand side yield, respectively,
\begin{eqnarray}
\nabla_\mu  \dF^{\mu j} &=&  \frac{1}{\alpha \sqrt{\gamma}} \left\{ \partial_t \sqrt{\gamma}  B^j + 
\partial_i  \sqrt{\gamma} \left[ \left(\alpha v^i - \beta^i\right) B^j - \left(\alpha v^j - \beta^j\right) B^i \right] \right\}\,\,,\\
\nabla_\mu g^{\mu j} \psi &=&  g^{\mu j} \partial_\mu \psi 
= \frac{1}{\alpha^2} \left[ \beta^j \partial_t \psi + \left(\alpha^2 \gamma^{i j} - \beta^i \beta^j \right) \partial_i \psi \right]\,\,.
\end{eqnarray}
Using~\eref{dphi-dt-eq} to eliminate the $\partial_t\psi$ term,
switching over to  $\Bcon^i$ as the primary magnetic field variable, and
combining derivatives to produce a flux-conservative form, we find after
some algebra, 
\begin{eqnarray}
&&\partial_t \Bcon^j  + \partial_i \left[ (\alpha
v^i-\beta^i) \Bcon^j - \alpha v^j \Bcon^i + \alpha
\sqrt{\gamma}\gamma^{i j} \psi 
 \right] 
 =
\\\nonumber
&& - \Bcon^i \partial_i \beta^j  
+ \psi \partial_i \left( \alpha \sqrt{\gamma} \gamma^{i j}  \right)\,\,,\label{dbcons-dt-eq}
\end{eqnarray}
which reduces to the standard evolution equation for $\psi\rightarrow 0$
and $\partial_i\Bcon^i\rightarrow 0$.  To evaluate the right-hand-side, we
make use of the identity 
\begin{eqnarray}
\partial_i(\sqrt{\gamma}\gamma^{ij}) =-\sqrt{\gamma}\gamma^{kl}\Gamma^j_{kl}=\sqrt{\gamma}\left[\frac{1}{2}\gamma^{ij}\gamma^{kl}\partial_i\gamma_{kl}-\gamma^{jk}\gamma^{il}\partial_i\gamma_{kl}\right]\,\,.
\end{eqnarray}

The characteristic velocities of the evolution
system for MHD without and with divergence cleaning differ.  In the
former, the largest-magnitude wave speed is the fast
magnetosonic wave speed. The inclusion of divergence cleaning introduces
two additional modes, corresponding to eigenvalues for the evolution of the
divergence cleaning field $\psi$ and the longitudinal component of the
magnetic field, (i.e., the case where $i=j$ inside the term
in brackets in~\eref{dbcons-dt-eq}), that 
in the flat spacetime case
decouple from the
remaining seven eigenvalues of the
system~\cite{Mignone:2010br,Penner:2011zz}.  In our scheme, both modes
have characteristic speed equal to the speed of light.
Our system
corresponds
to setting the hyperbolic damping speed $c_h$ of the scheme to be the
speed of light in the notation of Newtonian divergence cleaning
methods~\cite{Dedner:2002aa}.  This is in accordance with standard
practice in both Newtonian and relativistic calculations to choose a
characteristic speed for divergence cleaning that is as fast as
allowed without violating causality, or in the case of Newtonian
calculations, the Courant condition.  We note that for the
luminal-speed modes, the HLLE flux formula, \eref{eq:hlleflux} reduces
to the local Lax-Friedrichs form when evaluated on a Minkowski
background,
\begin{equation}
\label{eq:LFflux}
 F(U) = \frac{1}{2}\left[F(U^L) +F(U^R) - \widehat{\xi}(U^R - U^L)\right]\,\,,
 \end{equation}
where $\widehat{\xi} = \max(\widehat{\xi}_+,|\widehat{\xi}_-|)$.
Test runs performed with the
 HLLE Riemann solver, but without the luminal-speed
 velocities often develop large spurious oscillations, particularly in
 cases where strong rarefactions are present, e.g., the cylindrical
 blast wave evolution described in section~\ref{sec:cylexp} and the rotor
 test described in section~\ref{sec:rotor}.

In our current implementation, we use the hyperbolic divergence
cleaning technique only for flat spacetime tests. In this case the
divergence cleaning subsystem decouples from the remainder of the MHD
evolution system. In curved spacetime this decoupling is not as
obvious and we opt to use constrained transport techniques instead,
which we describe in the following.

\subsubsection{Constrained transport}\label{sec:constrained_transport}
As an alternative to divergence cleaning, and to simplify comparison
of the performance of the new code with existing GRMHD codes, we also
implemented a variant of a constrained transport
scheme~\cite{Evans:1988qd}. In constrained transport schemes one
carefully constructs a numerical update scheme such that the
divergence free constraint~\eref{eqn:divB_constraint}
is conserved to numerical round-off accuracy. Rather than implementing
the original scheme proposed
in~\cite{Yee:1966my,Evans:1988qd,Balsara:1999aa}, which relies on a
complicated staggering of the magnetic field components, we employ the
simplified scheme called ``flux-CT'' described
in~\cite{Toth:00,Giacomazzo:2007ti,Giacomazzo:2007fe}, a two-dimensional
version of which is used in the {\tt HARM} code~\cite{Gammie:2003rj}.
The scheme uses
cell-centred values of the magnetic field. 
For completeness of
presentation we reproduce the basic description of the scheme found
in~\cite{Giacomazzo:2007ti}, but refer the reader to the original
literature for more details.

\begin{figure}[t]
    \begin{center}
        \includegraphics{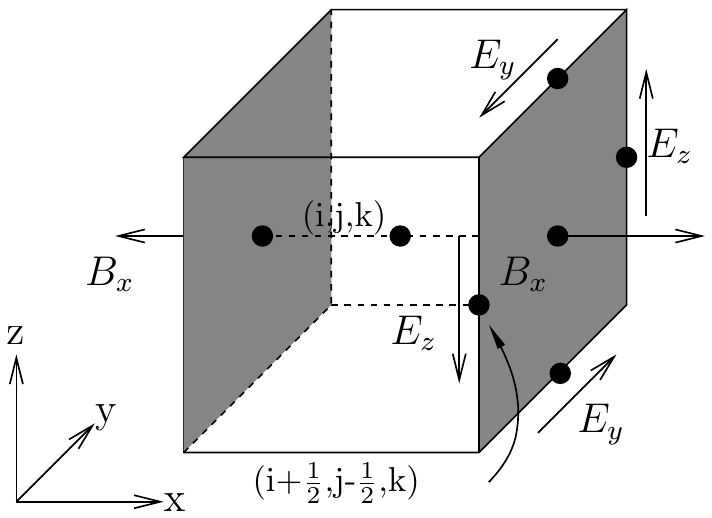}
    \end{center}
    \caption{Sketch of a simulation cell showing the location of face centred
    magnetic field and edge centred electric fields.}
    \label{fig:constrained-transport-simulation-cell}
\end{figure}

In constrained transport schemes the induction
equation~\eqref{dbcons-dt-eq} is written in terms of the electric
field $\vec \Econ$ at the edges of each face of a simulation cell (see
figure~\ref{fig:constrained-transport-simulation-cell}).

Employing Amp\`ere's law, the time derivative of the face-averaged magnetic
field component $\hat{\Bcon}^x$ is given by
\begin{eqnarray}
    \frac{\partial \hat{\Bcon}^x_{i+\frac12,j,k}}{\partial t} \, \Delta y \, \Delta z &=&
    - \Econ^y_{i+\frac12,j,k-\frac12} \, \Delta y - \Econ^z_{i+\frac12,j+\frac12,k} \, \Delta z
    + \Econ^y_{i+\frac12,j,k-\frac12} \, \Delta y
    \\\nonumber 
    &&+ \Econ^z_{i+\frac12,j-\frac12,k} \, \Delta z\,\,,
    \label{eq:constrained-transport-integrated-induction-equation}
\end{eqnarray}
with analogous equations for the other magnetic field components.  In
the ideal MHD approximation, the electric field components in
\eref{eq:constrained-transport-integrated-induction-equation} can be
expressed in terms of the fluxes
$\Bcon^k\tilde{v}^i-\Bcon^i\tilde{v}^k$ for the magnetic field $\vec
\Bcon$ given in~\eref{eq:definition-of-flux}.  Specifically, we use
the numerical fluxes of the induction equation to calculate these
electric field components.

Since the scheme of~\cite{Giacomazzo:2007ti} evolves the cell-averaged
magnetic field values, it computes the change in the cell-centred fields as
the average change in the face centred fields
\begin{eqnarray}
    \frac{\partial \Bcon^x_{i,j,k}}{\partial t} &=& 
    \frac 12 \left(\frac{\partial \hat{\Bcon}^x_{i-\frac12,j,k}}{\partial t} + \frac{\partial \hat{\Bcon}^x_{i+\frac12,j,k}}{\partial t}\right) 
,
\end{eqnarray}
which is the actual evolution equation implemented in the code.

In this formalism, the conserved divergence operator is given by
\begin{eqnarray}
    \left(\nabla \cdot \Bcon\right)_{i+\frac12,j+\frac12,k+\frac12} &=&
    \hphantom{+}
    \frac{1}{4 \Delta x} \sum_{j^\prime=j}^{j+1} \sum_{k^\prime=k}^{k+1} 
    \left( \Bcon^x_{i+1 , j^\prime  , k^\prime  } - \Bcon^x_{i , j^\prime  , k^\prime  } \right)  \nonumber\\&&
    + 
    \frac{1}{4 \Delta y} \sum_{i^\prime=i}^{i+1} \sum_{k^\prime=k}^{k+1} 
    \left( \Bcon^y_{i^\prime , j+1  , k^\prime  } - \Bcon^y_{i^\prime , j  , k^\prime  } \right)  \nonumber\\&&
    + 
    \frac{1}{4 \Delta z} \sum_{i^\prime=i}^{i+1} \sum_{j^\prime=j}^{j+1} 
    \left( \Bcon^z_{i^\prime , j^\prime  , k+1  } - \Bcon^z_{i^\prime , j^\prime  , k  } \right)\,\,,  
    \label{eq:constrained-transport-corner-centered-divergence}
\end{eqnarray}

which can be interpreted as a cell corner-centred definition of a divergence
operator.

\section{Tests}\label{sec:tests}

\subsection{Monopole tests}\label{sec:monopole}

In order to test the divergence cleaning formalism and to explore its
properties, we implement a number of tests that initialise the MHD
configuration with numerical monopoles. Each case uses a
uniform-density, uniform-pressure fluid which is initially at rest,
and we assume Minkowski spacetime.  The ambient magnetic field is
assumed to be zero. The initial magnetic configurations include
``point'' monopoles, where $B^x\ne 0$ is set at a single point in the
Center of the domain, and Gaussian monopoles for which
\begin{equation}
B^x=\left\{\begin{array}{rl}
e^{-r^2/R_G^2}-e^{-1}; & r<R_G\,\,,\\
0; & r\ge R_G\,\,,
\end{array}\right.
\end{equation}
where $R_G$ is the radius of the compactly supported monopole. In the
supplementary material we present additional results on the performance
of the algorithm when dealing with high frequency constraint violations.

\begin{figure}[t]
 \includegraphics[width=0.9\textwidth]{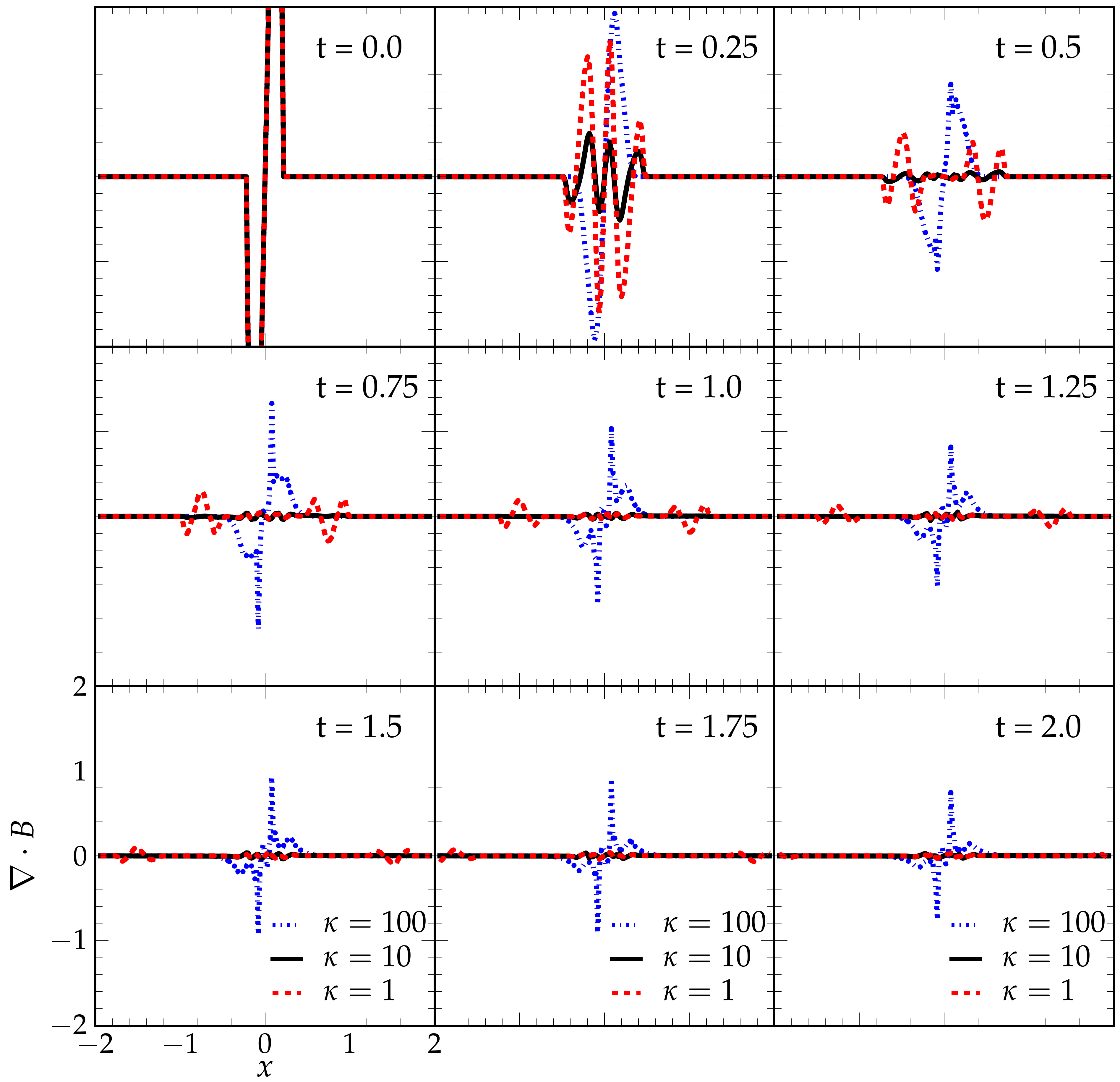}
 \caption{Behaviour of our divergence cleaning scheme,
   demonstrated by monopole damping and advection for a magnetic field 
   with an initially  Gaussian $B^x$ profile of radius $R_G=0.2$.  
   We show results for  different values of the divergence cleaning damping
   parameter $\kappa$.  For $\kappa=100$ (dot-dashed blue curve), the
   system is quite stiff and damping is substantially slower than in
   other cases and more of the divergence remains localised near the
   initial position, while for $\kappa=1$ (dashed red curve) the
   damping is only slightly faster but spreads more rapidly away
   across the grid.  For $\kappa=10$ (solid green curve), we find a
   nearly ideal choice of rapid damping and advection of the error
   away from the source, and recommend this value as a default for
   general calculations.\label{fig:monopole_gauss}}
\end{figure}

In Figs.~\ref{fig:monopole_gauss}, we 
show the evolution of Gaussian and 3-dimensional alternating Gaussian
monopole initial data.  In each case, the grid is chosen to be
$200^3$, spanning a coordinate range from $-2$ to $2$ in each
dimension, and we set $R_G=0.2$. The simulations are performed using
second-order Runge-Kutta (RK2) time integration and TVD-based
reconstruction with an monotonized central limiter and a
Courant-Friedrichs-Lewy (CFL) factor of 0.25 (i.e., $\Delta t/\Delta x=0.25$
here).  The fluid, assumed to follow a $\Gamma=5/3$ ideal-gas law, i.e.\ to
follow the condition $P=(\Gamma-1)\rho\epsilon$, was set to an initially
stationary state with density and internal energy given by $\rho=1.0$ and
$\epsilon=0.1$, respectively.  We vary 
the divergence cleaning parameter $\kappa$ that appears in the driving term in
\eref{general-div-cleaning-eq}, choosing values $\kappa=1,~10$, and $100$.

There is a markedly different evolution of the divergence of the
magnetic field in time, following a predictable pattern.

The choice $\kappa=1$ yields the slowest damping rate, which allows
the wave-like behaviour of the divergence
cleaning field to radiate away the divergence of the $B$-field away
from its original location as it damps downward.  By contrast,
$\kappa=100$ yields a much stiffer system of equations, resulting in a
relatively slow damping rate, particularly for the lower-frequency
terms in the initial data, with very low divergence in the magnetic
field spreading through the numerical grid.  The intermediate case,
$\kappa=10$, yields the closest analogue to a critically damped system
in that the amplitude of the divergence decreases most rapidly, while
errors are efficiently radiated away across the grid.  In general, we
would recommend values of $\kappa\sim 10$ to be used as a default.

\subsection{Planar MHD Shocktubes}\label{sec:shocktubes}

Historically, the simplest tests for an MHD scheme are shocktubes on a
static flat-space Minkowski metric.  They are important to demonstrate
the ability of the new code to capture a variety of MHD wave
structures.  Despite being simple setups they do provide a stringent
tests for the algorithm. For our one-dimensional shocktube tests, we
set up ``Left'' and ``Right'' MHD states on either side of a planar
interface, with parameters drawn from the five cases considered
in~\cite{Balsara:2001aa} and summarised in
Table~\ref{table:shocks}.  These have been widely used by several
groups to establish the validity of numerical MHD codes.  These
parameter choices include generalisations of familiar cases long used
to test non-relativistic MHD codes, e.g., case ``Balsara1'', which is
a generalisation of the Brio-Wu shock tube problem~\cite{Brio:1988aa}.

To validate the new code, we evolve each of these shocks in each of the
coordinate directions (i.e., ``Left'' states for domains of the grid
satisfying $x<0$, $y<0$, and $z<0$ and ``Right'' states for $x>0$,
$y>0$, and $z>0$, respectively), finding excellent agreement (limited
only by numerical precision) in each direction, as expected.  We also
evolve shocks with mid-planes oriented obliquely to the coordinate
planes, choosing $x+y=0$ (``2D diagonal'') and $x+y+z=0$ (``3-d
diagonal'') as the mid-planes.  Whereas shocks in any of the
coordinate planes initially satisfy the divergence free constraint to
roundoff error, and
maintain this state indefinitely given the symmetry in the setup,
diagonal shocks yield non-zero divergence of the magnetic field across
the shock front due to finite resolution effects.
The simplest way to understand this is to realise that for a 1-dimensional
domain (say along the $x$ direction), the $x$ component of the magnetic field
$B^x$ is necessarily constant in space and time as a consequence of the
induction equation~\eqref{dbcons-dt-eq}.  The slab symmetry of the test
on the other hand
implies that all fields, in particular the $B^y$ and $B^z$ field, are
independent of $y$ and $z$. These two effects imply that 
$\partial_i B^i = 0$ to roundoff precision initially and for all times, since
the finite volume scheme preserves this property.
Diagonal shocks, on the other hand, break this symmetry by having 
field components depend on two coordinates introducing divergence constraint
violations of order of the truncation 
error. For non-flat geometries, the non-linear coupling of GRMHD equations 
would generate violations on the order of the truncation error in all 
directions. This multidimensional test allows us to verify that these
constraint violations are controlled by employing divergence cleaning.

In all cases, the coordinate origin was chosen so that no grid cell was
centred directly on the shock front, so that  each cell was unambiguously
located in the left or right states.  In each run we used RK2 time integration
with a CFL factor of $0.8$ for 1-d shocks and $0.4$ for 2D diagonal cases,
with $\Delta x=1/1600$ and $\Delta x=\Delta y = 1/(800\sqrt{2})$ for the 1-d
and 2D diagonal cases to ensure equal resolution in the shock direction.  For
all results shown below, we use the HLLE Riemann solver and TVD reconstruction
of the primitive variables with a monotonized central limiter.  Boundary
conditions are implemented by ``copying'' all data to grid edges from the
nearest points in the interior located on planes parallel to the shock front.
Divergence cleaning is turned on for the 2D case with $\kappa=10$.
Table~\ref{table:shocks} lists the parameters describing each shock test that
is performed, and we note that all of them assume an ideal-gas law EOS with the
given value of $\Gamma$, here $\Gamma=2$ for the ``Balsara1'' shock test and
$\Gamma={5}/{3}$ for the others. These tests are also included as options
in the code release within the {\tt GRHydro\_InitData} thorn.  Exact solutions
for each case are computed using the open-source available code of
\cite{Giacomazzo:2005jy,Giacomazzo:Riemann_code}.  In all cases, we see very
good agreement with the expected results that are comparable to results of
similar codes in the
literature~\cite{Balsara:1998,Balsara:2001aa,DelZanna:2002rv,
Anton:2005gi,Giacomazzo:2005jy,Anderson:2006ay,Giacomazzo:2007ti,
DelZanna:2007pk}.
We conclude that divergence cleaning does not interfere with generating
physically meaningful GRMHD evolution results in the presence of shocks, and
that the new code performs as well as other codes based on the same numerical
techniques~\cite{Penner:2011zz}. 

\begin{table}
\caption{Shock tube parameters. For each of these shock tube tests, originally
compiled in \protect\cite{Balsara:2001aa}, we list for both the left ``$L$''
and right ``$R$'' states the fluid density $\rho$, specific internal energy
$\epsilon$, fluid 3-velocity $\vec{v}\equiv v^i$, and tangential magnetic field
$B^t$, as well as the (uniform) normal magnetic field magnitude $B^n$, the
adiabatic index $\Gamma$ of the ideal-gas law equation of state, and the time
$T_{\text{ref}}$ at which results are plotted.  Our values are quoted for shocks
moving in the $x$-direction, i.e., the shock front is oriented in the $y-z$
plane, so $B^t$ is equivalent to $(B^y,B^z)$ and $B^n=B^x$.  For diagonal
shocks, all quantities are rotated appropriately along the diagonal of 
$x-y$ plane. Results for cases marked by an asterisk$^*$ are shown in the
supplementary material only.} 
\label{table:shocks}
\lineup%
\begin{indented}
\item[]\begin{tabular}{@{}lcrcccccccccc}
\br
Name & $\rho_L$ & \multicolumn{1}{c}{$\epsilon_L$} & $\vec{v}_L$ & $B^{\rm t}_{L}$ & $\rho_R$ & $\epsilon_R$ & $\vec{v}_R$ \\
\mr
Balsara1     &  1.0\0 & \01.0\0\0 & $\vec{0}$ & (1.0,0) & 0.125 & 0.8\0 & $\vec{0}$ \\
Balsara2$^*$ &  1.0\0 & 45.0\0\0 & $\vec{0}$ & (6.0,6.0) & 1.0\0\0 & 1.5\0 & $\vec{0}$ \\
Balsara3$^*$ &  1.0\0 & 1500.0\phantom{00} & $\vec{0}$ & (7.0,7.0) & 1.0\0\0 & 0.15 & $\vec{0}$ \\
Balsara4     &  1.0\0 & 0.15\0 & $(0.999,0,0)$ & (7.0,7.0) & 1.0\0\0 & 0.15 & (-0.999,0,0) \\
Balsara5$^*$ &  1.08 & 1.425 & $(0.4,0.3,0.2)$ & (0.3,0.3) & 1.0\0\0 & 1.5\0 & (-0.45,-0.2,0.2) \\
\br
\end{tabular}
\item[]\begin{tabular}{@{}lcccccccccccc}
\br
Name & $B^{\rm t}_{R}$ & $B^{\rm n}$ & $\Gamma$ & $T_{\text{ref}}$\\
\mr
Balsara1\phantom{$^*$} (continued) &  (-1.0,0) & \00.5 & 2 & 0.4\0\\
Balsara2$^*$ (continued) & (0.7,0.7) & \05.0 & ${5}/{3}$ & 0.4\0\\
Balsara3$^*$ (continued) & (0.7,0.7) & 10.0 & ${5}/{3}$ & 0.4\0\\
Balsara4\phantom{$^*$}  (continued) & (-7.0,-7.0) & 10.0 & ${5}/{3}$ & 0.4\0\\
Balsara5$^*$ (continued) & (-0.7,0.5) & \02.0 & ${5}/{3}$ & 0.55\\
\br
\end{tabular}
\end{indented}

\end{table}

\begin{figure}
 \includegraphics[width=0.9\textwidth]{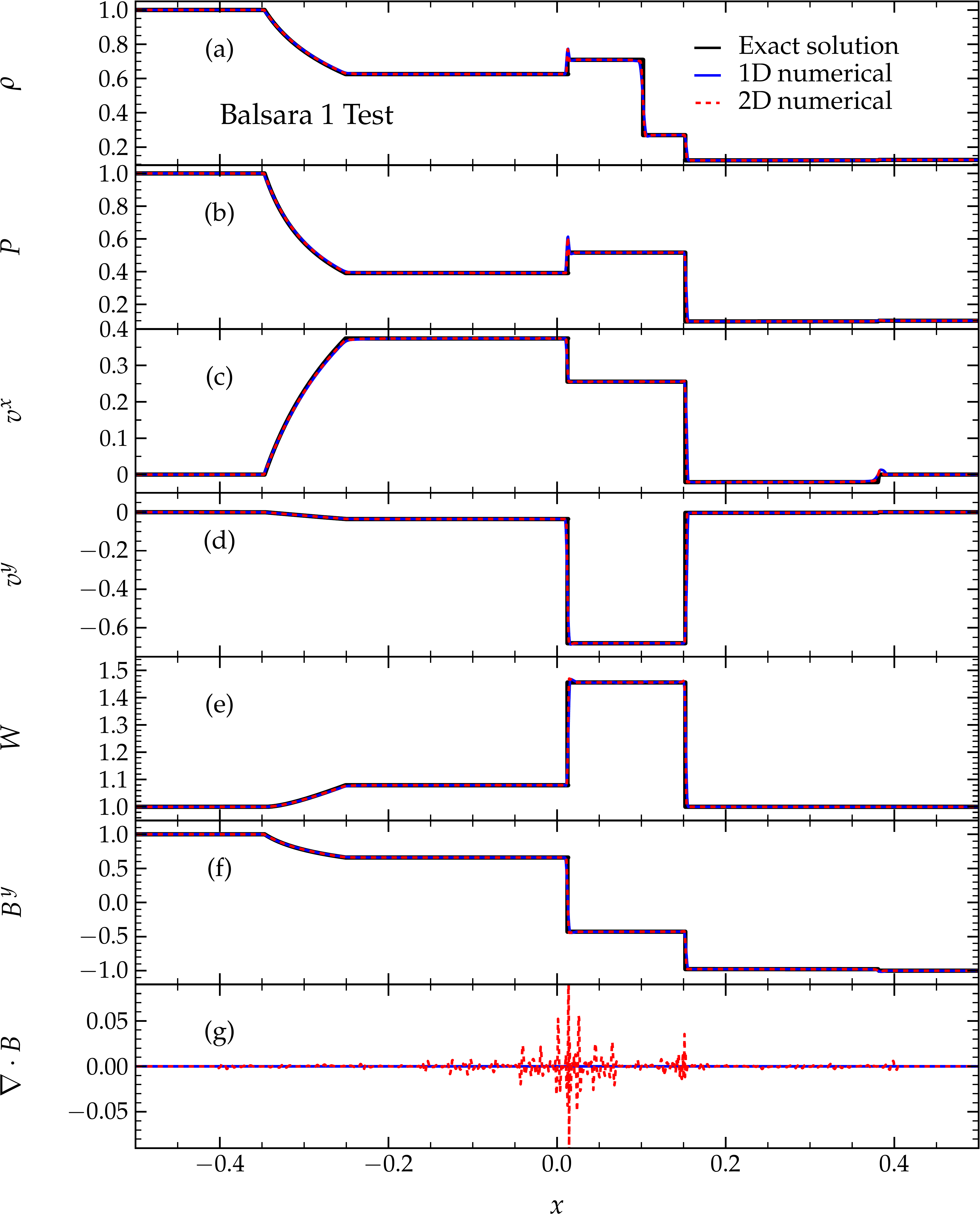}
 \caption{Evolution of the
   Balsara1 shocktube, performed with divergence cleaning, showing 
   1D and 2D diagonal cases as solid
   (blue) and dashed (red) curves, respectively. From top to bottom,
   we show the density $\rho$, gas pressure $P$, normal and tangential
   velocities $v^x$ and $v^y$, the Lorentz factor $W$, the tangential
   magnetic field component $B^y$, and the numerical divergence of the
   magnetic field for the 2D case (for 1D shocks, the divergence is
   uniformly zero by construction). Parameters for the initial shock
   configuration are given in Table \protect\ref{table:shocks}.  All
   results are presented at $t=T_{\text{ref}}$, given by the final column of
   Table \protect\ref{table:shocks}. The results agree well with those
   reported in~\cite{Giacomazzo:2007ti,Balsara:2001aa} with most plots
   being indistinguishable and only the Lorentz factor plot showing a
   slight overshoot on the left-hand side of the shock. A more detailed
   description of the test setup and parameters can be found in the main
   text in section~\ref{sec:shocktubes}.}
 \label{fig:shock1}
\end{figure}

\begin{figure}
 \includegraphics[width=0.9\textwidth]{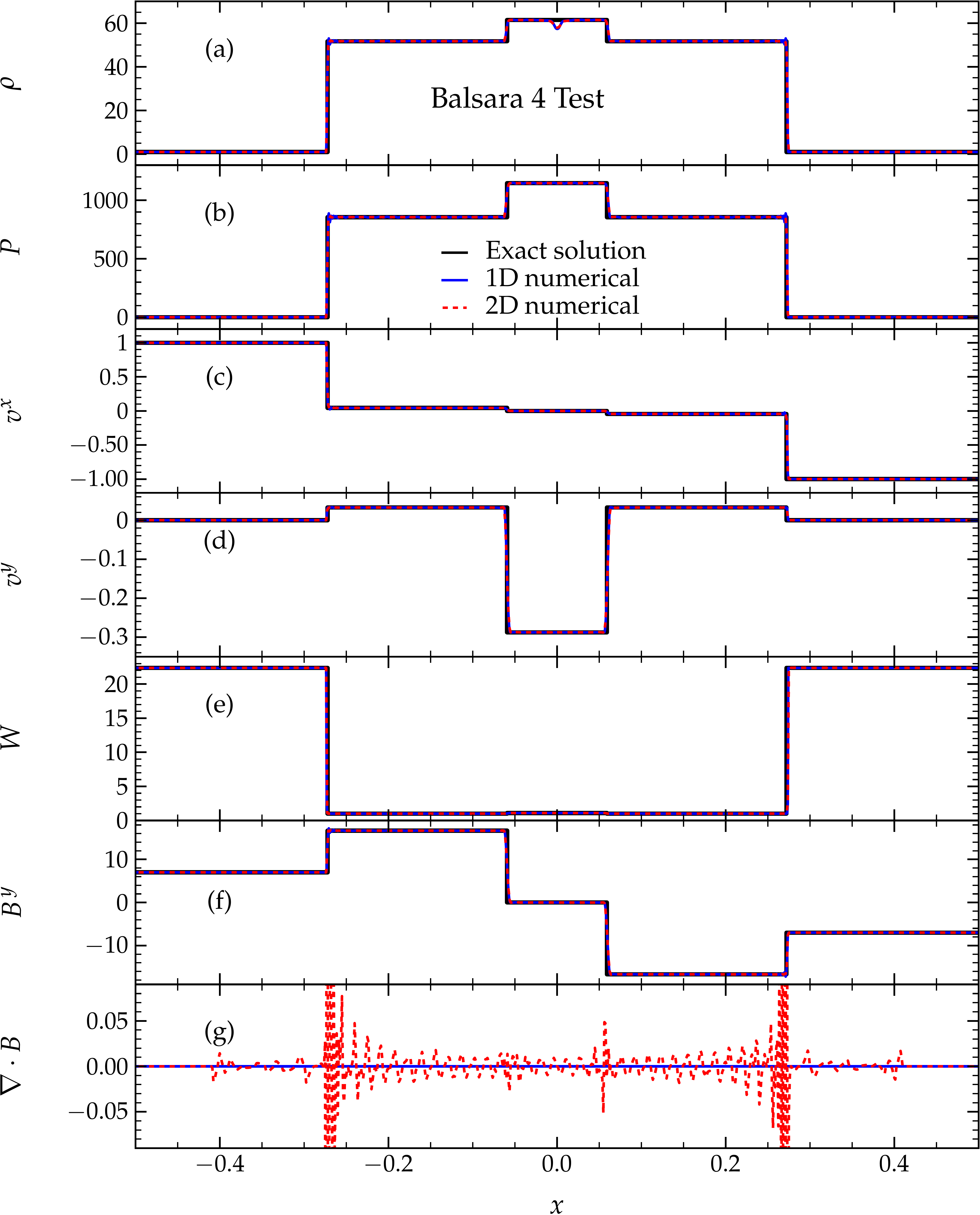}
 \caption{Evolution of the Balsara4 shocktube case, performed with
   divergence cleaning, with all conventions as in
   figure~\protect\ref{fig:shock1}. The results code reproduces the exact
   result very well, with deviations from the exact solution similar to those
   found inn~\cite{Balsara:2001aa}. A more detailed
   description of the test setup and parameters can be found in the main
   text in section~\ref{sec:shocktubes}.}
 \label{fig:shock4}
\end{figure}

In figure~\ref{fig:shock1} we show results for the relativistic generalisation of
the Brio \& Wu shock test develop in~\cite{Balsara:2001aa}. The initial
shock develops into a left-going fast rarefaction, a left-going compound
wave, a contact discontinuity, a right-going slow shock and a
right-going fast rarefaction. We find that the code captures
all elementary waves and is in good agreement with the exact solution
of~\cite{Giacomazzo:2007ti}. In panel (g) of~\ref{fig:shock1} we show
the constraint violation as measure by a 
2\textsuperscript{nd} order centred finite difference stencil
for $\partial_i  \Bcon^i$. 
Figure~\ref{fig:shock4}
displays the results of the relativistic MHD collision problem (problem 4
of~\cite{Balsara:2001aa} in which two very relativistic (Lorentz factor
$22.37$) streams collide with each other. We find deviations from the exact
solution that are on the same level as in~\cite{Giacomazzo:2007ti,Balsara:2001aa}.

\subsection{Cylindrical Shocks} \label{sec:cylexp}

A more stringent multidimensional code test is provided by a
cylindrical blast wave expanding outward in two dimensions.  We take
the parameters for this test problem from~\cite{Komissarov:1999aa}.
The density profile is determined by two radial parameters, $r_{\rm in}$
and $r_{\rm out}$, and two reference states, such that
\begin{eqnarray}
\rho(r) = \left\{\begin{array}{lll}
\rho_{\rm in} & ; & r\le r_{\rm in}\,\,,\\
\exp\left[\frac{(r_{\rm out}-R)\ln\rho_{\rm out}+(r-r_{\rm in})\ln\rho_{\rm in}}{r_{\rm out}-r_{\rm in}}\right] & ; & r_{\rm in}<r<r_{\rm out}\,\,,\\
\rho_{\rm out} & ; & r\ge r_{\rm out}\,\,,
\end{array}\right.
\end{eqnarray}
with an equivalent form for the pressure gradient.  The initial fluid
velocity is set to zero, and the initial magnetic is uniform
in the domain.  We use the following shock
parameters:
\begin{eqnarray}
r_{\rm in}&=&0.8,~r_{\rm out}=1.0;~\rho_{\rm in}=10^{-2},~\rho_{\rm out}=10^{-4};
~P_{\rm in}=1.0,\\\nonumber
P_{\rm out}&=&3 \times 10^{-5};~B^i = (0.1,0,0)\,\,.
\end{eqnarray}
All tests use $\Gamma$-law equation of state with adiabatic index $\Gamma =
4/3$. 
We use a $200\times200\times 8$ grid, spanning the coordinate range
$[-6,6]$ in the $x$ and $y$-directions, which is the setup used in~\cite{Komissarov:1999aa}. 
To verify that our results are indeed convergent with resolution and
to compare to~\cite{Etienne:2010ui}, we
run two more simulations using $400$ and $500$ points as well.
All tests shown use the divergence cleaning scheme with $\kappa = 5$,
HLLE Riemann solver,
TVD reconstruction using a monotonized central limiter. No explicit
dissipation is added to the system of equations. 

This test is known to push the limits of many combinations of MHD
evolution techniques, and often fails for specific combinations of
reconstruction methods and Riemann
solvers~\cite{Etienne:2010ui,Shibata:2005gp}.  In our
tests we only explore a small range of settings: we verify that the
test succeeds when we swap TVD reconstruction with a 2\textsuperscript{nd}
order ENO scheme or use constraint transport instead of
divergence cleaning. In the later case we add Kreiss-Oliger
dissipation~\cite{Kreiss:1973aa} of
order 3 and strength parameter $\epsilon = 3$~\cite{Kreiss:1973aa} to 
the magnetic field
variables to stabilise the system.

We find this test to be among the most sensitive one of our tests.
Including the two light-like modes of the divergence cleaning field (see
section~\ref{sec:numerical_divclean}) is crucial to obtain correct results.
When these are not included, the
code crashes quickly due to the growth of spurious oscillations in the
magnetic field in the rarefaction region that develops
behind the shock, particularly along the diagonals where the shock
front expands obliquely to the grid (as was seen in tests of a several
other codes~\cite{Shibata:2005gp,Etienne:2010ui,DelZanna:2007pk}).  We
note that implementing the local Lax-Friedrichs flux (see
\eref{eq:LFflux}) instead of HLLE also stabilises the evolution.

\begin{figure}
 \includegraphics[width=0.95\textwidth]{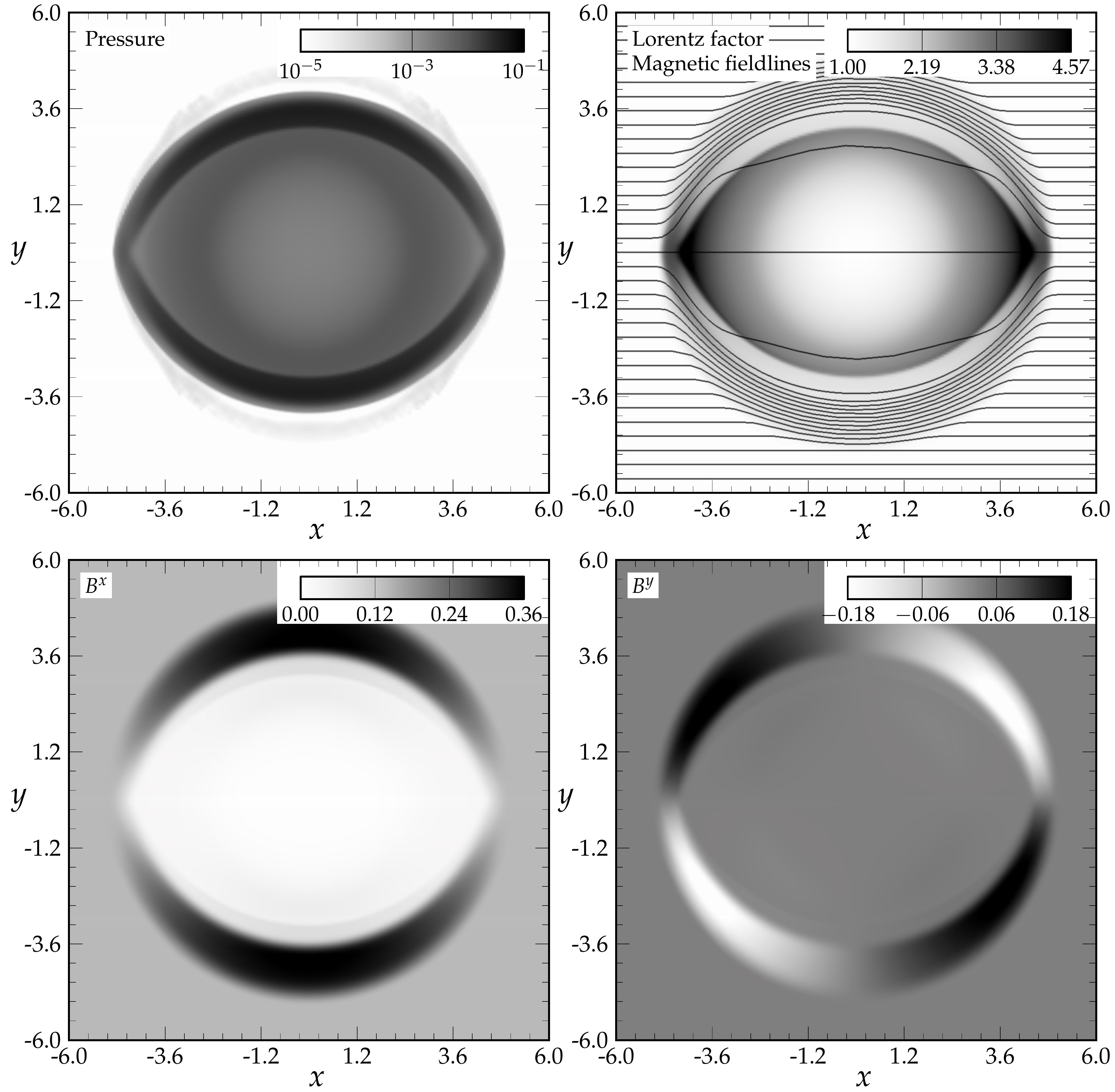}
 \caption{Evolved state of the cylindrical explosion test problem, with
parameters based on the test presented in \protect\cite{Komissarov:1999aa}.  
At $t=4$, we show the following quantities: top panels: Gas pressure $P$ and
Lorentz factor $W$, along with numerically determined magnetic field lines;
bottom panels, $B^x$ and $B^y$.  The numerical resolution is $\Delta x=0.06$.
Profiles are almost indistinguishable from the solutions found in figure~10 of
of~\protect\cite{Komissarov:1999aa}. The very low pressure region outside of
the shock from is slightly more extended in~\protect\cite{Komissarov:1999aa}.
It is worth noting that~\protect\cite{Komissarov:1999aa} apply some
numerical resistivity to control artifacts, while our divergence cleaning
scheme has dissipative terms present in the divergence cleaning equations but
do not add explicit resistivity,
hence our results in low pressure regions might well differ slightly.
}\label{fig:cylexp-one}
\end{figure}
  
\begin{figure}
\includegraphics[width=1.0\textwidth]{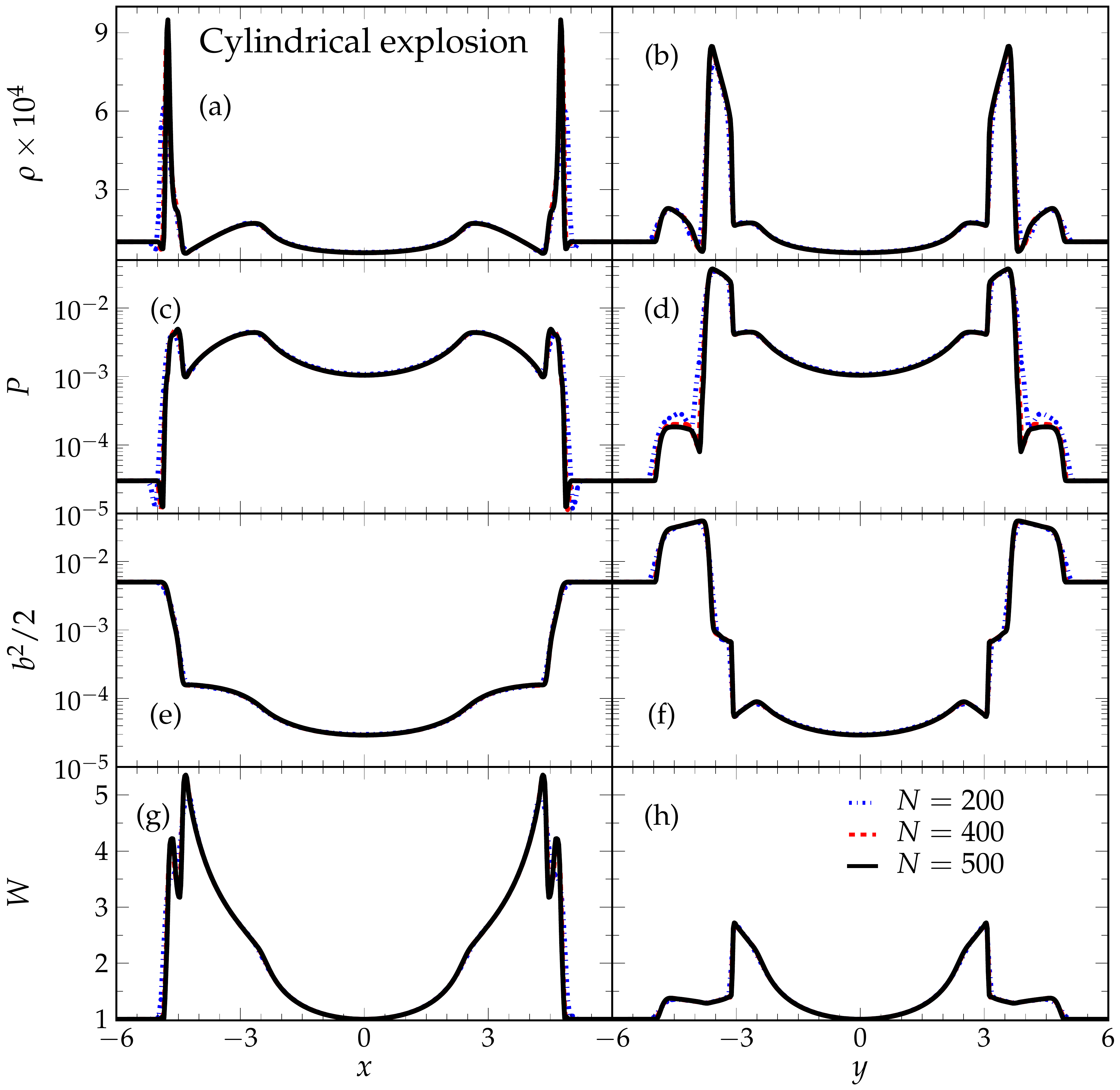}
\caption{One-dimensional slices along the $x-$ and $y-$axes for the
  evolved state of the cylindrical explosion test. Displayed are the
  rest mass density, gas pressure, magnetic pressure and Lorentz
  factor at $t=4$. The red solid lines corresponds to a resolution of
  $\Delta x = 0.06$, the black dashed one corresponds to a resolution
  of $\Delta x = 0.03$ and the blue dotted on corresponds to a
  resolution of $\Delta x = 0.024$.  These results may be compared to
  figure~7 of \protect\cite{Shibata:2005gp} and particularly figure~7 of
  \protect\cite{Etienne:2010ui}, noting that their parameters differ
  from ours, which were chosen to match the test as presented by
  \protect\cite{Komissarov:1999aa}, particularly by the presence of a
  narrow transition region whose effects are visible as a secondary shock in
  the outer edges of the shock front. Comparing the plots, both codes
  agree on the overall structure of the result, with the two-dimensional
  pseudocolor plots in~\ref{fig:cylexp-one} being indistinguishable and the
  one-dimensional slices clearly showing the same structure of the
  shocks. A more detailed discussion of the test setup and results can
  be found in the text.}\label{fig:cylexp_slices}
\end{figure}

Two-dimensional profiles are shown in figure~\ref{fig:cylexp-one} for
$P,~W_{\rm Lorentz},~B^x$, and $B^y$, along with numerically
determined magnetic field lines, at $t=4$.  These results agree well
with those shown in figure~10 of~\cite{Komissarov:1999aa}.
One-dimensional slices along the $x-$ and $y-$axes for the rest mass
density, gas pressure, magnetic pressure and Lorentz factor at $t=4$
are shown in figure~\ref{fig:cylexp_slices} for three
different numerical resolutions.  Our results show good qualitative
agreement with those presented previously by~\cite{Shibata:2005gp} and
particularly by~\cite{Etienne:2010ui}, once one accounts for
rescalings associated with different choices for the initial shock
parameters.  
We find that the benefit of
increasing the resolution is largest
in the region within the shock front, which is consistent with results
obtained by other groups~\cite{Etienne:2010ui,Shibata:2005gp}.

\subsection{Magnetic rotor}\label{sec:rotor}
As a second two-dimensional test problem we simulate the magnetic
rotor test first described in~\cite{Balsara:1999aa} for classical MHD
and later generalised to relativistic MHD in~\cite{DelZanna:2002rv}.
The setup consists of a cylindrical column (the \emph{rotor}) of
radius $r_{\text{in}}=0.1$ and density $\rho_{\text{in}}=10$ embedded
in a medium of lower density $\rho_{\text{out}}=1.0$. Initially the
pressure inside the column and in the medium are equal
$P_{\text{in}}=P_{\text{out}}=1$ and the cylinder rotates with uniform
angular velocity of $\Omega=9.95$ along the cylinder axis, so that the
fluid 3-velocity reaches a maximum value of $v_{max}=0.995$ at the
outer edge of the cylinder.  The fluid outside the cylinder is
initially at rest.  The equation of state used is a $\Gamma$-law, with
$\Gamma=5/3$.  The cylinder is threaded by an initially uniform
magnetic field of magnitude $B^x=1.0$ along the $x$ direction,
covering the entire space; cylinder and exterior. We use TVD reconstruction
with the $\operatorname{minmod}$ limiter, the HLLE Riemann solver, RK2 time
stepping with CFL factor 0.25. The
magnetic field is evolved using the divergence cleaning technique using a
damping factor $\kappa = 5.0$.

At the beginning of the simulation, a strong discontinuity is present
at the edge of the cylinder since we do not apply any
smoothing there. During the simulation, magnetic braking slows down
the rotor while the magnetic field lines themselves are dragged from
their initial horizontal orientation. At the end of the simulation, at
$t=0.4$, the field lines in the central region have rotated by nearly
90 degrees while at large radii the orientation of the field lines
remains unchanged.

\begin{figure}
 \includegraphics[width=0.95\textwidth]{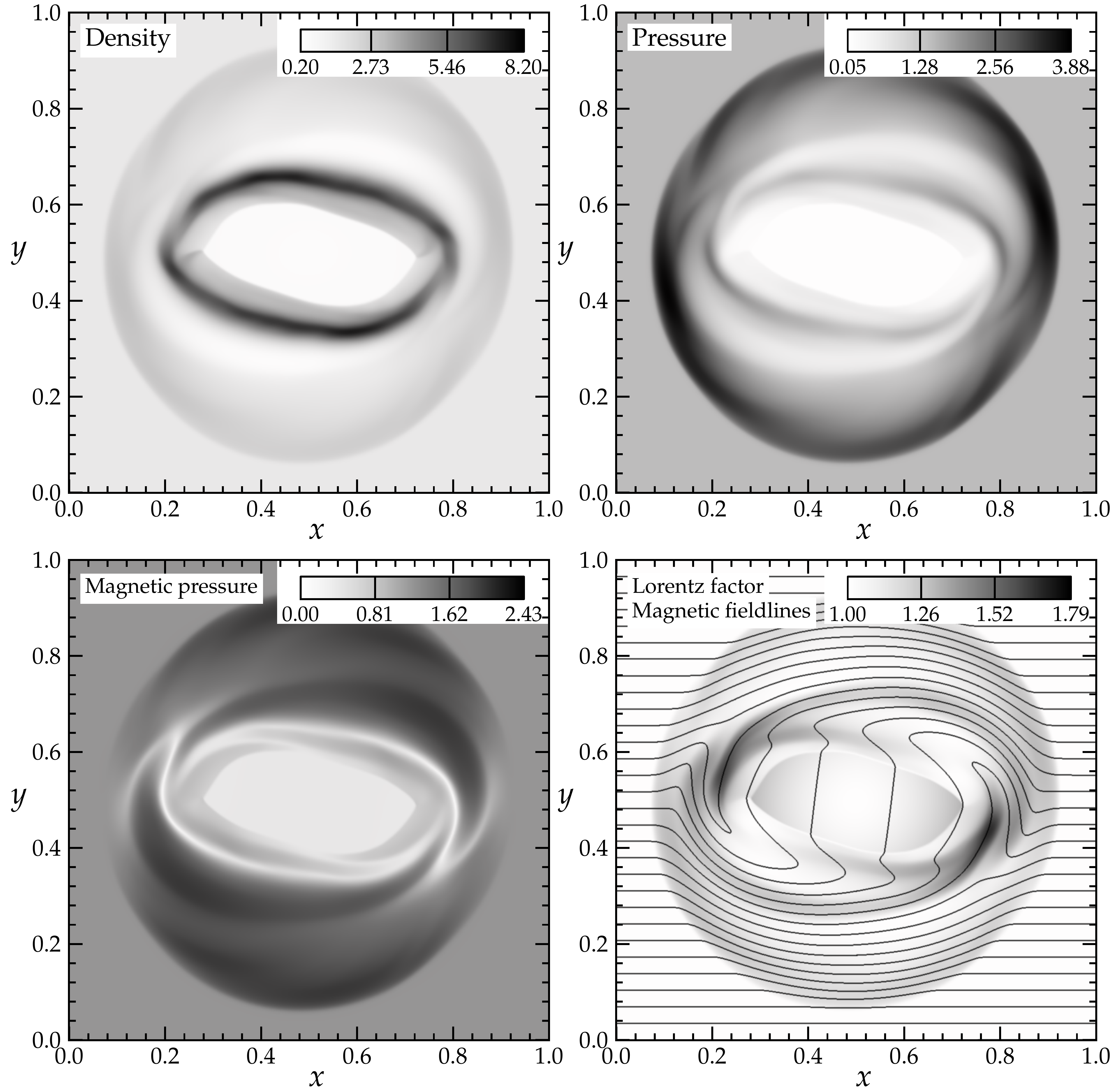}
 \caption{Evolved state of the magnetic rotor test problem, with parameters
based on the test presented in \protect\cite{DelZanna:2002rv}.  At $t=0.4$, we
show the following quantities: top panels: Density $\rho$ and pressure $P$;
bottom panels, magnetic pressure $P_m\equiv \frac{b^2}{2}$ and Lorentz factor
$W$, along with numerically determined magnetic field lines.  Our results are
in close agreement with those shown in
\protect\cite{DelZanna:2002rv,Shibata:2005gp,Etienne:2010ui}.}
\label{fig:rotor-one}
\end{figure}

Our results, shown in figure~\ref{fig:rotor-one} for a grid spacing
$\Delta x={1}/{400}$, compare well with those presented in figure~5
of~\cite{DelZanna:2002rv} and figure~8 of~\cite{Etienne:2010ui}. The
density profile at the end of simulation is reproduced, showing
slightly less noisy behaviour at the location
of the expanding high density shell than the results reported 
in~\cite{DelZanna:2002rv}.
 In addition, the magnetic field
lines displayed in figure~5 of~\cite{DelZanna:2002rv} are also reproduced
very well.  
A slight over-density near the left and right
corners of the (now low density) rotor is present in our simulations
that is absent in~\cite{DelZanna:2002rv}.

\begin{figure}
\includegraphics[width=0.95\textwidth]{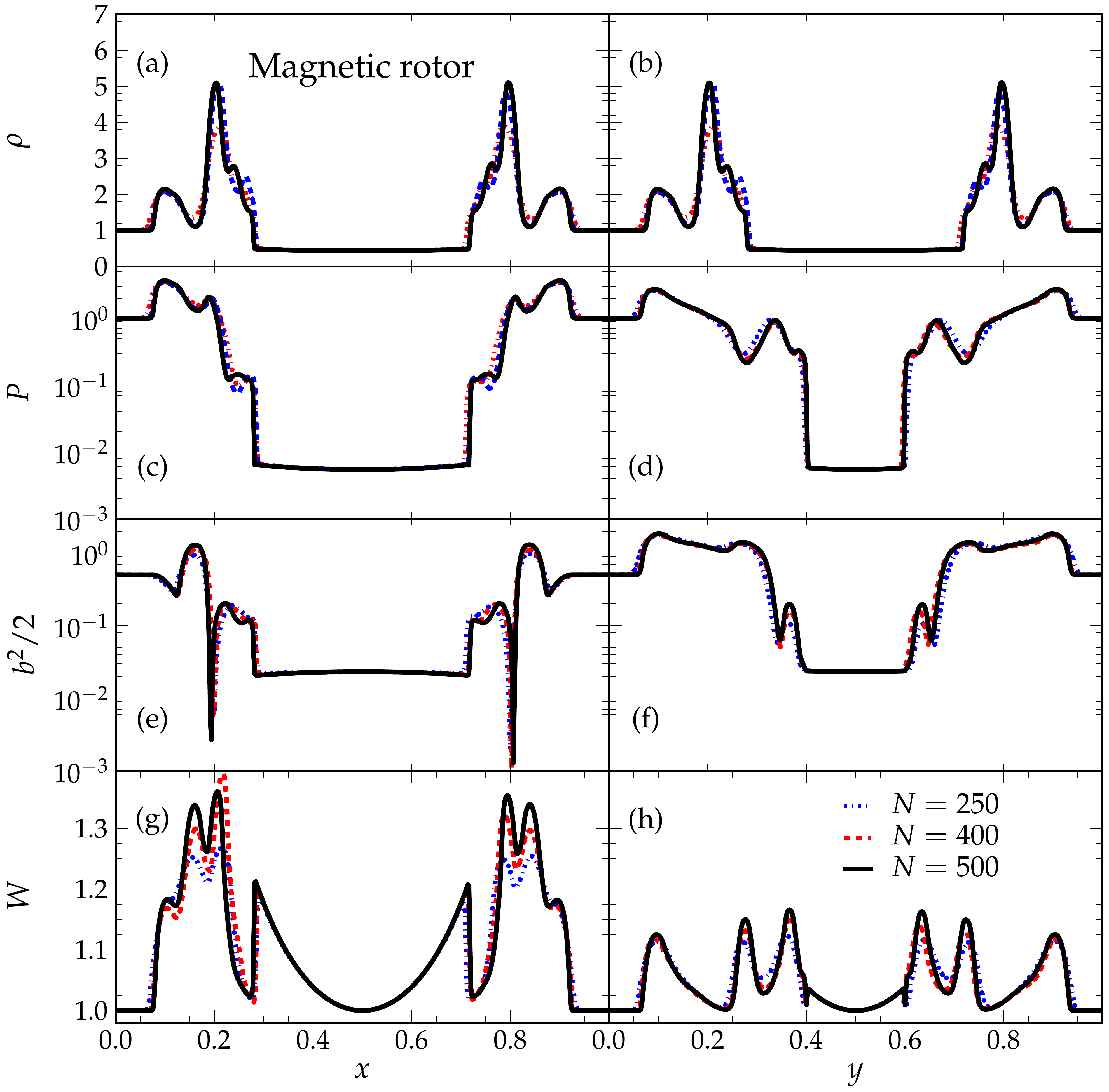}
\caption{Evolved state of the magnetic rotor test problem, after $t=0.4$. We
  display we display data for three different resolutions using $250$,
  $400$, and $500$ cells each to cover the domain. The data was
  extracted along the $x=0.5$ and $y=0.5$ axis respectively.  Our
  results are in very good agreement with figure~9 of
  \protect\cite{Etienne:2010ui}.}
\label{fig:rotor-two}
\end{figure}

Figure~\ref{fig:rotor-two} depicts one-dimensional slices through the
simulation domain at the end of the run. We show results for three
different resolution, using $250$, $400$ and $500$ points to cover the
domain.  We find excellent agreement with previous work at similar
resolutions~\cite{Etienne:2010ui}.

\subsection{Alfv\'en Wave}

The propagation of low-amplitude, circularly-polarised Alfv\'en waves
has an exact solution that is useful for testing the performance of
the MHD sector of the code~\cite{Toth:00} and provides a scenario for
clean convergence tests, since the solution is smooth.  The initial
configuration is a uniform-density fluid with a velocity profile given
by
\begin{equation}
v^x=0;~v^y=-v_A A_0 \cos(kx);~v^z=-v_A A_0\sin(kx)\,\,,
\end{equation}
and corresponding magnetic field configuration
\begin{equation}
B^x={\rm const.};~B^y=A_0 B^x\cos(kx);~B^z=A_0 B^x\sin(kx)\,\,,
\end{equation}
where we choose $B^x=1.0$, an amplitude parameter $A_0=1.0$, $\rho=1.0$. A
$\Gamma$-law EOS with $\Gamma=5/3$ was used, as well as TVD as reconstruction
method, a 2nd order Runge-Kutta (RK)~\cite{Runge:1895aa,Kutta:1901aa}
time integration with the Courant factor set to
$0.2$, and the HLLE Riemann solver, as well as divergence cleaning.

\begin{figure}
 \includegraphics[width=0.95\textwidth]{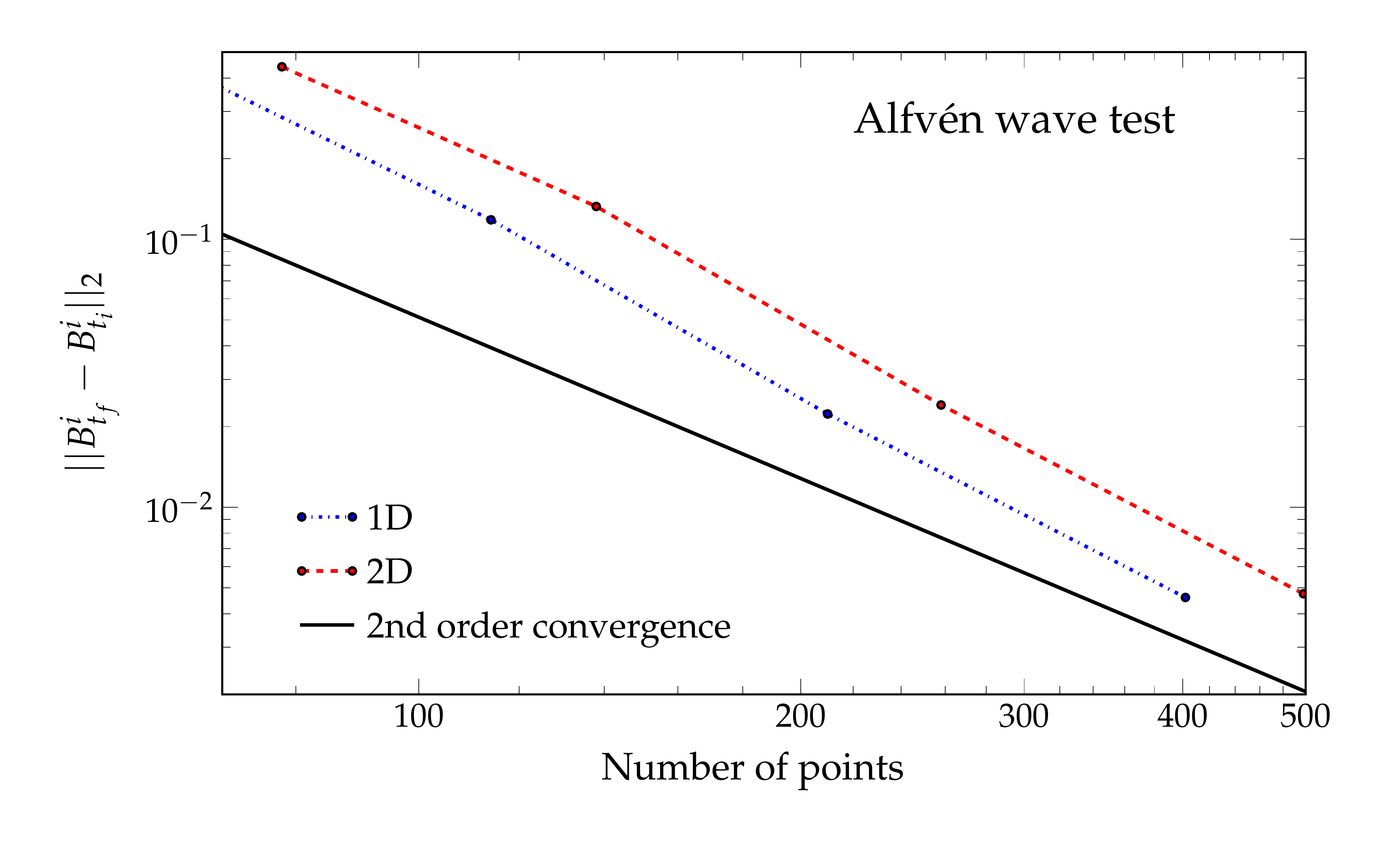}
\caption{Shown is the $L_2$ norm of the difference between the evolved solution
and the initial data of the Alfv\'en wave test. Convergence is observed for the
one-dimensional (1D, blue, dotted line) and two-dimensional (2D, red, dashed
line) test, along with scaling expected for second-order convergence (solid
line). In each case, as a function of $N_x$, the number of points in the
$x$-direction, we plot the $L_2$ norm of the $B$-field along a line drawn in
the $x$-direction through the centre of the computational domain, measuring the
absolute difference between the initial and final values of $B^i$ after 5 full
wave cycles for both the one-dimensional and two-dimensional cases.
The resolutions used here were $(16, 32, 64, 128)$ and 
$(20\times15, 40\times30, 80\times60, 160\times120)$ points in 1D and
2D respectively.
\label{fig:alfvenwave_converge}}
\end{figure}

The Alfv\'en speed is given by the expression
\begin{equation}
v_A^2 = \frac{2(B^x)^2}{\rho h+(B^x)^2(1+A_0^2)}\left[1+\sqrt{1-\left(\frac{2A_0(B^x)^2}{\rho h+(B^x)^2(1+A_0^2)}\right)^2}\right]^{-1}\,\,,
\end{equation}
and the wavevector $k=2\pi/L_x$ for one-dimensional cases.  For
two-dimensional cases, we rotate the coordinates so that wave fronts
lie along diagonals of the grid.  In all cases, periodic boundary
conditions are assumed.  Several values for the pressure have been
used for testing purposes by other groups, spanning the range
$P=0.1$~\cite{Toth:00} to $P=1.0$~\cite{Beckwith:2011iy}, but here we
choose $P=0.5$, which yields the convenient result $v_A=0.5$. In time,
we expect the wave to propagate across the grid, such that if it
travels an integer number of wavelengths we should, in the ideal case,
reproduce our initial data exactly.  In
figure~\ref{fig:alfvenwave_converge}, we show the convergence results
for our code for both 1-dimensional and 2-dimensional cases, finding
the expected second-order convergence with
numerical resolution when plotting the $L_2$ norm of the difference
between the evolved solution and the initial data, evaluated along the
$x-$axis (a full two-dimensional comparison produces the same result).
We note that while the overall error magnitude appears larger than
that shown for one-dimensional and two-dimensional waves in similar
works, e.g.,~\cite{Beckwith:2011iy}, our results are shown after
five periods, and we find comparable
accuracy to~\cite{Beckwith:2011iy} for the first cycle.

\subsection{Loop advection}

In this test, originally proposed in~\cite{DeVore:1991aa} and
presented in a slightly modified form
by~\cite{Toth:1996aa,Stone:2008mh,Beckwith:2011iy}, a region with a
circular cross section is given a non-zero azimuthal magnetic field of
constant magnitude (the ``field loop''). The magnetic field is
initialised to zero outside this region.  The entire fluid
configuration is given a uniform velocity and periodic boundary
conditions are imposed.  The analytical solution is for the field loop
to move with constant velocity across the grid, with no magnetic field
evolution.  With $r_c\equiv \sqrt{x^2+y^2}$ as the cylindrical radius
we have an initial magnetic field
\begin{equation}
B_x,~B_y=\left\{\begin{array}{rl}
-A_{loop}y/r_c,~A_{loop}x/r_c; & r<R_{loop}\,\,,\\
0; & r\ge R_{loop}\,\,.
\end{array}\right.
\end{equation}
In practice, this configuration results from the choice of the vector
potential $\vec{A}=(0,0,\max[0,A_{loop}(R_{loop}-r_c)])$, though we
note that we here set the $B$-field values explicitly in the initial
data, rather than by finite differencing of the $A$-field, an option
also included in the code.

Here, following~\cite{Beckwith:2011iy}, we choose parameters
$A_{loop}=10^{-3},~\rho=1.0,~P=3.0$ for the initial magnetic field and
hydrodynamic configuration.  We consider two different models for the
direction of the field loop: a ``two-dimensional'' case in which it is
oriented vertically, and a ``three-dimensional'' case for which we
rotate the direction of the tube so that it is oriented along the
$x-z$ face diagonal of our rectangular grid. For the initial velocity
field of the two-dimensional configurations, we set
$v^x={1}/{2},~v^y={1}/{24}$ and consider two different cases for the
vertical velocity field parallel to the loop: one in which $v^z=0$, so
that the motion is perpendicular to the field loop, and one in which
$v^z={1}/{24}$ which additionally tests the ability of the
conservative-to-primitive variable inversion scheme to maintain this quantity
unchanged over time.  For the three-dimensional case, we set
$v^x=0.2\sqrt{2},~v^y=0.2,~v^z=0.1$, and then rotate the tube and all vectors
by $\pi/4$ radians in the $x-z$-plane so that the loop runs along the $x-z$
face diagonal.  In all cases, velocity components are chosen to yield integer
intervals over which the flux tube should propagate across the grid, $T=24$ for
the two-dimensional cases and $T=5$ for the three-dimensional case.  Runs were
performed using RK2 timestepping, TVD reconstruction with an monotonized
central limiter, the HLLE Riemann solver, divergence cleaning active with
$\kappa=10$, and  CFL factors of 0.4 for the 2-d cases and 0.25 for the 3-d
case.

\begin{figure}
 \includegraphics[width=0.85\textwidth]{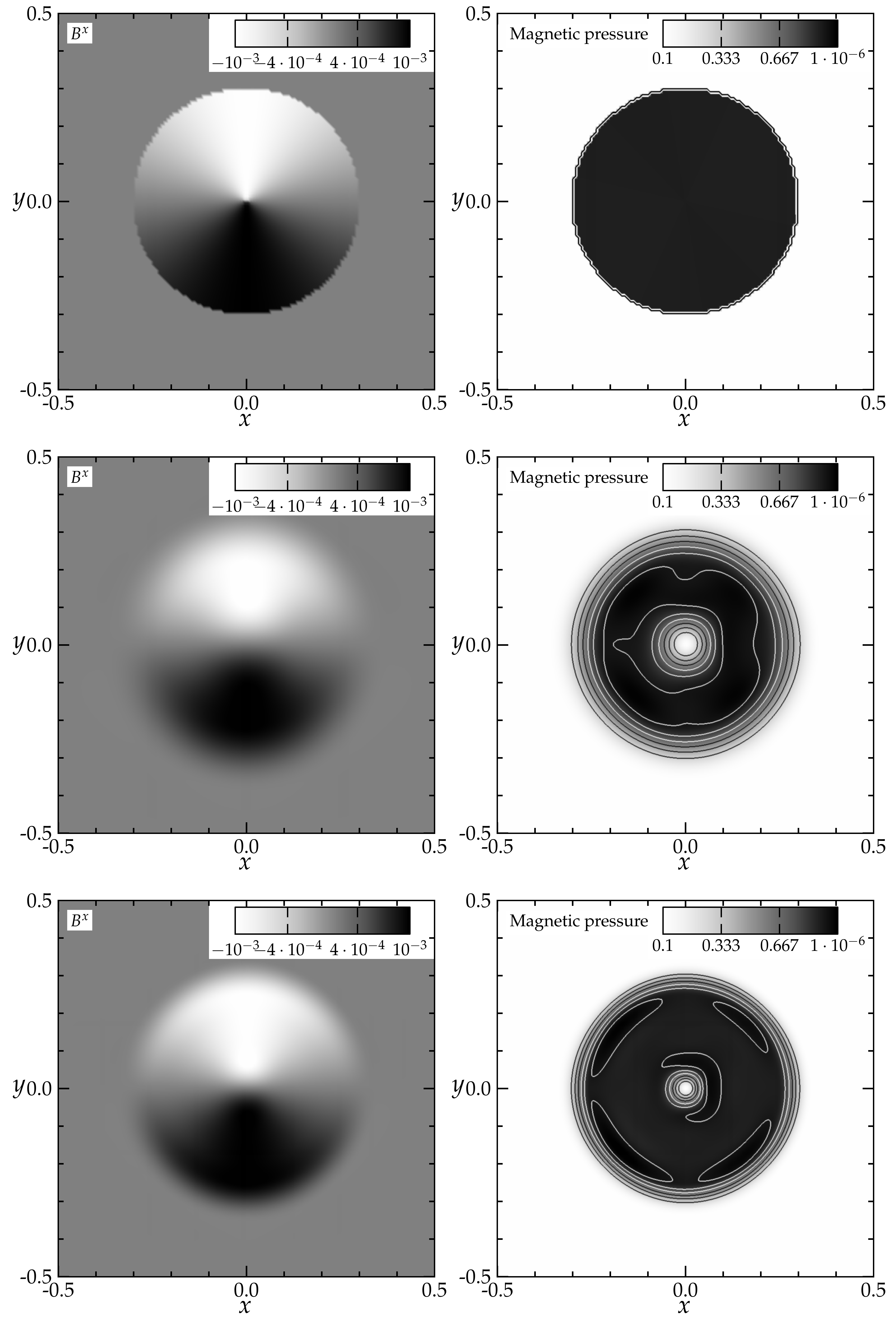}
 \caption{Results of the two-dimensional advected loop test with $v^z = 1/24$.
   In the top panels, we show the initial state of $B^x$ in greyscale,
   along with the (uniform) magnetic pressure.  In the centre and
   bottom rows, we show the result of low ($\Delta x={1}/{128}$) and
   high ($\Delta x = 1 / 256$) resolution simulations after one period
   has passed ($T=24$). The loss of magnetic pressure in the centre of
   the loop is greatly reduced in the higher-resolution
   simulation.}\label{fig:advloop-one}
\end{figure}

\begin{figure}
 \includegraphics[width=0.829\textwidth]{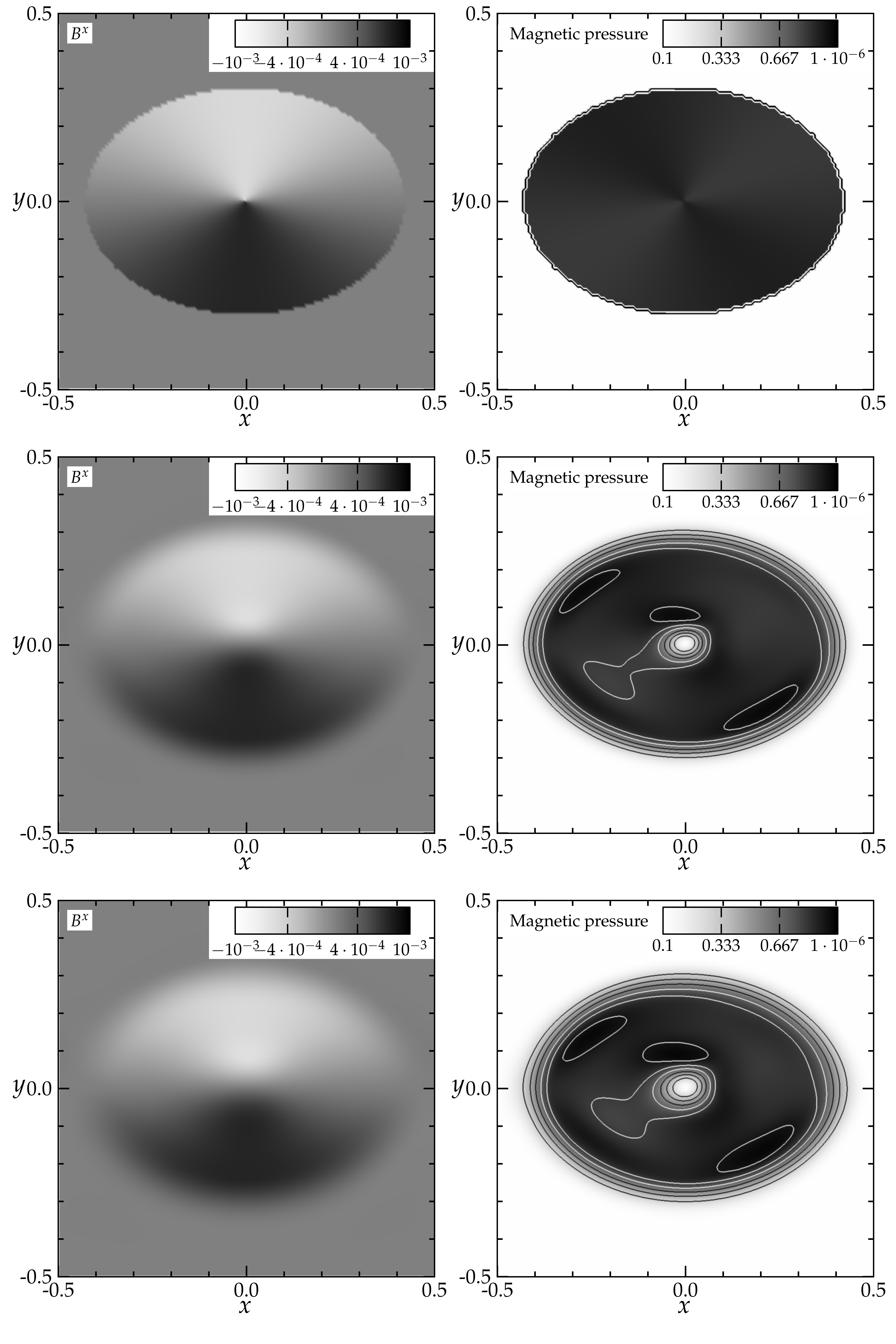}
 \caption{Results of the three-dimensional advected loop test with $v^z\ne 0$. 
   In the top panels, we show the 2d $z=0$ slices through the 
   initial state of $B^x$ in greyscale, along
   with the (uniform) magnetic pressure.  In the centre and bottom
   rows, we show the results after one ($T=5$) and two ($T=10$)
   periods, respectively.  The flux tube formed by the moving flux loop is
   oriented diagonally in the domain, hence the slice is elliptical rather than
   circular.  The grid spacing used is $\Delta x =
   {1}/{128}$.  Our result is extremely similar in how well it preserves 
   the magnetic flux and pressure of the loop to the results shown in 
   figure 4 of~\protect\cite{Beckwith:2011iy}.}\label{fig:advloop-two}
\end{figure}

In the top panels of figure~\ref{fig:advloop-one}, we show the initial
state of the flux tube of the two-dimensional case, with $B^x$ on the
left and the magnetic pressure $P_m = {b^2}/{2}$ on the right.  In the
centre panels, we show the results of a low-resolution simulation,
with $\Delta x={1}/{128}$, and in the bottom panels a high-resolution
simulation with $\Delta x={1}/{256}$. Both simulations are evolved for
one grid crossing time, $T=24$, with $v^z = 1/24$.  We find no
substantial differences in the results after one period between $v^z =
0 $ and $ v^z = 1/24$. While the observable decrease in the central
magnetic pressure is the inevitable result of the simulation, we find
the effect is significantly decreased by the increase in resolution.
We find similar results for the three-dimensional loop, whose
evolution is depicted by figure~\ref{fig:advloop-two}, and demonstrates 
comparable uniformity in the pressure and field strength for a given
resolution.  From top to bottom, we show the initial state and the states after
one and two periods ($T=5$ and $T=10$, respectively).  The initial pressure,
which should remain invariant, is maintained throughout the body of the loop,
but decreases in the central region and near the edges, as expected (this
result can be compared with figure~4 of~\cite{Beckwith:2011iy}, with which we
find excellent agreement).

\subsection{Bondi Accretion}

Bondi accretion is a steady-state solution for spherically-symmetric,
adiabatic accretion onto a black hole. The Einstein Toolkit
implementation of the general relativistic analogue of Bondi's
original solution~\cite{Bondi:1952ni} follows the methodology laid out
in~\cite{Shapiro:1983du}, assuming a fixed background spacetime of a
Schwarzschild black hole of mass $M_{\rm BH}$. The solution is determined by the
continuity equation
\begin{equation}
\dot{M} = 4\pi \hat{\rho} \hat{u} \hat{r}^2\,\,,
\end{equation}
and the integrated Euler equation for a polytropic equation of state,
\begin{equation}
\dot{Q} = \hat{h}^2 ( 1 - \frac{2 M_{\rm BH}}{\hat{r}} + \hat{u}^2 )\,\,,
\end{equation}
where $\hat{}$ denotes a variable in Schwarzschild coordinates, $h$ is
the specific enthalpy, $u =-u_i r^i$, and $c_s$ the speed of sound.  
Adding a \emph{radial} magnetic field to the system does not alter the Bondi
solution.  Following~\cite{Etienne:2011ea}, we set
the radial magnetic field by prescribing a primitive magnetic field vector of
\begin{equation}
B^i = \frac{B_0 M_{\rm BH}^2 x^i}{\sqrt{\gamma} r^3}\,\,,
\label{eqn:bondi-magnetic-field}
\end{equation}
where $r = \sqrt{x^2 + y^2 + z^2}$ is the coordinate radius and $B_0$
controls the magnetic field strength. A magnetic field vector of this form by
construction satisfies the divergence free
constraint~\eref{eqn:divB_constraint}.

A particular solution is chosen by specifying the BH mass $M_{\rm
  BH}$, the mass accretion rate $\dot{M}$, the location of the sonic
point in Schwarzschild coordinates (i.e., the areal radius)
$\hat{r}_s$, the magnetic field strength $B_0$ and the adiabatic index
$\Gamma$ of the equation of state. Table~\ref{tab:magn_bondi_pars}
lists the values of the parameters used in our simulations. We explore
various values for $B_0$ and $\dot M$ but keep the sonic radius
$\hat{r} = 8\,M_{\rm BH}$ and adiabatic index $\Gamma = 4/3$ fixed.

By requiring the solution to remain smooth
at the sonic
point, the sound speed and thus velocity of the fluid at that point
must satisfy $\hat{u}_s = c_s = M_{\rm BH}/(2\hat{r}_s)$.  From this
equation we find the polytropic constant $K$
for the EOS and subsequently from the continuity equation the conserved
quantity $\dot{Q}$.  The
solution in the remaining domain is then determined by integrating
outwards (or inwards) from the sonic point before converting to the
desired coordinate system of the evolution. We choose to work in ingoing
Kerr-Schild coordinates which are horizon penetrating and therefore
allow us to model the flow of matter across the event horizon.

To avoid numerical issues due to the singularity inside of the event horizon
we use a transition function
\begin{equation}
\chi(r) = 
  \cases{
  \frac{1}{2} \left[1+\tanh\left(\tan\left(\frac{\pi r}{M_{\rm BH}}-\frac{\pi}{2}\right)\right)\right] & $r < M_{\rm BH}$ \\
  1 & otherwise\\
  }
\label{eqn:bondi_transition_function}
\end{equation}
to damp magnetic field $B^i \rightarrow \chi \vec B^i$ and velocity
$v^i \rightarrow \chi v^i$ to zero near the singularity. Similarly we reduce
the mass density, pressure and internal energy density to values corresponding
to our atmosphere level $\rho_{\text{atmo}}$ which is $~10^{-8}$ of the
density at the event horizon. 

Our numerical setup follows~\cite{Giacomazzo:2007ti}, i.e.\ we
simulate a rectangular box of size $[0,11] \times [0,11] \times
[-11,11]$ using resolution of $0.11$ for the low resolution and
$0.073$ for the high resolution run. Since the system is spherical, we
used a 90 degree rotational symmetry around the $z$ axis to reduce the
computational costs. We use a HLLE Riemann solver, TVD
reconstruction using a $\operatorname{minmod}$ limiter. We integrate in
time using a third-order Runge-Kutta (RK3) integrator with Courant
factor $0.25$. For this test we employed our constrained transport
scheme which we found to be more stable in the strongly magnetically
dominated regimes explored by us in this test. After each RK
substep we re-set the grid inside of $\hat{r} < 1<M_{\rm BH}$ and at
the outer boundary to the analytic initial solution, providing
Dirichlet type boundary data.

We note that due to our choices of numerical timestepping method, stencil size
and resolution grid points outside of the event horizon $\hat{r} = 2\,M_{\rm
BH}$ are not affected by grid points inside of $\hat{r} = 1\,M_{\rm BH}$. Thus
our method is effectively excising the interior of $r<1\,M_{\rm BH}$ from the
simulated region.

\begin{figure}
 \includegraphics[width=0.85\textwidth]{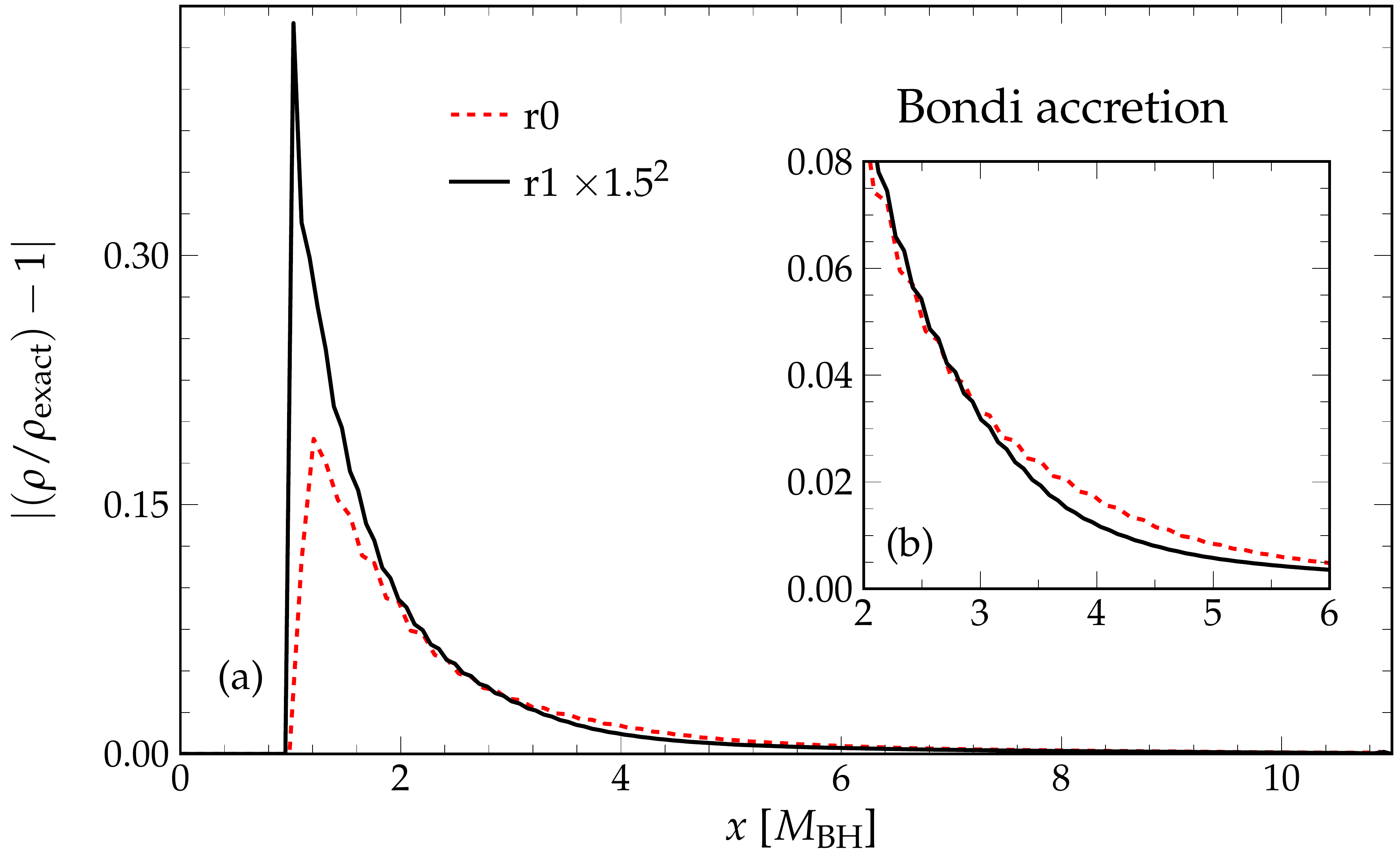}
 \caption{Bondi inflow after 100$M_{\rm BH}$ of evolution compared
   with the semi-analytic solution for varying numerical
   resolution. We observe 2\textsuperscript{nd} order convergence in
   the region outside of the event horizon at $r=2\,M_{\rm
     BH}$. The red dashed line shows the errors in $\rho$ compared to the
   analytical solution for the low resolution run r0 with resolution $0.11$ while
   the black solid line shows the error in the high resolution run r1 using a
   resolution of $0.073$ scaled so that for second order convergence the curves
   coincide with each other. This inset shows a zoom in of the region around the 
   event horizon where matter is moving fastest and the magnetic field is
   strongest, i.e.\  where we expect the largest errors. Our errors are
   almost identical to those found by~\cite{Giacomazzo:2007ti}.}
          \label{fig:mag_bondi}
\end{figure}

To verify the accuracy of our code, we evolve the analytical initial data for
$T=100\,M_{\rm BH}$ and compare the data along the $z=0.055$,
$y=0$ line to the analytic initial data. We display the result of this comparison in
figure~\ref{fig:mag_bondi}. We observe 2\textsuperscript{nd} order convergence
with increasing resolution in the region outside of the event horizon at
$r=2\,M_{\rm BH}$. Our results compare very well with those of~\cite{Giacomazzo:2007ti}, figure~6. 
 
To adequately test the stability of the
steady-state solution, we repeat this test and vary both the accretion rate
and the magnetic field
strength. We also include a pure hydrodynamic simulation for each accretion
rate.
In Table~\ref{tab:magn_bondi_pars}, we list the set of parameters and their
\begin{table}
 \caption{List of tests performed for the magnetised Bondi
infall. The first columns shows the run name, the second the magnetic field
strength parameter, the third the (unitless) mass accretion rate, the fourth
the ration of magnetic pressure to gas density at the horizon, the fifth the
resolution and the last column lists the maximum
relative error in $\rho$ outside of $r = 2M$ for each test
run. }\label{tab:magn_bondi_pars}
\begin{indented}
\item[]\begin{tabular}{lr@{.}lr@{.}lr@{.}lr@{.}lc}
\br
Run & 
\multicolumn{2}{c}{$B_0$} & 
\multicolumn{2}{c}{$\dot M$} & 
\multicolumn{2}{c}{$b^2/\rho$} &
\multicolumn{2}{c}{$\Delta x$} &
$\left|(\rho / \rho_{\rm exact}) - 1\right|_{\infty}$ \\
\mr
A &  0&0  & 12&57 &  0&0  & 0&11  & $9.70\cdot10^{-5}$ \\
B &  0&0  & 12&57 &  0&0  & 0&073 & $4.60\cdot10^{-5}$ \\
C & 11&39 & 12&57 & 25&0  & 0&11  & $7.42\cdot10^{-2}$ \\
D & 11&39 & 12&57 & 25&0  & 0&073 & $3.92\cdot10^{-2}$ \\
E & 11&39 & 25&13 & 12&5  & 0&11  & $3.88\cdot10^{-2}$ \\
F &  5&7  & 12&57 &  6&26 & 0&11  & $2.21\cdot10^{-2}$ \\
G &  5&7  & 25&13 &  3&13 & 0&11  & $1.37\cdot10^{-2}$ \\
\br
\end{tabular}
\end{indented}
\end{table}
values that we vary for this test along the supremum norm of the
relative error on the grid outside of the event horizon after
$t=100\,M_{\rm BH}$ of evolution. 

We found that higher order reconstruction method such as \codename{PPM} and
methods employing divergence cleaning performed less well in this test. In the
case of divergence cleaning we suspect that our inability to specify initial
data with vanishing \emph{numerical} divergence is partially the source of
these difficulties. Even in the constrained transport case, we observe a
certain amount of initial drift away from the analytic solution in the field
values. Eventually the solution settles down and we the amount drift to
decrease with increasing resolution. In the case of divergence cleaning, our
imperfect boundary data serves as a constant source for constraint violations,
possibly spoiling the result. Similarly the interpolation inherent in adaptive
mesh refinement techniques seems to serve as a source of constraint violation
and prevent us from simulating this system using AMR and either divergence
cleaning or constraint transport. 

\subsection{Magnetised TOV Star}
\begin{figure}
 \includegraphics[width=0.95\textwidth]{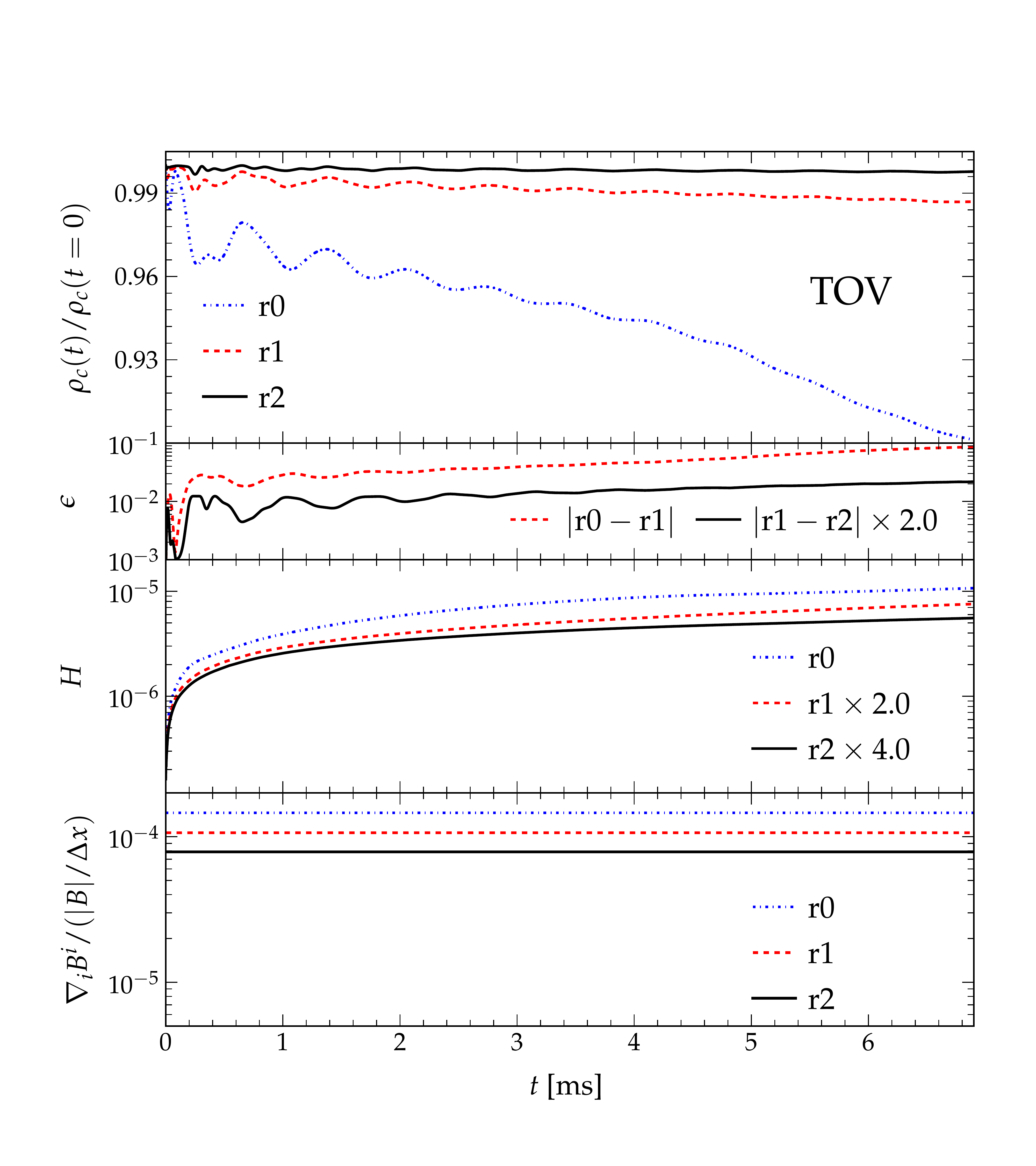}
\caption{Results of the magnetised TOV simulation.
  First panel: Normalised maximum rest-mass density $\rho_{\rm
    max}/\rho_{\rm max,0}$. Second panel: $L_2$-norm of the
  Hamiltonian constraint $||H||_2$ for the three resolutions $r0$,
  $r1$, and $r2$. Third panel: Absolute difference $\epsilon$ in
  normalised maximum rest-mass density between low ($r0$) and medium
  ($r1$), and medium ($r1$) and high ($r2$) resolutions are shown
  (centre panel). Fourth panel: $L_2$-norm of the normalised
  divergence of the magnetic field $|\nabla_i B^i / (|B|\, \cdot\, \Delta
  x^{-1})|_2$, where $\Delta x$ is the finest grid spacing,
  for each of the the three resolutions $r0$, $r1$, and $r2$ and 
  $|\vec{B}|_{\infty}$ is the maximum magnetic field strength on the grid.
  The
  differences and the $L2$-norm of the Hamiltonian constraint and the
  divergence of magnetic field are scaled for first order
  convergence.}
\label{fig:tov_conv}
\end{figure}

In this section we consider the stability of a stationary, poloidally
magnetised spherically symmetric neutron star model, obtained via the
solution of the Tolman-Oppenheimer-Volkhoff
equation~\cite{tolman:1939}.  We follow~\cite{Duez:2005cj} and set up
a poloidal magnetic field on top of the independently solved fluid
configuration.  This setup provides a more challenging test for the
code than the previous ones since it probes both the GR and the MHD
evolution. While the initial configuration is stationary, we evolve
both the spacetime and the GRMHD equations to test the ability of the
new code to maintain such a solution. The hydrodynamic initial data
is a TOV star generated using a polytropic EOS with
polytropic constant $K=100$, adiabatic index $\Gamma = 2.0$ and
initial central density of $1.28\cdot 10^{-3} M_{\odot}$. The fluid
is evolved with a $\Gamma$-law EOS and $\Gamma = 2$. As
in~\cite{Liu:2008xy}, we introduce a poloidal field that we add on top
of the TOV initial data.  We choose the vector potential $A_\varphi =
A_b\,\varpi^2 (1 - {\rho_0}/{\rho_0^{\rm max}})^{n_p}
\max{(P-P_{\rm cut},0)}$ where $\varpi \equiv \sqrt{(x-x_\star)^2 +
  y^2}$. Here $A_b$ determines the strength of the
initial magnetic field, $n_p$ the offset of the maximum of
the magnetic field strength in respect to the maximum density
$\rho_{max}$ of the fluid, and $P_{\rm cut}$ the pressure
outside of which the magnetic field is set to zero. In this test we
confine the magnetic field to the interior of the star by setting
$P_{\rm cut} = 10^{12} P_{\rm atmo}$, much higher than the atmospheric
pressure.  The parameter $A_b$ is set to unity, which, together with
$n_p=0$, provides a magnetic field with a maximum strength of
$8.5\times10^{-6}$ in geometrised units which corresponds to
$2.5\times10^{15}~\rm{G}$.
 
We place the TOV star with radius $8.13 M_{\odot}$ and gravitational
mass $1.40M_{\odot}$ at the origin of a grid with 5 levels of mesh
refinement with cubical extents $320M_{\odot}$, $120M_{\odot}$,
$60M_{\odot}$, $30M_{\odot}$, and $15M_{\odot}$.  The extent of the
finest level is chosen in order to cover the entire star.  We perform
three simulations with coarsest resolutions of $8 M_{\odot}$
($r0$). $4 M_{\odot}$ ($r1$), and $2 M_{\odot}$ ($r2$). The
corresponding resolutions of the finest grids are $1.0 M_{\odot}$
($r0$), $0.5M_{\odot}$ ($r1$) and $0.25M_{\odot}$ ($r2$). We
simulate for 6.9 ms using oPPM reconstruction, the HLLE Riemann
solver, and evolve the magnetic field with constrained
transport. 
We use 4th-order Runge-Kutta time integration with a Courant factor of
0.25. We set the $\Gamma$-Driver gauge condition~\cite{Alcubierre:2002kk}
parameter to $\eta = 0.5$ and add Kreiss-Oliger dissipation to the
spacetime variables with a dissipation coefficient of $\epsilon =
0.1$. 

Figure~\ref{fig:tov_conv} shows the fractional change of the maximum
rest-mass density $\rho$ for all three resolutions, the differences
between low ($r0$) and medium ($r1$), and medium ($r1$) and high
($r2$) resolutions in maximum rest mass density and the $L_2$-norm of the
Hamiltonian constraint $||H||_2$. The differences in rest-mass density
and the Hamiltonian constraint are scaled for first order convergence.
While the TOV solution is stationary, numerical perturbations excite
oscillations in the central density of the star, commonly observed in
simulations of TOV stars, which must converge away with increasing
resolution. We can analyse the convergence properties of our
simulations more carefully by considering the second panel of
figure~\ref{fig:tov_conv}, which shows the absolute differences
between low ($r0$) and medium ($r1$), and medium ($r1$) and high
($r2$) resolutions in the normalised maximum rest-mass density. The
differences are scaled by a factor of $2$ for first-order
convergence. We observe between first and second order convergence,
which is the expected behaviour, since the oPPM scheme reduces to
first-order accuracy at the surface of the star.
The third panel of figure~\ref{fig:tov_conv} shows the $L_2$-norm of
the evolution of the Hamiltonian constraint $||H||_2$ for the
resolutions $r0$, $r1$ and $r2$.  The medium and high resolutions are
rescaled for first-order convergence. We again observe between
first-order and second-order convergence, consistent with what is
expected. The fourth panel of figure~\ref{fig:tov_conv} shows the
$L_2$-norm of the normalised divergence of the magnetic field as a
function of time for the resolutions $r0$, $r1$ and $r2$. Again, we
observe between first-order and second-order convergence. Note that the
lines are constant, since constrained transport preserves the initial
constraint violation to round-off.

\subsection{Rotating Core Collapse}
\begin{figure}
\begin{center}
\includegraphics[width=0.95\textwidth]{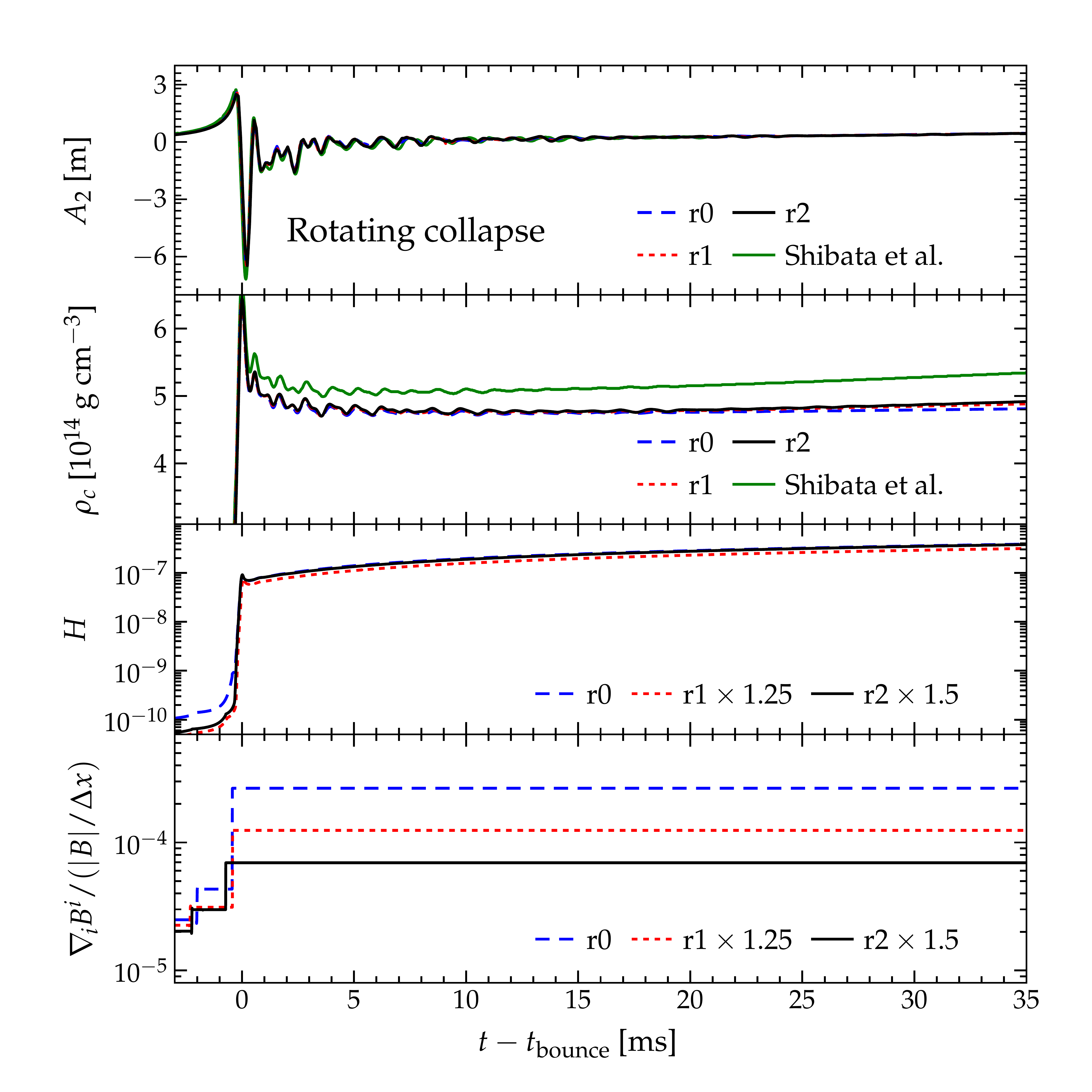}
\caption{First panel: Convergence of the gravitational wave signal
  $A_2$ computed via the quadrupole formula as in~\cite{Shibata:2006hr}. 
  and gravitational wave signal from~\cite{Shibata:2006hr} for 
  the corresponding model. Second panel: Central rest-mass density
  $\rho_c$. Third panel: $L_2$-norm of the Hamiltonian constraint
  $||H||_2$.  Fourth panel: $L_2$-norm of the normalised divergence of
  the magnetic field $|\nabla_i B^i / (|B|\cdot\Delta x^{-1})|_2$.
  All four quantities are shown for the three resolutions $r0$, $r1$,
  and $r2$. The $L_2$-norm of the Hamiltonian constraint and 
  the normalised divergence of the magnetic
  field in panels three and four are scaled for first order
  convergence. The constraints initially exhibit second-order
  convergence, which drops to first order after core bounce generates
  a shock. The divergence of the magnetic field is kept
  constant by the constrained transport scheme, unless a change in the
  grid structure occurs. This causes the jumps in the time evolution
  of $|\nabla_i B^i / (|B|\cdot\Delta x^{-1})|_2$ since the newly created
  grids are populated using non-constraint-preserving operators.}
\label{fig:A2_conv}
\end{center}
\end{figure}
 
\begin{figure}
 \includegraphics[width=0.95\textwidth]{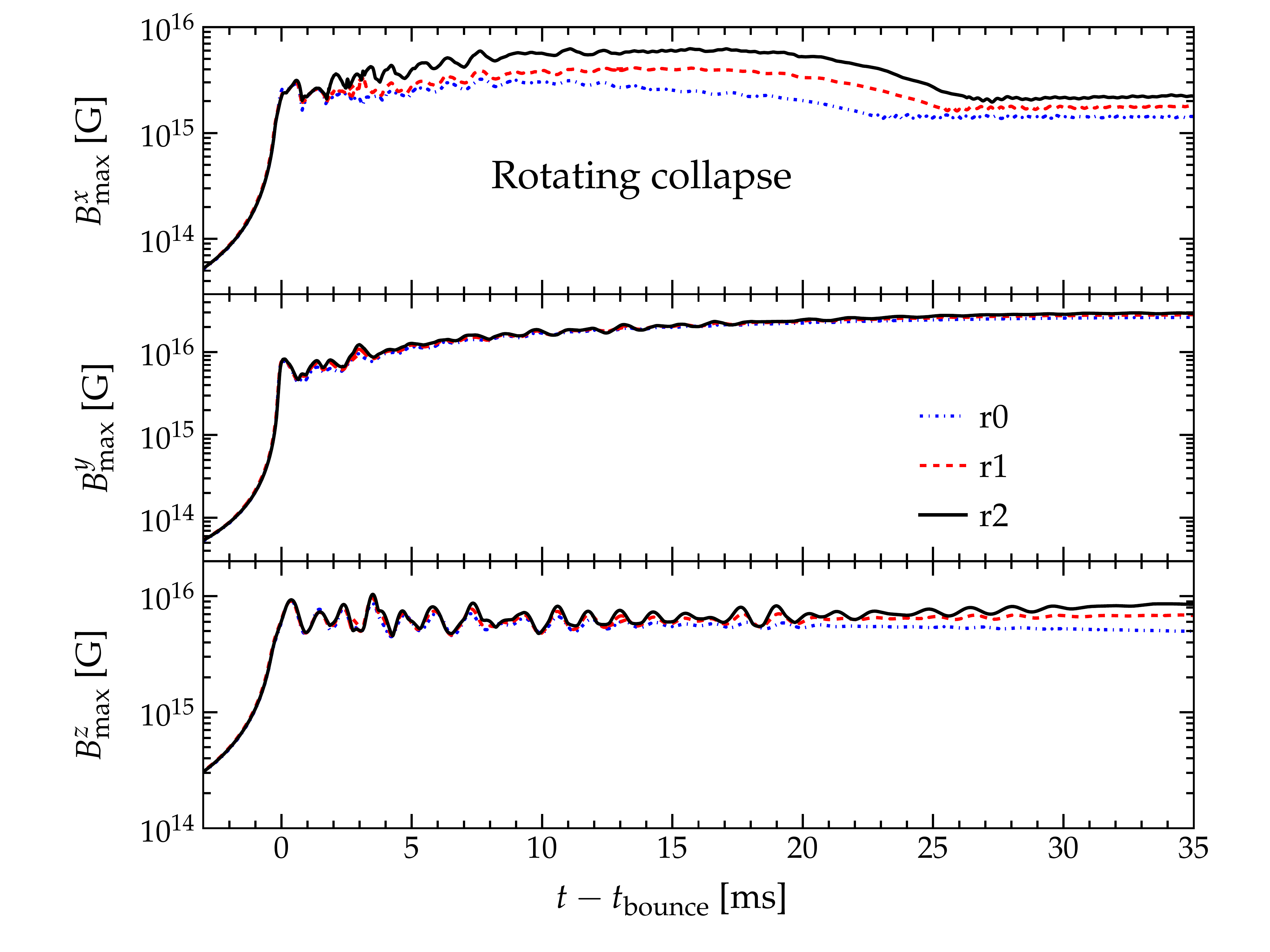}
\caption{Convergence of the maximum value of the magnetic field
  components $B^i$ for the three different resolutions $r0$, $r1$ and
  $r2$ in the rotating core collapse simulation. Differences between
  the three resolutions are most obvious for $B^x$ (top panel), but
  also present for $B^y$ (middle panel) and $B^z$ (bottom panel). This
  is however to be expected, as differences in the maximum of the
  magnetic field components may arise from local turbulent behaviour
  and depend sensitively on resolution. We observe a similar overall
  evolution of the magnetic field maxima as in~\cite{Shibata:2006hr}. 
  We note that the  differences in evolution for $B^x_{\mathrm{max}}$ and
  $B^y_{\mathrm{max}}$ result from the fact that we choose to plot
  the maximum value and not the maximum of the absolute value. As our 
  domain is symmetric under rotations by 90 degrees about the $z$-axis, 
  there is a sign flip between $B^x$ and $B^y$ in any given quadrant 
  of our domain, i.e. $B^x_{\mathrm{max}} = -B^y_{\mathrm{min}}$.}
\label{fig:A2_bmax}
\end{figure}

As a final test, we present results for the collapse of a rotating
magnetised stellar core with simplified microphysics. We analyse a
number of quantities to demonstrate the ability of our code to
simulate magnetised stellar collapse in full GR.  For this, we
consider model A1 of~Shibata~\emph{et~al}.~\cite{Shibata:2006hr}, who
carried out simulations in axisymmetry. While our code is 3D, we can
limit non-axisymmetric dynamics to even $\ell$ and $m$ (in terms of
spherical harmonics) that are multiples of $4$ by evolving only in one
octant of the 3D domain and enforcing rotational symmetry on the two
inner faces of the octant and reflection symmetry in the z-direction.
We model the precollapse star as a $\Gamma={4}/{3}$-polytrope with a
central density of $\rho_c = 10^{10}\,\mathrm{g}\,\mathrm{cm}^{-3}$
and a polytropic constant of $K=4.897\cdot 10^{14} [cgs]$. This
corresponds to an ultra-relativistic degenerate electron gas at
electron fraction $Y_e = 0.5$.  We choose the initial rotation profile
as $u^tu_{\phi} =\varphi_d^2\left(\Omega_c-\Omega\right)$, where
$\Omega_c$ is the central angular velocity, $\Omega$ is the angular
velocity, and $\varphi_d$ is a constant describing the degree of
differential rotation. Model A1 of~\cite{Shibata:2006hr} is uniformly
rotating ($\varphi_d \rightarrow \infty$) and has a ratio of polar to
equatorial radius $R_p/R_e = 0.667$. The initial data were generated
using the RNS code~\cite{stergioulas:95}.  During evolution, we employ a
hybrid EOS~\cite{janka:93,dimmelmeier:02}, consisting of a two-piece
polytropic cold pressure component $P_\mathrm{P}$ and a thermal pressure
component $P_\mathrm{th}$,
\begin{equation}
P=P_\mathrm{P} + P_\mathrm{th}\,\,.
\end{equation}
Following~\cite{Shibata:2006hr}, in the cold part, the adiabatic index
jumps from $\Gamma_1 = 1.3$ to $\Gamma_2 = 2.5$ at $\rho \ge
2\times10^{14}\,\mathrm{g}\,\mathrm{cm}^{-3}$ to mimic the stiffening
of the nuclear EOS at core bounce. The thermal component of the hybrid
EOS is a $\Gamma$-law with $\Gamma_\mathrm{th} = \Gamma_1$. The
specific internal energy is given by $\epsilon = \epsilon_P +
\epsilon_\mathrm{th}$ and consists of a polytropic and a thermal
part. We induce collapse by following the prescription outlined
in~\cite{Shibata:2006hr}, that is we set the internal energy according
to $\epsilon = 3K \rho^{1/3}$ and then determine the pressure as $P
= 3K(\Gamma_1 - 1)\rho^{4/3}$. This reduces the pressure from its
initial value by a factor of $(1-3(\Gamma_1-1))$ (i.e.\ 10\% for
$\Gamma_1=1.3$).  Following~\cite{Shibata:2006hr}, we assume a
poloidal initial magnetic field computed from the vector potential
\begin{equation}
A_{\phi} = A_b \varpi^2 \max \left[\rho^{1/n_b}-\rho_{cut}^{1/n_b},0\right]\,\,,
\end{equation}
with $n_b = 1$, $\rho_\mathrm{cut} = 10^{-4}\rho_c$ and $\varpi =
\sqrt{x^2+y^2}$. The parameter $A_b$ is chosen so that the maximum
initial magnetic field strength is $7.2\times 10^{12}G$.  For
convenience, we summarise the model parameters in
Table~\ref{table:rotcol}.
\begin{table}[t]
\caption{Initial parameters and properties of model A1.}
\label{table:rotcol}
\begin{indented}
\item[]\begin{tabular}{lll}
\br
Central rest-mass density $[\mathrm{g}\,\mathrm{cm}^{-3}]$ & $\rho_c$ & $1.0\times10^{10}$ \\
ADM mass $[M_\odot]$      & $M_{\rm ADM}$      & $1.503$ \\
Baryonic mass $[M_\odot]$ & $M_B$             & $1.503$ \\
Ratio of rotational kinetic to potential energy & $T/W$ & $8.9\times10^{-3}$ \\
Equatorial radius [km] & $R_e$      & $2267$ \\
non-dimensional angular momentum parameter & $J/M_{\rm ADM}^2$ & $1.235$ \\
Ratio of polar to equatorial radius & $z/R_e$ & $0.667$ \\
Differential rotation parameter & $\tilde{A}$ & $\infty$ \\
Angular velocity at the rotation axis & $\Omega_c$ & $4.11$ \\
Maximum initial magnetic field strength [G] & $|B^i|_{max}$ & $7.2\times 10^{12}$\\
Maximum initial ratio of magnetic and fluid pressure & $(P_\mathrm{mag}/P)_\mathrm{max}$ & $2.1\times 10^{-4}$\\ 
Ratio of magnetic to rotational kinetic energy & $E_\mathrm{mag}/T_\mathrm{rot}$ & $1.9\times 10^{-3}$ \\
\br
\end{tabular}
\end{indented}
\end{table}
Our numerical setup initially consists of 4 levels of nested grids
each refined by a factor of $2$ in resolution.  As the star collapses,
and the central density increases, we progressively add 5 additional
refinement levels, following the approach taken in
\cite{ott:07prl}. The density thresholds are chosen to be
$[8.0\times 10^{10}$, $3.2\times 10^{11}$, $1.28\times10^{12}$,
  $5.12\times10^{12}$,
  $2.05\times10^{13}]\,\mathrm{g}\,\mathrm{cm}^{-3}$.  The final grid
structure consists of 9 levels of nested grids with cubical extents
[$5669.8\, \mathrm{km}$, $3061.7\,\mathrm{km}$, $2438\,\mathrm{km}$,
  $1559.2\,\mathrm{km}$, $283.5\,\mathrm{km}$, $212.6\,\mathrm{km}$,
  $144.7\,\mathrm{km}$, $59.1\,\mathrm{km}$, $23.6\,\mathrm{km}]$. The
extent of the finest level is chosen such that it covers the entire
compact remnant (the ``proto-NS'') after core bounce.  The inner-most
refinement level for our fiducial setup ($r1$) has a linear resolution
of $369.1\,\mathrm{m}$. We consider two more resolutions, $r0$ with
fine grid resolution $461.4\,\mathrm{m}$ and $r2$ with fine grid
resolution $295.3\,\mathrm{m}$. This allows us to study the resolution
dependence of our results.  We use vertex-centred AMR, constrained
transport, and TVD reconstruction with the van Leer monotonized central
limiter. While {\tt GRHyro} includes higher-order reconstruction
schemes, we use TVD in these simulations to match the numerical
methods used in~\cite{Shibata:2006hr} as closely as possible.  The
time integration is performed using RK4 with a Courant factor of
$0.2$. The damping coefficient of the $\Gamma$-driver gauge condition
for the spacetime evolution is set to $\eta = 1/2$. We add
5th-order Kreiss-Oliger dissipation to the spacetime variables with a
dissipation coefficient $\epsilon = 0.1$. The atmosphere is chosen to
be a factor $10^{10}$ smaller than the central density. Shibata et
al.~\cite{Shibata:2006hr} extract the gravitational wave signal from
rotating core collapse and bounce via a quadrupole formula and define
a wave amplitude
\begin{equation}
A_2 = \ddot{I}_{xx} - \ddot{I}_{zz}\,\,,
\end{equation}
with the definitions of $I_{xx}$ and $I_{xx}$ given
in~\cite{Shibata:2006hr}. Since the gravitational wave signal is a
good tracer of the overall dynamics of rotating core collapse, we
extract $A_2$ from our simulations for comparison
with~\cite{Shibata:2006hr}.

In figure~\ref{fig:A2_conv}, we show convergence for the central
density $\rho_c$, the gravitational wave signal $A_2$, the $L_2$-norm
of the Hamiltonian constraint $||H||_2$, and the normalised divergence
of the magnetic field $|\nabla_i B^i / (|B|\cdot\Delta
x^{-1})|_2$. The Hamiltonian constraint in the third panel of
figure~\ref{fig:A2_conv} and the normalised divergence of the magnetic
field in the fourth panel are scaled for first-order
convergence. Before core bounce, second order convergence is achieved,
but reduced to first order after bounce.  Our numerical scheme reduces
to first-order accuracy at shocks, contact discontinuities and local
maxima, i.e. at the centre of the proto-NS, which dominates the
Hamiltonian constraint violation. The gravitational wave signal $A_2$,
the central rest-mass density $\rho_c$, the $L_2$-norm of the
Hamiltonian constraint and the normalised divergence of the magnetic
field are shown for the three resolutions $r0$, $r1$, and $r2$. In the
top two panels, we also show the gravitational wave signal and central
density evolution as extracted and digitised from figures 2 and 17
of~\cite{Shibata:2006hr}. We observe excellent agreement for the
gravitational wave amplitude $A_2$ with the results
of~\cite{Shibata:2006hr}.  The central density evolution
of~\cite{Shibata:2006hr} reaches a peak value of $6.5\cdot
10^{14}\,\mathrm{g}\,\mathrm{cm}^{-3}$ at bounce while we find a
maximum bounce value of $6.4 \cdot
10^{14}\,\mathrm{g}\,\mathrm{cm}^{-3}$.  The post-bounce evolution
differs by an approximately constant offset of $\simeq 6\%$, however
the features in the time evolution itself are very similar.  We
attribute the observed deviations in maximum rest-mass density evolution
to differences in our numerical setup.
  
We find between second-order and first-order convergence for the
gravitational wave signal $A_2$, the central density $\rho_c$, and the
$L_2$-norm of the Hamiltonian constraint $||H||_2$, which is within 
the expected behaviour of the code. The convergence in
both central density and maximum rest-mass evolution are less clean
than for the Hamiltonian constraint due to their oscillatory
nature. The fourth panel of figure~\ref{fig:A2_conv} shows the
evolution of the $L_2$-norm of the normalised divergence of the
magnetic field $|\nabla_i B^i / (|B|\cdot\Delta x^{-1})|_2$. The
constrained transport scheme used in this test maintains the
divergence present in the initial data to round-off precision.
Constraint violations appear only when additional grids are added and
filled by non-constraint-preserving interpolation.

Figure~\ref{fig:A2_bmax} shows the maximum of the magnetic field
components $|B^i|_\mathrm{max}$ as a function of time for the three
resolutions $r0$, $r1$, and $r2$. We observe resolution-dependent
differences in the magnetic field evolution in the postbounce phase.
This is particularly obvious for $B^x$, but also visible in $B^z$.
This behaviour is expected, since the evolution of the magnetic field
is very sensitive to resolution due to effects of turbulence and
instabilities, most importantly the magnetorotational instability
(MRI)~\cite{balbus:91,balbus:98}.  A similar behaviour for the
resolution dependence of $B^i$ is shown in figure 6 of
~\cite{Shibata:2006hr}. Our results for the postbounce magnetic-field
evolution differ quantitatively from \cite{Shibata:2006hr}, which is
expected due to our different numerical setup. While our setup
consists of different layers of mesh refinement with varying
resolution throughout our 3D domain,~\cite{Shibata:2006hr} employ a
uniform axisymmetric grid with a coordinate singularity at the
rotation axis.

\section{Conclusions}\label{sec:conclusions} 

We have presented the implementation and a set of rigorous tests of
ideal general-relativistic magnetohydrodynamics (GRMHD) in the
\GRHydro code, which is part of the Einstein Toolkit. All tests and
results presented in this paper can be reproduced with the code and
input files available at \url{http://einsteintoolkit.org}.

\GRHydro is the first open-source code that provides for fully
self-consistent GRMHD evolutions on dynamical general-relativistic
spacetimes. Its development includes many contributions from a wide
array of collaborators involved with other relativistic codes,
particularly the original version of the general-relativistic
hydrodynamics code {\tt Whisky}~\cite{Baiotti:2004wn}, from which
\GRHydro started, and the publicly available fixed-background {\tt
  HARM} code~\cite{Gammie:2003rj}, from which we have adapted a number
of routines and techniques.  Combined with other functionality from
the Einstein Toolkit, the code may be used to perform GRMHD evolutions
of a broad range of relativistic astrophysical systems. We expect
these to primarily consist of magnetised isolated and binary neutron
stars and collapsing stellar cores, but should the proper initial data
be constructed, accretion disks and tori, white dwarfs, and a number
of other configurations could be simulated as well.

Our implementation of the equations of GRMHD also features the ability
to eliminate spurious divergence in the magnetic field through two
different methods: hyperbolic divergence cleaning and constrained
transport.  While these are clearly sufficient for many problems, one
of the responsibilities in implementing a community-based code is to
incorporate techniques developed by other groups to solve similar
problems.  Along these lines, we foresee work in the future to
incorporate vector potential treatments in which the $B$-field is
computed from the values of an underlying vector potential $A$ that
serves as a fundamental evolved variable.  Such a scheme has recently
been implemented by Etienne~\emph{et
  al.}~\cite{Etienne:2010ui,Etienne:2011re} and Giacomazzo~\emph{et
  al.}~\cite{Giacomazzo:11a}. The former work also included
appropriate electromagnetic gauge conditions for evolutions of compact
object binaries involving neutron stars and other relativistic systems
of interest.

The \GRHydro code now includes several improved higher-order
techniques for reconstruction of states at cell interfaces for use in
high-resolution shock capturing, including three that were described
in the original Einstein Toolkit code paper~\cite{Loffler:2011ay},
TVD, PPM, and ENO, as well as three new methods that have been
implemented and released since, ePPM
\cite{mccorquodale:11,reisswig:13a},
WENO5~\cite{shu:98,Tchekhovskoy:2007zn} and MP5~\cite{suresh:97} (see
\ref{sec:recon-comp}).

In the near future, we will extend the new GRMHD version of \GRHydro
to the generalised multipatch infrastructure of \cite{reisswig:13a},
which will be released as part of the Einstein Toolkit and will enable
simulations on a combination of Cartesian and curvilinear
grids. Further work will be directed towards implementing improved
approximate Riemann solvers for GRMHD. While the Roe~\cite{Roe:1981ar}
and Marquina~\cite{Donat:1996cs,Aloy:1999ne} solvers implemented in
\codename{GRHydro} for purely hydrodynamic evolutions are likely to be
too computationally expensive, requiring an eigenmode breakdown of an
$8\times8$ matrix system at every grid cell, the HLLC solver, which
recovers the contact wave dropped by the HLLE approximation, may be an
attractive and computationally efficient
improvement~\cite{toro:99,Mignone:2006aa}. Additionally, we will
introduce improved techniques for performing conservative-to-primitive
variable conversions to handle cases in which the current routines are
known to fail, particularly those where the Lorentz factor becomes
very large or the fluid becomes dominated by magnetic
pressure~\cite{Mignone:2007fw}.

In the long term, we expect the \GRHydro code and the Einstein Toolkit to
evolve into a toolkit not just for numerical relativity, but for the
broad computational relativistic astrophysics community. Towards this
end, we will include additional physics capabilities in future
versions of the Toolkit. These will include better treatments of
finite-temperature and composition dependence, nuclear reaction
networks, radiation transport of neutrinos and photons, and
improvements over the current ideal GRMHD approximation, following the
most recent developments in resistive
GRMHD~\cite{Dionysopoulou:2012zv,Palenzuela:2012my}.

\ack

We thank Bruno Giacomazzo for use of his exact Riemann solver code, as
well as helpful conversations on a number of topics related to the
work here.  We also thank E.~Abdikamalov, M.~Campanelli, M.~Duez,
Z.~Etienne, B.~Farris, U.~Gamma, P.~Laguna, L.~Lehner, C.~Lousto,
C.~Palenzuela, V.~Paschilidis, J.~Penner, and Y.~Zlochower for useful
conversations related to various aspects of the presented work.

The Einstein Toolkit is directly supported by the National Science
Foundation in the USA under the grant numbers 0903973 / 0903782 / 0904015
(CIGR\@) and 1212401 / 1212426 / 1212433 / 1212460 (Einstein Toolkit).
Related grants contribute directly and indirectly to work in support
of the Einstein Toolkit, including NSF AST-1028087, NSF PHY-1214449, NSF
AST-1212170, 
CAREER NSF PHY-0969855, 
NSF OCI-0905046, 
DMS-0820923, 
NASA 08-ATFP08-0093, an NSERC grant to Erik Schnetter,
and Deutsche Forschungsgemeinschaft
grant SFB/Transregio~7 ``Gravitational-Wave Astronomy''. Christian Ott
acknowledges support from the Alfred P. Sloan Foundation. Roland Haas
acknowledges support by the Natural Sciences and Engineering Council
of Canada. Christian Reisswig acknowledges support by NASA through
Einstein Postdoctoral Fellowship grant number PF2-130099 awarded by
the Chandra X-ray center, which is operated by the Smithsonian
Astrophysical Observatory for NASA under contract NAS8-03060. Tanja Bode
acknowledges support from NSF grants 0941417/12058564.

Portions of this research were conducted with computational resources
provided by Louisiana State University (allocations hpc\_cactus,
hpc\_numrel and hpc\_hyrel), by the Louisiana Optical Network
Initiative (allocations loni\_cactus and loni\_numrel), by the
National Science Foundation through XSEDE resources (allocations
TG-PHY060027N, TG-ASC120003, TG-PHY100033, TG-MCA02N014, and
TG-PHY120016), the Georgia Tech FoRCE cluster, and the Caltech compute
cluster Zwicky (NSF MRI award PHY-0960291).

The multi-dimensional visualizations were generated with the open-source
\codename{VisIt} visualisation package~\cite{Childs:2005ACS,VisIt:web}.
All other figures were
generated with the \codename{Python}-based \codename{matplotlib}
package~\cite{Hunter:2007aa}.

\appendix
\section{Notation in \texorpdfstring{{\tt HARM}}{HARM} and Valencia formulations}\label{app:notation}

The conservative to primitive inversion routine found in the Einstein
Toolkit was originally developed as part of the {\tt HARM} GRMHD
code~\cite{Noble:2005gf}, and reflects its notation internally. For
historical reasons the development of GRMHD in the Einstein Toolkit
follows the Valencia formulation of
equations~\cite{Marti:1991wi,Banyuls:1997zz,Ibanez:2001:godunov,Font:2007zz,Anton:2005gi},
and this notation was used throughout the first paper summarising the
Einstein Toolkit~\cite{Loffler:2011ay}. To convert from one set of
notation to the other, we note that the equivalence of various
physical quantities in Table~\ref{table:harmvalencia}.

\begin{table}
\caption{Overview of differences in variable notation between {\tt HARM} and Valencia.}\label{table:harmvalencia}
\begin{indented}
\item[]\begin{tabular}{p{0.4\linewidth}lll}
\br
Quantity & Valencia/ET & {\tt HARM}\\
\mr
Pressure & $P$ & $p$\\
Energy density & $\rho h\equiv \rho(1+\epsilon)+P$ & $w$\\
3-velocity for a normal observer & $\displaystyle v^i\equiv \frac{u^i}{\gamma}+\frac{\beta^i}{\alpha}$ & $v^i$ \\
Lorentz factor & $W$ & $\gamma$\\
Projected velocity &  $Wv^i$ & $\displaystyle \tilde{u}^i\equiv j^i_\mu u^\mu = u^i+\frac{W\beta^i}{\alpha}$\\
Momentum density for a normal observer & $\mathcal{S}_\mu\equiv\alpha T^0_\mu$ & $Q_\mu$\\
 Projected momentum density & $\tilde{\mathcal{S}}_\mu\equiv (\delta_\mu^\nu+n_\mu n^\nu)\mathcal{S}_\nu$ & $\tilde{Q}_\mu$\\
Auxiliary Con2prim variable & $Q\equiv \rho hW^2$ & $W$\\
\br
\end{tabular}
\end{indented}
\end{table}

\section{Comparison of reconstruction methods}
\label{sec:recon-comp}

We briefly compare the new reconstruction methods
(section~\ref{sec:numerical_reconstruction}) that have been
implemented in {\tt GRHydro}.  In~\cite{reisswig:13a} a detailed
comparison between ePPM and oPPM for cell-centred and vertex-centred
AMR was performed.  These authors found that ePPM is superior to oPPM
in all cases they considered, and especially in those using
cell-centred AMR.  Here, we compare oPPM, ePPM, WENO5, and MP5
reconstruction using the example of an magnetised dynamically
evolved TOV star using vertex-centred AMR.  We use 7 levels of mesh
refinement with cubical extent
$[400M_\odot,200M_\odot,50M_\odot,30M_\odot,19M_\odot,12M_\odot]$ and
employ octant symmetry. The finest level has a resolution of $\Delta x
= 0.2M_\odot$ and contains the entire star. We use the HLLE Riemann
solver.  We evolve spacetime with a $\Gamma$-driver gauge parameter of
$\eta=0.5$ and 5th-order Kreiss-Oliger dissipation with $\epsilon=0.1$
is added to the right-hand side of spacetime variable.  We integrate
in time with a fourth-order Runge-Kutta integrator and a Courant
factor of $0.4$.  We construct the initial initial data by solving the
TOV equations using a polytropic EOS with $\Gamma=2$ and scale
parameter $K=100\,M_\odot$.  The central density is set to
$\rho_c=1.28\times10^{-3}\, M_\odot^{-2}$. During the evolution, we
use a $\Gamma$-law EOS with $\Gamma=2$.

\begin{figure}
 \includegraphics[width=0.8\textwidth]{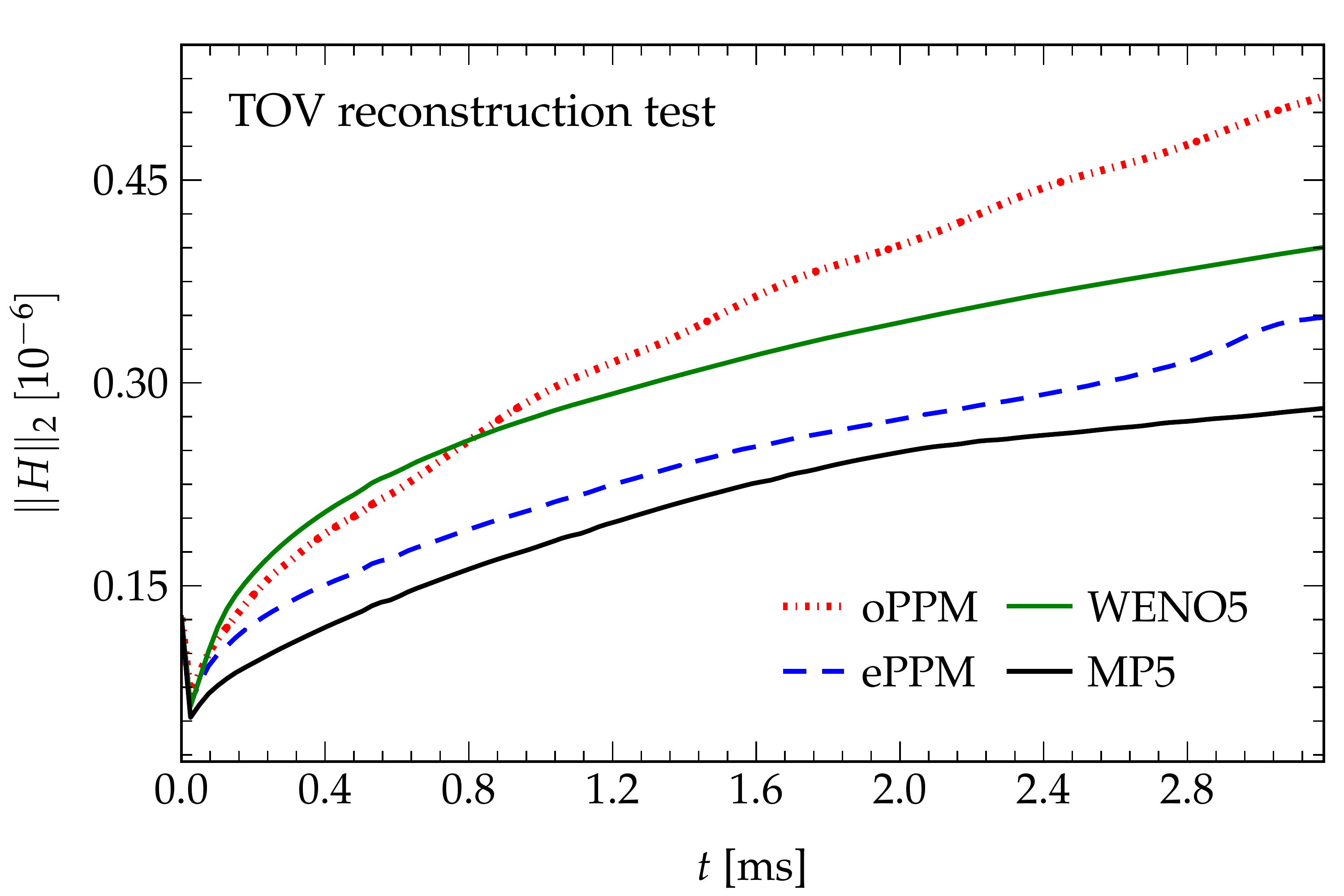}
\caption{Comparison of oPPM, ePPM, WENO5, and MP5 reconstruction based
  on the example of a dynamically evolved TOV star.  We show the
  $L_2$-norm of the Hamiltonian constraint $|| H ||_2$. MP5 performs
  best, while oPPM performs worst in this simple test.}
\label{fig:tov-recon}
\end{figure}

In figure~\ref{fig:tov-recon}, we show the $L_2$-norm of the
Hamiltonian constraint $|| H ||_2$ as a function of time.  All
simulation settings are identical, except for the reconstruction
method.  We find that MP5 reconstruction performs better than ePPM or
WENO5 in this simple test. WENO5 performs worse than ePPM and MP5. The
worst results are obtained with oPPM which reduces to first order at
the smooth density maximum at the center of the star.  In the
vertex-centred case, however, the error due to oPPM reconstruction is
not as large as in the cell-centred case~\cite{reisswig:13a}, since
in the former, the central density maximum is approximately located on
a grid point.  With oPPM, the dominant source of error is thus
generated at the center of the star.  The other reconstruction methods
maintain high order at the smooth density maximum.  With them, the
dominant source of error is generated at the steep density drop at the
surface of the star.  We emphasise, however, that this test is only
one of many possible tests. A more detailed comparison would be
necessary for more general conclusions.

\section{Full set of planar MHD shocktubes}
\label{sec:allshocktubes}

The five test cases collected in~\cite{Balsara:2001aa} are widely used
to test new MHD codes. We verified that our code passes all tests and
compare the results to the exact solution of~\cite{Giacomazzo:2007ti}.
The main text~\ref{sec:shocktubes} contains detailed description of our
test setup.

\begin{figure}
 \includegraphics[width=0.9\textwidth]{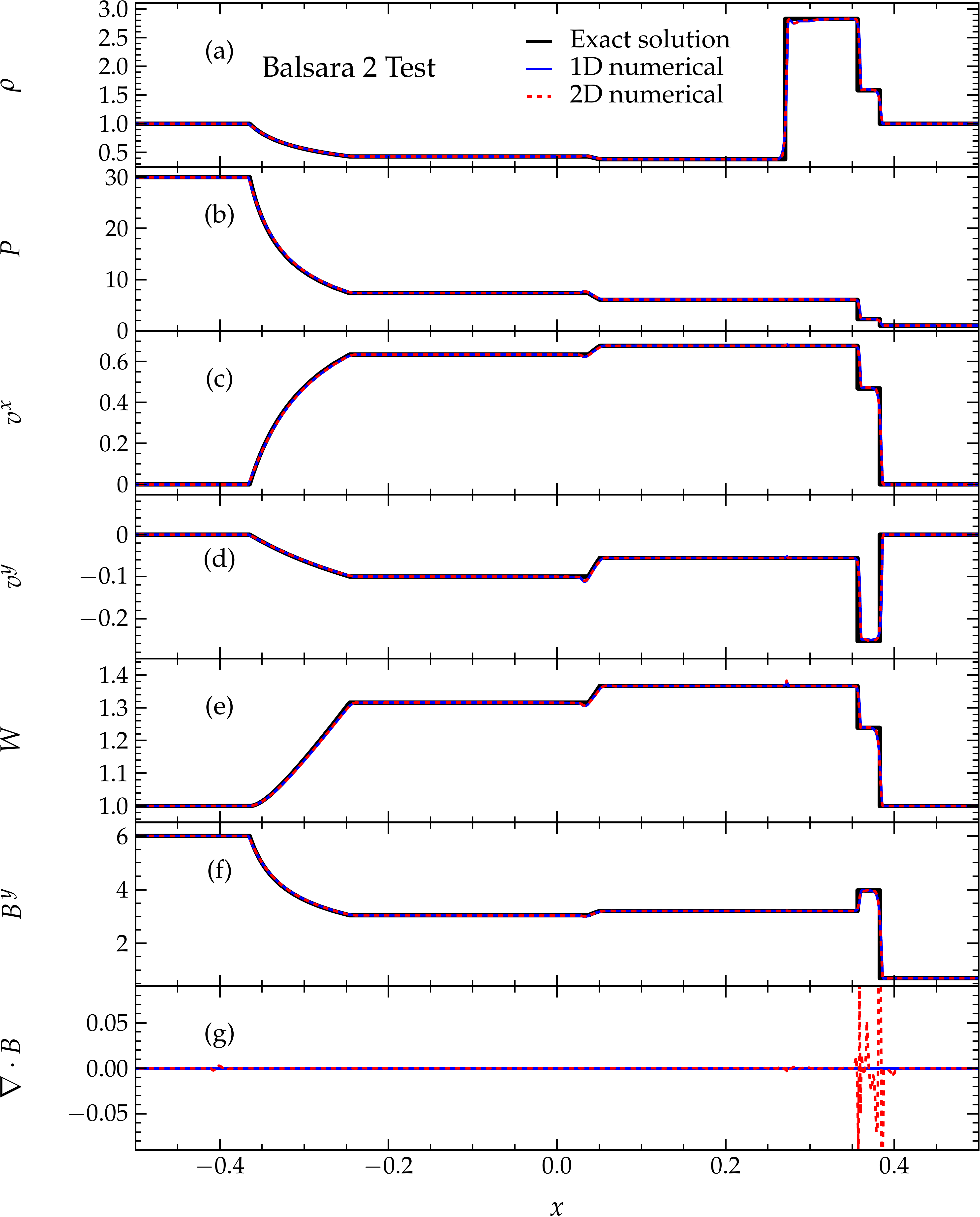}
 \caption{Evolution of the Balsara2 shocktube case, performed with
   divergence cleaning, with all conventions as in
   figure~\protect\ref{fig:shock1}. The numerical code reproduces the
   analytical result very well, with only some oscillations visible
   in the density $\rho$ near the shock front, capturing the sharp shock
   features at least as well as~\cite{Balsara:2001aa}.  A more detailed
   description of the test setup and parameters can be found in the main
   text in section~\ref{sec:shocktubes}.}
 \label{fig:shock2}
\end{figure}

\begin{figure}
 \includegraphics[width=0.9\textwidth]{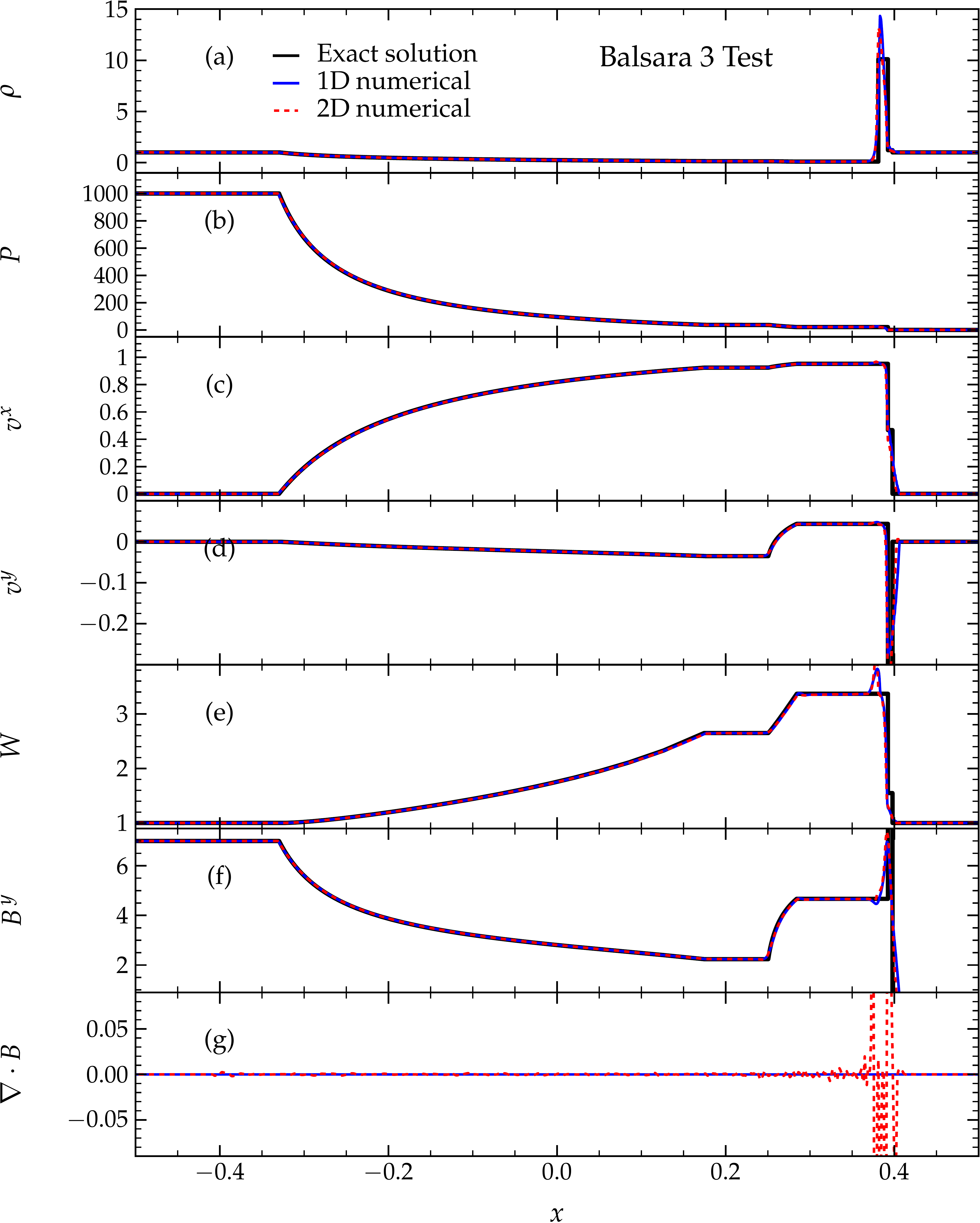}
 \caption{Evolution of the Balsara3 shocktube case, performed with divergence cleaning, with all conventions as in figure~\protect\ref{fig:shock1}.
 This test proved more demanding
   on our code, in particular the narrow right moving fast shock is not
   resolved well using 1600 grid points, resulting in a sizeable overshoot near
   the shock. This feature
   is common to other codes~\cite{Balsara:2001aa} independent of the specific
   formulation used. A more detailed
   description of the test setup and parameters can be found in the main
   text in section~\ref{sec:shocktubes}.}
 \label{fig:shock3}
\end{figure}

\begin{figure}
  \includegraphics[width=0.9\textwidth]{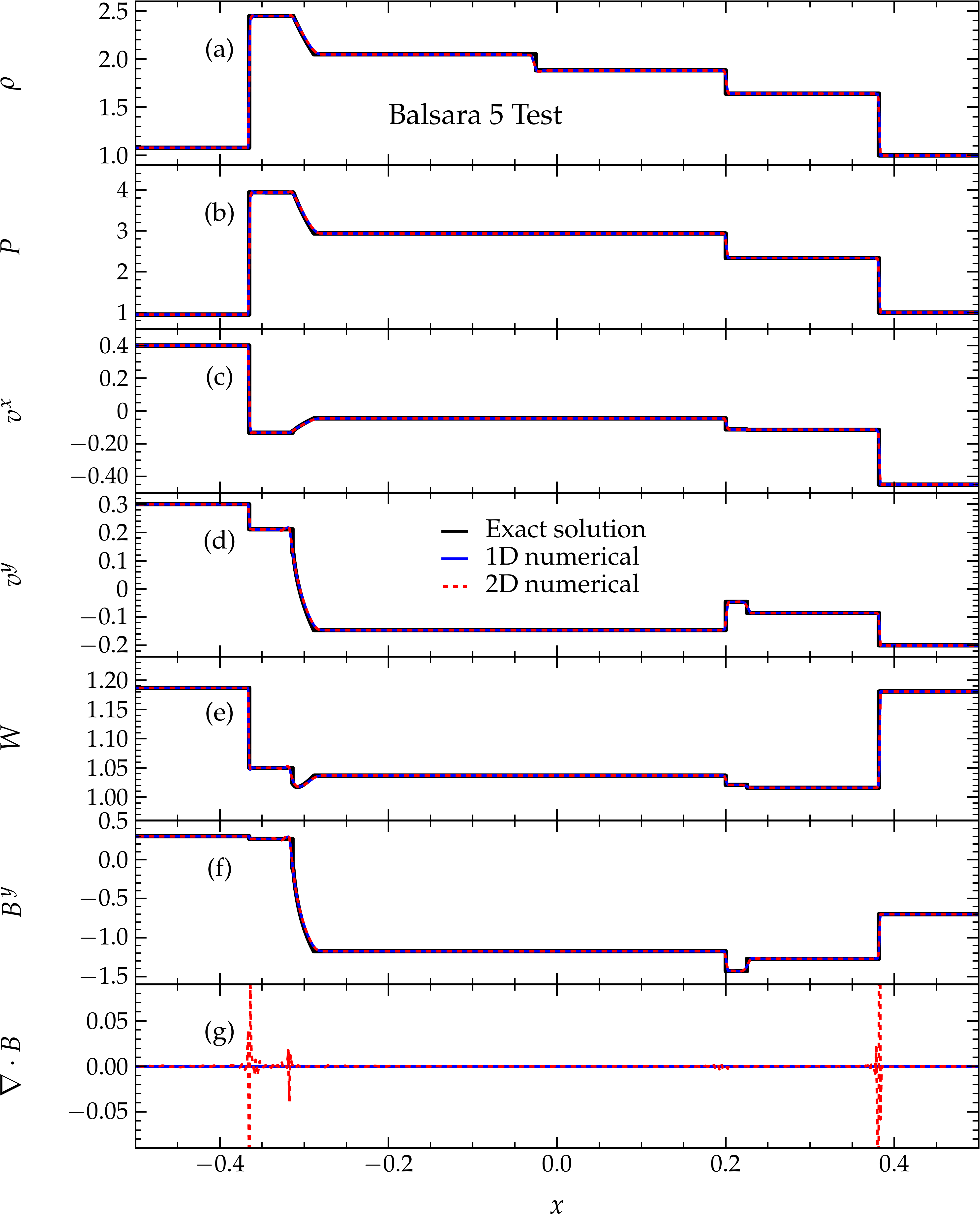}
 \caption{Evolution of the Balsara5 shocktube case, performed with
   divergence cleaning, with all conventions as in
   figure~\protect\ref{fig:shock1}. A more detailed
   description of the test setup and parameters can be found in the main
   text in section~\ref{sec:shocktubes}.}
 \label{fig:shock5}
\end{figure}

In figures~\ref{fig:shock2} and~\ref{fig:shock3}
we present results for the  moderate and strong blast wave tests
of~\cite{Balsara:2001aa} which prove much more challenging to the code due to
the Lorentz contraction of the fast moving shock.
Finally in figure~\ref{fig:shock5}
we present the non-coplanar test problem of~\cite{Balsara:2001aa}.

\section{Monopole tests modelling high frequency noise}
\label{sec:altmonopole}

We also include initial data in our tests that models high-frequency
numerical noise
in the magnetic field. Here, the spatial dependence of $B^x$
is set as in section~\ref{sec:monopole} and the other components of
the $B$-field
are set to the same amplitude or to zero, with non-zero components 
 multiplied by $(-1)^{N(i,j,k)}$, where $N(i,j,k)$ is
a function of the indices $i$, $j$, and $k$ of the grid cell in the
$x$, $y$, and $z$-directions, to produce an alternating pattern.  For
a one-dimensional alternating pattern we set $N(i,j,k)=i+j+k$, and
only set $B^x$ to be non-zero and alternating.  For a two-dimensional
test, we use the same function for $N(i,j,k)$ and apply it to $B^x$ to
make it alternate, but set $B^y$ to the original Gaussian state and
$B^z=0$.  For a three-dimensional test, we set $N(i,j,k)=i+j$, so the
alternation is two-dimensional in character, but set $B^x=B^y=B^z$ and
allow all three to alternate.

\begin{figure}[t]
 \includegraphics[width=0.9\textwidth]{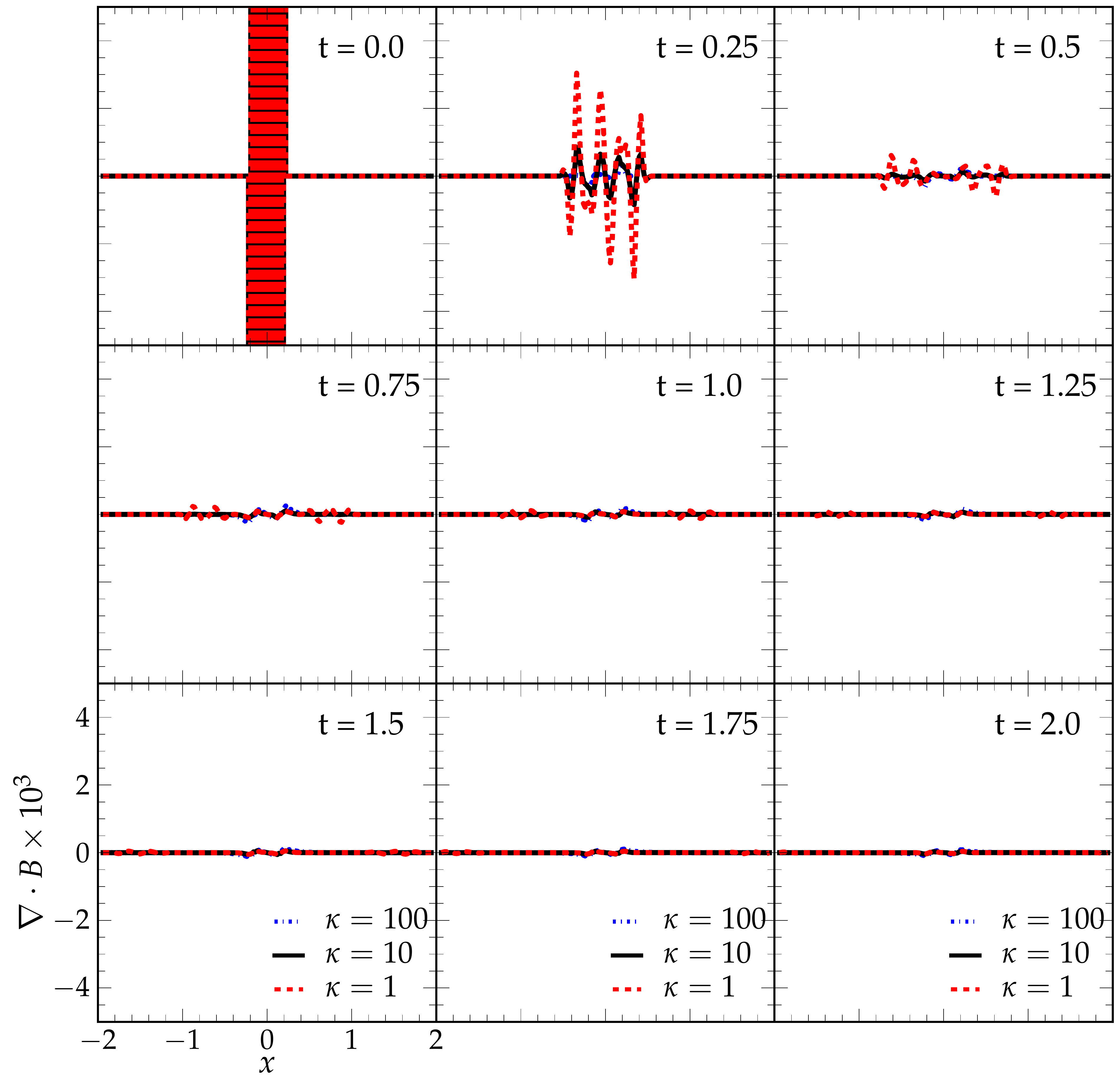}
 \caption{Behaviour of our divergence cleaning scheme,
   demonstrated by monopole damping and advection for a three-dimensional 
   alternating Gaussian configuration.  Conventions are the same as in
   figure~\ref{fig:monopole_gauss}.  We still see more significant
   divergences travelling further from their initial location for
   $\kappa=1$, but damping is markedly faster for the case
   $\kappa=100$ when dealing with high-frequency noise as compared to
   the lower-frequency modes characterising the Gaussian monopole,
   seemingly because of cancellation effects from neighbouring points.
   The choice $\kappa=10$ still yields the best results for
   numerically eliminating spurious divergence of the magnetic field.
\label{fig:monopole_altgauss}} 
\end{figure}

Figure~\ref{fig:monopole_altgauss} shows the performance of the
divergence cleaning algorithm in removing high frequency constraint
violating noise such as noise generated during the numerical evolution.
Clearly the algorithm is very efficient in damping out the high
frequency constraint violations, reducing them much more quickly than
the simple Gaussian pulse present in figure~\ref{fig:monopole_gauss}.

\clearpage
\providecommand{\newblock}{}

\end{document}